\documentclass[reprint,prl,aps]{revtex4-2}
\usepackage{graphicx}
\usepackage{color}
\usepackage{dcolumn}
\usepackage{bm}
\usepackage{amssymb}
\usepackage{color,amsmath}
\usepackage{mathrsfs}
\usepackage{soul,xcolor} 
\setstcolor{red} 
\usepackage{epstopdf}
\usepackage{gensymb}
\usepackage{lineno}
\usepackage[hypertexnames=false]{hyperref}
\hypersetup{hidelinks}
\usepackage{xurl}
\newcommand{\eff}{\mathrm{eff}}
\usepackage{soul}

\begin{document}

\title{Current Induced Hidden States in Josephson Junctions}

\author{Shaowen Chen$^{1\dagger*}$}
\author{Seunghyun Park$^{1\dagger}$}
\author{Uri Vool$^{1,2}$}
\author{Nikola Maksimovic$^{1}$} 
\author{David A. Broadway$^{3}$}
\author{Mykhailo Flaks$^{3}$}
\author{Tony X. Zhou$^{1\ddagger}$} 
\author{Patrick Maletinsky$^{3}$}
\author{Ady Stern$^{4}$} 
\author{Bertrand I. Halperin$^{1}$}
\author{Amir Yacoby$^{1*}$}

\affiliation{$^{1}$Department of Physics, Harvard University, Cambridge MA, 02138, USA}
\affiliation{$^{2}$Max Planck Institute for Chemical Physics of Solids, 01187 Dresden, Germany}
\affiliation{$^{3}$Department of Physics, University of Basel, Klingelbergstrasse 82, Basel CH-4056, Switzerland}
\affiliation{$^{4}$Weizmann Institute of Science, Rehovot, 76100, Israel}
\affiliation{$^{\dagger}$These authors contributed equally to this work,}
\affiliation{$^{*}$shaowenchen@g.harvard.edu (S.C.); yacoby@g.harvard.edu (A.Y.)}
\affiliation{$^{\ddagger}$Present address: Northrop Grumman Mission Systems, Linthicum
MD, 21090, USA}

\maketitle

\addcontentsline{toc}{section}{Main Text}

\textbf{Josephson junctions enable dissipation-less electrical current through metals and insulators below a critical current. Despite being central to quantum technology based on superconducting quantum bits and fundamental research into self-conjugate quasiparticles, the spatial distribution of super current flow at the junction and its predicted evolution with current bias and external magnetic field remain experimentally elusive. Revealing the hidden current flow, featureless in electrical resistance, helps understanding unconventional phenomena such as the nonreciprocal critical current, i.e., Josephson diode effect. Here we introduce a platform to visualize super current flow at the nanoscale. Utilizing a scanning magnetometer based on nitrogen vacancy centers in diamond, we uncover competing ground states electrically switchable within the zero-resistance regime. The competition results from the superconducting phase re-configuration induced by the Josephson current and kinetic inductance of thin-film superconductors. We further identify a new mechanism for the Josephson diode effect involving the Josephson current induced phase. The nanoscale super current flow emerges as a new experimental observable for elucidating unconventional superconductivity, and optimizing quantum computation and energy-efficient devices.}

\subsection*{Introduction}

Characterization and control over the super current flow is critical for Josephson junctions (JJs) \cite{josephson1962possible,de1964boundary,josephson1964coupled}, which have become a building block in quantum and classical technology \cite{devoret2013superconducting,fedorov2014fluxon,kjaergaard2020superconducting,wang2010coherent,welp2013superconducting,kirtley1999scanning,kirtley2010fundamental,holmes2013energy} while remained a rich area of exploration into fundamental particles \cite{grosfeld2011observing,beenakker2013search,lutchyn2018majorana} and unconventional superconductivity \cite{pal2022josephson,jeon2022zero,nadeem2023superconducting}. Compared to spectroscopic probes that measures the amplitude of the superconducting (SC) wavefunction \cite{roditchev2015direct}, the super current flow encodes the SC phase. Mapping the spatial distribution of super current has revealed the pairing symmetry of unconventional superconductors \cite{tsuei1994pairing,hilgenkamp2003ordering}, and recently identified screening current as the source of SC diode effect in SC/ferromagnet structures \cite{gutfreund2023direct}. In addition, the local super current flow affects device parameters such as the impedance of SC circuits and anharmonicity of SC qubits due to the change in kinetic inductance \cite{tinkham1996introduction}. Despite the scientific and technological relevance, direct visualization of the Josephson current flow and its response to external tuning knobs such as bias current and magnetic field remain experimentally beyond reach \cite{gross1994low,roditchev2015direct,dremov2019local,hovhannisyan2021lateral,grebenchuk2020observation,stolyarov2022revealing}. This is mostly due to the sensitive nature of the JJ which responds to small perturbations and the nanoscale spatial resolution needed to resolve the evolution of the super current flow. To date, JJ characterization has primarily relied on indirect measurements such as the critical current that separates the dissipation-less (zero electrical resistance) and resistive states. However, this only provides insight into the resistive state while the ground state below the critical current stay hidden. 

Here we quantitatively visualize the current flow in a JJ device with nano-scale resolution. The spatial distribution of Josephson current flow can be modulated by varying the SC phase difference between two sides of the junction. In any JJ, the SC phase difference is governed by three factors: (i) external magnetic field; (ii) external bias current; (iii) self-field or SC phase gradient induced by the finite Josephson current density. 
Our measurements reveal the evolution of Josephson current flow with all three factors, including features associated with the change of the number of current loops at the junction known as the Josephson vortex (JV). In particular, factors (i) and (ii) can affect (iii), altering the super current flow even without detectable transport features. We find two previously unidentified effects of the Josephson current induced phase from factor (iii). First, hidden ground states with different number of JVs are found within the zero resistance state, which can be electrically switched below the critical current. Second, a new mechanism for the Josephson diode effect is established based on the second harmonic phase terms induced by the Josephson current when time-reversal and inversion symmetry are broken.

The measurement setup is shown in Fig. \ref{fig:1}a. We employ a diamond tip containing a single nitrogen-vacancy (NV) center to map the local magnetic field generated by the current flow \cite{vool2021imaging}. The results are obtained from two devices with junction width $W = 0.15$ and $0.2~\mu$m, length $L = 1.5 ~\mu$m and thickness $t=35$ nm. The SC electrodes are measured to be in the thin film limit $L \ll \lambda_p$, where $\lambda_p$ is the Pearl length (Supplementary Fig. \ref{fig:S_Pearl}). This suggests the factor (iii) contribution in our device comes from the Josephson current induced phase associated with the kinetic inductance of the SC film, instead of the self-field effect. The junctions are diffusive (electron mean free path $l_\mathrm{mfp} < W$) and over-damped (no hysteresis during bias current sweeps). Throughout the paper, we refer to the transverse (longitudinal) direction as $x(y)$, and the direction perpendicular to the plane as $z$. The origin $x=y=0$ is set to the center of JJ.

\subsection*{Results}

\subsubsection{Current Induced Phase}

The Josephson current density can be modeled by the sinusoidal current-phase relation \cite{tinkham1996introduction}
\begin{equation}
    J_y(x) = J_c \sin{[\phi(x)]},
\label{eqn:1}
\end{equation}
where $J_c$ is the Josephson critical current density (assumed constant for now), and $\phi(x)$ is the phase difference across the JJ at position $x$. $\phi(x)=\phi_e(x) + \phi_{\mathrm{bias}}$, where $\phi_e(x)$ arises from the external magnetic field $B_z$ (factor i), and $\phi_{\mathrm{bias}}$ is the additional phase difference due to the injected bias current (factor ii). The strength of Josephson current induced phase (factor iii) is regulated by $J_c$. For small $J_c$, the Josephson penetration length $\lambda_J \approx \sqrt{{\Phi_0 Lt}/{4\pi\mu_0J_c \lambda_L^2}}  \gg L$ \cite{tolpygo1996critical}, the Josephson current induced phase can be neglected (``weak junction'' limit). $\Phi_0$ is the flux quantum, $\lambda_L$ is the London penetration length.

In the weak junction limit, external $B_z$ controls the number of JV. The transport critical current $I_c$ oscillates and reaches zero at nodes $B_z=\pm B_n$ ($n$ is integer). It is known as the ``Fraunhofer map'' \cite{rowell1963magnetic,clem2010josephson}. In each lobe where $B_{n-1}<|B_z|<B_{n}$, there are $n$ JVs at the junction; in the central lobe there is 0 JV (Fig. \ref{fig:1}b); the only way to change the number of JV is by sweeping $B_z$ through the nodes $B_n$ (Supplementary Fig. \ref{fig:S_extfluxprofile}). 
In weak junctions the first term $\phi_e (x)$ is induced by the screening current $J_x$ in the SC electrodes, and scaled by $\phi_{e0} \equiv \phi_e|_{x=L/2} \approx 1.7 B_z L^2 / \Phi_0$ \cite{clem2010josephson} (Supplementary Eqn. \ref{eqn:nonlocalphie}). The second term $\phi_{\mathrm{bias}}$ changes from $-\pi/2+n\pi$ to $\pi/2+n\pi$ when $I_\mathrm{bias}$ sweeps from $-|I_c|$ to $+|I_c|$ (Fig. \ref{fig:1}c-d), which can be viewed as moving the JV along the $x$-direction.

In ``strong junctions'' ($\lambda_J \ll L$), $\phi(x)$ is altered by the Josephson current induced phase and lacks analytical solutions. Qualitatively, the screening current $J_x$ deviates from the weak junction limit by an amount proportional to the Josephson current $J_y$ (Fig. \ref{fig:1}e), due to the continuity of current. This leads to additional phase gradient $\partial \theta/ \partial x \propto \mathcal{L}_k J_x$, which changes $\phi_e(x) = \theta(x)|_{y=W/2} - \theta(x)|_{y=-W/2}$. Here $\mathcal{L}_k$ is the kinetic inductance, which is inversely proportional to the superfluid stiffness. $\theta$ is the SC phase. Specifically, the larger $J_x$ in the 1-JV state leads to enhanced $\phi_{e0}$, compared to the 0-JV state at the same external magnetic field (Fig. \ref{fig:1}f). $I_\mathrm{bias}$ can further change the Josephson current and its induced phase. Thus in strong junctions, the total $\phi(x)$ and current flow need to be solved self-consistently. We find that our device is close to the weak-junction limit at $T=7$ K, but is in an intermediate range at $T=4$ K.

\subsubsection{Visualizing Josephson Current Flow}

To optimize magnetic field sensitivity, we utilize the ``ac magnetometry" protocol, synchronizing NV control pulses with the signal (Fig. \ref{fig:1}g). In this protocol, different bias currents $I_1$ and $I_2$ is applied to the junction during two halves of each cycle. The magnetic field generated by $I_k$ rotates the prepared NV spin superposition state along the equator of the Bloch sphere by an angle $\varphi_k = 2\pi \gamma_e b_{\mathrm{nv}} \tau/2$, where $\gamma_e$ is the gyromagnetic ratio of electron spin, $b_{\mathrm{nv}}$ is the magnetic field projected along NV axis, and $\tau$ is cycle duration. After the sequence, the accumulated angle is $\varphi_1 - \varphi_2$ so each measurement records the \textit{difference} between two selected scenarios. $b_\mathrm{nv}$ is converted to vector components of the magnetic field $b_{x,y,z}$ using a Fourier method (see Methods). The current vector $(j_x, j_y)$ is then reconstructed with the Fourier method using in-plane components of the magnetic field. We find similar results with regularization and machine learning methods (see \ref{SI:recondetail}). 

We start at $T=7$ K where the transport result suggests a weak junction (Fig. \ref{fig:1}b). To visualize the evolution of Josephson current, we use two types of sequences. In the first sequence, we highlight the effect of $\phi_\mathrm{bias}$ by taking the difference between finite and zero $I_{\mathrm{bias}}$ (schematics in Fig. \ref{fig:2}a). The expected current profiles $j_y(x)$ are shown in Fig. \ref{fig:2}a, by subtracting the relevant lines in Fig. \ref{fig:1}c. The sign of $I_{\mathrm{bias}}$ determines the direction of the profile shift and the amplitude determines the amount of the shift. Fig. \ref{fig:2}b-c show measurements using $I_{\mathrm{bias}} \approx \pm|I_c|$ in the sequence while no JV is in the junction. As expected, current features are seen at opposite side of the junction for $\pm I_{\mathrm{bias}}$. $j_y(x)$ at the junction also shows the lateral movement when sweeping $I_{\mathrm{bias}}$ (Supplementary Fig. \ref{fig:E_iphilines}). $\phi_\mathrm{bias}$ can be acquired by fitting $j_y(x)$ to the calculated profiles from Fig. \ref{fig:2}a (Supplementary Eqn. \ref{eqn:deltaphifit}). The result agrees with the sinusoidal current phase relation (Fig. \ref{fig:2}d).

Next we show the effect of magnetic flux on the Josephson current by taking the difference between $\pm I_{\mathrm{bias}}$ (schematics in Fig. \ref{fig:2}e). Here the expected signals are cosine-like and only grow in amplitude with $I_\mathrm{bias}$ (Fig. \ref{fig:2}e). For the same $I_{\mathrm{bias}}$ direction, the signal flips sign for 0- and 1-JV states (switching from red to blue branch in Fig. \ref{fig:1}b). Measurements using this sequence are shown in Fig. \ref{fig:2}f-g, where we use $I_{\mathrm{bias}} \approx |I_c|$ in both cases. Results measured at $I_{\mathrm{bias}}\approx 0.5|I_c|$ show the same shape with half the amplitude (Supplementary Fig. \ref{fig:E_0.5Ic}a-b). We repeat the measurement at various magnetic fields, and the current profile reversal can be seen when external $B_z$ crosses the node $B_0 \approx 1.5$ mT, i.e., when JV number changes by 1 (Supplementary Fig. \ref{fig:E_7Kallmap}). Notably, the current flow at $x=0$ is parallel (anti-parallel) to $I_{\mathrm{bias}}$ when the junction contains even (odd) number of JVs. In the 2-JV state, the current flow at $x=0$ and bias current are both positive (Supplementary Fig. \ref{fig:E_phi2}), like the 0-JV state (Fig. \ref{fig:2}f).

Our measurements provide quantitative details of the current flow compared to previous methods \cite{dynes1971supercurrent,mayer1993magnetic,holm1995novel,dremov2019local}. 
First, we confirm the SC is in the thin film limit ($L \ll \lambda_p$). The absolute value of $\lambda_p \approx 13~\mu$m at $T=7$ K is directly measured from the stray field (Supplementary Fig. \ref{fig:S_Pearl}). In comparison, indirect measurements of $\lambda_p$ range from 1 to 5 $\mu$m (see Methods), and thus are unable to determine whether the SC is in the thin film regime.
Second, the JV extends into the SC electrode on both sides by about 350 nm (Supplementary Fig. \ref{fig:E_0.5Ic}c), consistent with the effective area expected from the nodes $B_n$. However, the measured $j_y(x)$ profile does not match the expectation under external $B_z$. For example, at $B_z=1.1$ mT the $j_y(x)$ in Fig. \ref{fig:2}f is expected to be negative at $x=\pm L/2$, but stays positive in the experiment, suggesting a smaller than expected $\phi_{e0}$ (Supplementary Fig. \ref{fig:E_7Kallmap}f). 

We introduce an effective phase difference $\phi_{\eff}$ as a fitting parameter for the measured $j_y(x)$, replacing the theoretically predicted $\phi_{e0}$ (Supplementary Eqn. \ref{eqn:phiefffit}). Fig. \ref{fig:2}h shows $\phi_{\eff}$ is lower (higher) than the phase induced by the external field $\phi_{\mathrm{ext}}$ when $B_z < B_0$ ($B_z>B_0$). The discrepancy is a direct consequence of the Josephson current induced phase, as expected in strong junctions. While we find $\lambda_J$ is comparable to $L$ when calculated using experimentally measured parameters, no strong junction feature is observed in the ``Fraunhofer map" (Fig. \ref{fig:1}b). The Josephson current induced phase only causes small changes to $I_c$, and the effect cancels out when tracking $I_c$ over large range of $B_z$ \cite{dynes1971supercurrent}. However, such effect is still pertinent to designing SC devices such as JJ arrays because the inductance of each junction is affected by the super current flow \cite{kuzmin2023tuning}.

\subsubsection{Electrically Switching JV Below $|I_c|$}

The ``Fraunhofer map'' changes when measured at $T=4$ K. The nodes of $|I_c|$ at $B_n$ are lifted although sharp kinks remain (Fig. \ref{fig:3}a). This could be caused by a combination of reasons, including an asymmetric critical current density, $J_c(x) \neq J_c(-x)$ and the strong junction effect due to the increased $|I_c|$ \cite{mayer1993magnetic,hilgenkamp2002grain,boris2013evidence}. However, the precise mechanism remains difficult to dissect owing to the zero resistance below $I_c$. Here the origin of non-zero local minima of $|I_c|$ is revealed by mapping the current flow. A differential magnetic field $\Tilde{b}_{\mathrm{nv}}$ is measured using a sequence that senses the small ac bias ($\Tilde{i}_{\mathrm{ac}}$) response while fixing the dc bias ($I_{\mathrm{dc}}$), shown by the schematic drawing above Fig. \ref{fig:3}b. The $\Tilde{b}_{\mathrm{nv}}$ is measured around the kinks of $|I_c|$ while the NV is fixed over the center of the junction (Fig. \ref{fig:3}b main panel). Abrupt changes of $\Tilde{b}_{\mathrm{nv}}$ at large $|I_{\mathrm{dc}}|$ match the transport $I_c$ (circles in Fig. \ref{fig:3}b). This suggests that the junction is minimally perturbed by the NV magnetometer.

An additional sharp boundary of $\Tilde{b}_{\mathrm{nv}}$ below $I_c$ separates the 0- and 1-JV states. This is confirmed by the spatial maps in Fig. \ref{fig:3}d-e. In Fig. \ref{fig:3}d (0-JV state), the current is almost uniform despite being measured at higher external $B_z$ than Fig. \ref{fig:2}f, suggesting $\phi_{e0}$ is strongly suppressed by the Josephson current induced phase. Intriguingly, the 0- and 1-JV states can be reached at the same $B_z$ but different $I_\mathrm{bias}$ (Fig. \ref{fig:3}d-e). Measuring the difference between two such $I_\mathrm{bias}$, the current profile that corresponds to the JV number changing by 1 is observed (Fig. \ref{fig:3}f). This demonstrates precise JV number control using pure electric means while staying below $I_c$, which could be useful for low-dissipation memory and logic devices based on SC hybrid structures.

The phase boundary below $I_c$ supports Josephson current induced phase as the primary reason for the node-lifting in our device. The induced phase enables the co-stability of the 0- and 1-JV states originating from the existence of two local minima of the energy as a function of the order parameter $\Psi(\textbf{r})$ at the same external magnetic field, as shown by the time-dependent Ginzburg-Landau (TDGL) simulation (Fig. \ref{fig:3}c, for details see \ref{SI:TDGL}). The overlapping states with different number of JVs were predicted to arise from the self-field effect previously \cite{owen1967vortex,pagano1991magnetic,kuplevakhsky2006static,kuplevakhsky2007exact,kuplevakhsky2010exact}. However, we do not expect self-field to be the main effect here. The measured self-field of the current is insignificant in our device (less than 5\% of the external field), and the SC is still in the thin film limit at $T=4$ K ($L \ll \lambda_p$, see Methods). Furthermore, the boundary of the 0- and 1-JV phase diagram only extends from the 1-JV region (Fig. \ref{fig:3}b) in the experiment. This is independent of sweeping directions of $B_z$ or $I_{\mathrm{dc}}$ (Supplementary Fig. \ref{fig:E_bzac_extra}a), and similar behavior is observed between the 1- and 2-JV states (Supplementary Fig. \ref{fig:E_bzac_extra}b). The lack of hysteresis is quite unexpected. One possibility is that the JJ relaxes to the ground state with lower energy due to the elevated temperature and small perturbations of the measurement, although the detailed mechanism is an open question for future work. The simulated Gibbs free energy difference $\Delta \varepsilon = \varepsilon_\mathrm{0V}-\varepsilon_\mathrm{1V}$ shows the 1-JV state has lower energy than the 0-JV state in most, but not all of the overlap region (Fig. \ref{fig:3}c). In fact, including the self-field effect energetically favors the 0-JV over the 1-JV state, further deviating from the experimental results (Supplementary Fig. \ref{fig:S_TDGL_sfe}).

\subsubsection{Inversion Asymmetry and Josephson Diode Effect}

The transport $I_c$ is nonreciprocal when $I_{\mathrm{bias}}$ is applied in opposite directions at $T=4$ K, i.e. $|I_c^+| \neq |I_c^-|$ (Fig. \ref{fig:4}a). This is referred to as the ``Josephson diode effect'' \cite{moll2023evolution}. The asymmetry parameter, $\eta = \frac{|I_c^+|-|I_c^-|}{|I_c^+|+|I_c^-|}$, exceeds 10\% in our device. We identify a new mechanism for the diode effect comprising three ingredients, (i) time reversal symmetry breaking (by $B_\mathrm{z}$), (ii) inversion symmetry breaking, and (iii) Josephson current induced phase. The first two conditions are required by symmetry \cite{zhang2022general,nadeem2023superconducting}, while the third provides a mechanism whereby the Josephson current is not a simple sinusoidal function of $\phi_\mathrm{bias}$. As a result the critical current density is reached on opposite sides of the junction at $\pm I_{\mathrm{bias}}$, which leads to asymmetric $I_c^\pm$ (Fig. \ref{fig:4}b).
Interestingly, it can be shown theoretically that first two ingredients alone are not sufficient to generate the diode effect (\ref{SI:JDE_2JJ}). Combining the symmetry breaking with the Josephson current induced phase introduces higher harmonic terms with a phase offset in the current-phase relation (\ref{SI:JDE_2JJ}). Our results confirm all three components are necessary. For example, the current profile measured at $T = 7$ K is asymmetric, $j_y(x)\neq j_y(-x)$, suggesting broken inversion symmetry (Supplementary Fig. \ref{fig:S_asymm7Kfit}). However, the weaker current induced phase due to smaller $J_c$ is insufficient to generate a non-reciprocal global critical current response (\ref{SI:nondiode7K}).

The current flow measurement directly reveals the broken inversion symmetry even when the global $I_c^{\pm}$ is almost symmetric. The $\Tilde{b}_{\mathrm{nv}}$ map at $\pm I_{\mathrm{bias}}$ is measured at $B_z=0.5$ mT. Although $\eta$ is only about $2\%$, the current flow pattern is clearly asymmetric for $\pm I_\mathrm{bias}$; a loop appears near the left edge for $-I_{\mathrm{bias}}$ (Fig. \ref{fig:4}c) but not for $+I_{\mathrm{bias}}$ (Fig. \ref{fig:4}d). We model the non-uniform $J_c(x)$ with an uneven junction width $W$ ($W_1>W_2$) in the TDGL simulation, confirming the role of broken inversion symmetry (Fig. \ref{fig:4}e-f); if the junction is inversion symmetric, the ac current flow for $\pm I_{\mathrm{bias}}$ should be mirrored along the $x$-direction (Supplementary Fig. \ref{fig:S_asymmACDC}f-h). In fact, the broken inversion symmetry is also responsible for the skewed phase boundary for $\pm I_{\mathrm{bias}}$ in Fig. \ref{fig:3}b-c. In reality the non-uniform $J_c(x)$ could be due to variations of junction width, SC/normal barrier transparency or normal metal resistivity.

\subsection*{Discussion}

The JV discussed in our work should be distinguished from the Abrikosov vortex in type-II superconductors \cite{Abrikosov1957Magnetic}. While both move in the same direction with $I_\mathrm{bias}$ and exhibit normal cores, observed in spectroscopic studies \cite{hess1989scanning,roditchev2015direct}, only the JV configuration can be controlled by a small change of $I_\mathrm{bias}$ below $I_c$. Even when the current induced phase is weak, the JV can be precisely moved side-to-side by the small change of $I_\mathrm{bias}$ from $-|I_c|$ to $|I_c|$ at $B_z = B_n \pm \varepsilon, (\varepsilon \ll 1)$. In particular, $|I_c|$ should vanish at $B_z = B_n$, if the junction is symmetric about its midpoint. This control over the JV position enables us to observe the large alternating magnetic field signal at the JJ with minimal changes in $I_{\mathrm{bias}}$ (Fig. \ref{fig:2}g). 
Finally, the minimal energy cost associated with JV movement ($I_c\Delta \phi$) as $I_c \rightarrow 0$ supports JV control as an energy-efficient way of communication between qubits \cite{wallraff2003quantum}.

The new mechanism of the Josephson diode effect offers a blueprint to realize a scalable SC rectifier with any thin film SC. Conventional Josephson diodes that are driven by the self-field effect require large operating current because the geometric inductance is usually small, especially at the nano-scale \cite{goldman1967meissner, golod2022demonstration}. However, the kinetic inductance can dominate in SC with small superfluid stiffness (e.g. low superfluid density), making it possible to reduce the device size. This also enables electric tuning of the Josephson diode by injecting a small current $J_x$ to control the Josephson current induced phase.

Finally, spatial mapping of the current flow $\textbf{J}(x,y)$ presents an alternative observable to electrical transport in SC hybrid structures. By accurately measuring $\textbf{J}(x,y)$ with high sensitivity and spatial resolution, we pinpoint the origin of the Josephson diode effect in our device, which is otherwise hidden. Our approach opens up further avenues to unseal the mechanisms for the non-reciprocity in a broad range of SC systems, and symmetry breaking in gate-tunable superconductors based on van der Waals and moir\'e materials.
The measured current flow could be directly compared with simulations based on TDGL or quantum transport to diagnose SC circuits such as local transparency of the JJ barrier.
\section*{Methods}
\subsection*{Variable Temperature Scanning Setup}
Measurements were performed in a home-built variable temperature system with optical access. There are multiple nano-pillars containing NV centers on each diamond probe, and a goniometer with both pitch and yawn control is used to set the stand-off distance between the NV and the sample \cite{vool2021imaging}, which ranges between 130 to 180 nm throughout the study. The NV center is excited with 532 nm green laser (Coherent Sapphire), and read out with standard optical detected magnetic resonance (ODMR) technique using a 600 nm long-pass optical filter. The time-averaged power of the green laser pulses is less than 50 $\mu$W. The microwave (MW) drive is provided via on-chip transmission line next to the sample. MW is sourced from SGS-100A (Rohde \& Schwarz) and modulated with the built-in IQ mixer. MW pulses are then amplified by +40 db using 30S1G6C (AR Inc), and routed through another switch (RF lambda) to reduce noise from the amplifier. MW and bias current control sequences are generated by arbitrary wave generator AWG5014C (Tektronic). 

\subsection*{Device Fabrication and Electrical Characterization}
The SC and normal parts of the JJ are made of niobium nitride (NbN) and gold (Au) thin films, respectively. The JJs are fabricated on undoped Silicon substrate with 285 nm SiO$_2$ on top. Standard electron beam lithography method is used to define the device geometry using double-layer e-beam resist. The normal part of the junction is first formed with thermal evaporation (2 nm Ti/ 35 nm Au). A short Argon milling process is used right before sputtering SC electrodes (2 nm Ti/ 6 nm Nb / 30 nm NbN). The microwave strip line is formed with 2 nm Ti/ 60 nm Au. Four terminal resistance result was first measured with dc bias from Keithley 2400 and dc voltage with Keithley 2100, and then taken numerical derivative to acquire differential resistance shown in the main text.
 
\subsection*{Probe Fabrication and Typical Characteristics}
The diamond fabrication process follows Ref. \cite{zhou2017scanning}.
Specifically, ultra pure diamond with natural 13C abundance and [100] facet (electronic grade from Element Six) is diced into thin slabs about 50 $\mu$m thick. One side of the slab is etched by Argon/Chloride plasma to relieve surface strain, then implanted with 15N ions at a dose of $5 \times 10^{10} / \mathrm{cm}^2$ and acceleration energy of 6 keV (Innovion). Then the diamond is annealed in ultrahigh vacuum ($<3\times 10^{-8}$ Torr) at 800 $\degree$C for 24 hours to form NV centers. The diamond nano-pillars are defined with standard e-beam lithography and etched with O$_2$ plasma. On average we get 1 NV center per diamond pillar with this process. The NV depth from the surface is about 15-20 nm. Typical ODMR red photon count is 100k/s, contrast in pulsed measurement is 20 to 30$\%$, and the coherence time is $T_2^* \approx 1~\mu$s and $T_2 \approx$ 30 $\mu$s at 4K and the small magnetic field used in this study.

\subsection*{Detail about NV Magnetometry}
NV is a spin-1 system with low energy states $s=|0\rangle, |\pm 1\rangle$. The $|0\rangle$ is split in energy from $|\pm1\rangle$ by the zero field splitting (2.87GHz) and $|\pm1\rangle$ are further split by the Zeeman energy $E_Z = g\mu_B B_\mathrm{nv}$, here $g=2$ is the Land\'e g-factor for electron, $\mu_B$ is the Bohr magneton, and $B_\mathrm{nv}$ is the magnetic field along NV axis. In practice, we apply a external field of less than 50 G along the NV axis, and drive the $|0\rangle$ and $|-1\rangle$ states as a qubit using MW. As mentioned in the main text, ``ac'' magnetometry is used to filter out low frequency noise and maximize sensitivity by utilizing the longer $T_2$ coherence time. Specifically, the NV qubit is first prepared in the $|0\rangle$ state using a green laser pulse, and then driven into the superposition state $\frac{1}{\sqrt{2}}(|0\rangle + i |-1\rangle)$ by a $X_\frac{\pi}{2}$ MW pulse. We use two types of dynamic decoupling sequences, the spin echo (Hahn echo) with one $\pi$-pulse, and the Carr-Purcell-Meiboom-Gill (CPMG) with $n$-$Y_\pi$ pulses in the experiment \cite{biercuk2011dynamical,pham2012enhanced}. Between neighbouring $\pi$-pulses, the qubit rotates by an angle $\varphi =2 \pi \gamma_e b_\mathrm{nv} \tau_n $, where $\gamma_e=28$ GHz/T is the gyro-magnetic ratio of the electron spin, $b_\mathrm{nv}$ is the magnetic field generated by the current projected along NV axis, and $\tau_n$ is the evolution time between neighbouring MW pulses. The $\pi$-pulses reverse the qubit rotation direction, and the total angle is the difference of the accumulation in each half of the sequence. The frequency of NV control sequence and bias current modulation is $f=$ 100 to 500 kHz, corresponding to less than 1 nA bias current due to the AC Josephson effect $I = hf/2eR_N$ \cite{tinkham1996introduction} ($h$ is Planck's constant, $e$ is electron charge, $R_N \approx 1~\Omega$ is the normal state resistance of the JJ). This is 3 to 4 orders of magnitude smaller than the bias current applied to the JJ.

To extract the phase accumulation angle, the NV spin is projected to the $|0\rangle$ and $|-1\rangle$ states using four $\frac{\pi}{2}$-pulses $\frac{\pi}{2}_{\pm X/Y}$ and record the ODMR signal. The angle is then calculated from

\begin{equation} \label{eq_phasecalculation}
\varphi = \arctan{\frac{C_{\frac{\pi}{2}_X }-C_{\frac{\pi}{2}_{-X} }}{C_{\frac{\pi}{2}_Y}-C_{\frac{\pi}{2}_{-Y}}}}
\end{equation}

where $C_{\frac{\pi}{2}_{\pm X/Y}}$ are the photon counts from $\frac{\pi}{2}_{\pm X/Y}$ projections. The measurement sequences are averaged up to 100k times (about 10 seconds) at each point to extract the $B_\mathrm{nv}$.

\subsection*{Reconstructing Current Flow from Magnetic Field}

In this section we discuss the three methods used to reconstruct current flow $j_{x,y}$ from $b_\mathrm{nv}$. For all methods, we first convert the magnetic field projected along NV axis $b_\mathrm{nv}$, to Cartesian vector magnetic field $b_{x,y,z}$ using the source-free constraint for the stray field \cite{blakely1996potential,lima2009obtaining,casola2018probing},

\begin{equation}
\nabla \times b = 0
\end{equation}

Thus in the Fourier space the vector components are, 
\begin{equation}
\begin{aligned}
b_z(\textbf{k}) &= \frac{b_\mathrm{nv}(\textbf{k})}{\textbf{u}_\mathrm{nv}
\cdot \textbf{u}} \\
b_x(\textbf{k}) &= -i\frac{k_x}{k} b_z(\textbf{k})\\
b_y(\textbf{k}) &= -i\frac{k_y}{k} b_z(\textbf{k})
\end{aligned}
\end{equation}

here $\textbf{k}=(k_x,k_y)$ is the 2D wavevector, $k=\sqrt{k_x^2+k_y^2 }$, $\textbf{u}_\mathrm{nv}$ is the unit vector of the NV axis, $\textbf{u} = (-ik_x/k,-ik_y/k,1)$. The singularity point at $k=0$ is discarded during the reconstruction. Because the SC electrode is much longer than our measurement window in the $y$-direction, the $j_y$ outside the window on the top and bottom sides also contribute to the measured $b_\mathrm{nv}$. In practice we extend the measured $b_\mathrm{nv}$ with the top and bottom lines in the $y$ direction, and linearly extrapolates $b_\mathrm{nv}$ to zero in the $x$ direction. The padding size in each direction is 10 times of the measurement window, at which point increasing the size does not change the reconstruction result. More detail of the reconstruction process is discussed in \ref{SI:recondetail}.

\section*{Data availability}
All experimental and numerically simulated data included in this work are available at the Zenodo database \cite{ZenodoLink}.

\section*{References}
\bibliographystyle{naturemag}
\bibliography{references_JJ}

\begin{thebibliography}{10}
\expandafter\ifx\csname url\endcsname\relax
  \def\url#1{\texttt{#1}}\fi
\expandafter\ifx\csname urlprefix\endcsname\relax\def\urlprefix{URL }\fi
\providecommand{\bibinfo}[2]{#2}
\providecommand{\eprint}[2][]{\url{#2}}

\bibitem{josephson1962possible}
\bibinfo{author}{Josephson, B.~D.}
\newblock \bibinfo{title}{Possible new effects in superconductive tunnelling}.
\newblock \emph{\bibinfo{journal}{Phys. Lett.}} \textbf{\bibinfo{volume}{1}},
  \bibinfo{pages}{251--253} (\bibinfo{year}{1962}).

\bibitem{de1964boundary}
\bibinfo{author}{de~Gennes, P.}
\newblock \bibinfo{title}{Boundary effects in superconductors}.
\newblock \emph{\bibinfo{journal}{Rev. of Mod. Phys.}}
  \textbf{\bibinfo{volume}{36}}, \bibinfo{pages}{225} (\bibinfo{year}{1964}).

\bibitem{josephson1964coupled}
\bibinfo{author}{Josephson, B.}
\newblock \bibinfo{title}{Coupled superconductors}.
\newblock \emph{\bibinfo{journal}{Rev. of Mod. Phys.}}
  \textbf{\bibinfo{volume}{36}}, \bibinfo{pages}{216} (\bibinfo{year}{1964}).

\bibitem{devoret2013superconducting}
\bibinfo{author}{Devoret, M.~H.} \& \bibinfo{author}{Schoelkopf, R.~J.}
\newblock \bibinfo{title}{Superconducting circuits for quantum information: an
  outlook}.
\newblock \emph{\bibinfo{journal}{Science}} \textbf{\bibinfo{volume}{339}},
  \bibinfo{pages}{1169--1174} (\bibinfo{year}{2013}).

\bibitem{fedorov2014fluxon}
\bibinfo{author}{Fedorov, K.~G.}, \bibinfo{author}{Shcherbakova, A.~V.},
  \bibinfo{author}{Wolf, M.~J.}, \bibinfo{author}{Beckmann, D.} \&
  \bibinfo{author}{Ustinov, A.~V.}
\newblock \bibinfo{title}{Fluxon readout of a superconducting qubit}.
\newblock \emph{\bibinfo{journal}{Phys. Rev. Lett.}}
  \textbf{\bibinfo{volume}{112}}, \bibinfo{pages}{160502}
  (\bibinfo{year}{2014}).

\bibitem{kjaergaard2020superconducting}
\bibinfo{author}{Kjaergaard, M.} \emph{et~al.}
\newblock \bibinfo{title}{Superconducting qubits: Current state of play}.
\newblock \emph{\bibinfo{journal}{Annual Rev. of Cond. Matt. Phys.}}
  \textbf{\bibinfo{volume}{11}}, \bibinfo{pages}{369--395}
  (\bibinfo{year}{2020}).

\bibitem{wang2010coherent}
\bibinfo{author}{Wang, H.} \emph{et~al.}
\newblock \bibinfo{title}{Coherent terahertz emission of intrinsic josephson
  junction stacks in the hot spot regime}.
\newblock \emph{\bibinfo{journal}{Phys. Rev. Lett.}}
  \textbf{\bibinfo{volume}{105}}, \bibinfo{pages}{057002}
  (\bibinfo{year}{2010}).

\bibitem{welp2013superconducting}
\bibinfo{author}{Welp, U.}, \bibinfo{author}{Kadowaki, K.} \&
  \bibinfo{author}{Kleiner, R.}
\newblock \bibinfo{title}{Superconducting emitters of thz radiation}.
\newblock \emph{\bibinfo{journal}{Nat. Photon.}} \textbf{\bibinfo{volume}{7}},
  \bibinfo{pages}{702--710} (\bibinfo{year}{2013}).

\bibitem{kirtley1999scanning}
\bibinfo{author}{Kirtley, J.~R.} \& \bibinfo{author}{Wikswo~Jr, J.~P.}
\newblock \bibinfo{title}{Scanning squid microscopy}.
\newblock \emph{\bibinfo{journal}{Annual Rev. of Mat. Sci.}}
  \textbf{\bibinfo{volume}{29}}, \bibinfo{pages}{117--148}
  (\bibinfo{year}{1999}).

\bibitem{kirtley2010fundamental}
\bibinfo{author}{Kirtley, J.}
\newblock \bibinfo{title}{Fundamental studies of superconductors using scanning
  magnetic imaging}.
\newblock \emph{\bibinfo{journal}{Rep. on Prog. in Phys.}}
  \textbf{\bibinfo{volume}{73}}, \bibinfo{pages}{126501}
  (\bibinfo{year}{2010}).

\bibitem{holmes2013energy}
\bibinfo{author}{Holmes, D.~S.}, \bibinfo{author}{Ripple, A.~L.} \&
  \bibinfo{author}{Manheimer, M.~A.}
\newblock \bibinfo{title}{Energy-efficient superconducting computing - power
  budgets and requirements}.
\newblock \emph{\bibinfo{journal}{IEEE Transactions on Applied
  Superconductivity}} \textbf{\bibinfo{volume}{23}}, \bibinfo{pages}{1701610}
  (\bibinfo{year}{2013}).

\bibitem{grosfeld2011observing}
\bibinfo{author}{Grosfeld, E.} \& \bibinfo{author}{Stern, A.}
\newblock \bibinfo{title}{Observing majorana bound states of josephson vortices
  in topological superconductors}.
\newblock \emph{\bibinfo{journal}{PNAS}} \textbf{\bibinfo{volume}{108}},
  \bibinfo{pages}{11810--11814} (\bibinfo{year}{2011}).

\bibitem{beenakker2013search}
\bibinfo{author}{Beenakker, C.}
\newblock \bibinfo{title}{Search for majorana fermions in superconductors}.
\newblock \emph{\bibinfo{journal}{Annu. Rev. Condens. Matter Phys.}}
  \textbf{\bibinfo{volume}{4}}, \bibinfo{pages}{113--136}
  (\bibinfo{year}{2013}).

\bibitem{lutchyn2018majorana}
\bibinfo{author}{Lutchyn, R.~M.} \emph{et~al.}
\newblock \bibinfo{title}{Majorana zero modes in superconductor--semiconductor
  heterostructures}.
\newblock \emph{\bibinfo{journal}{Nat. Rev. Mat.}}
  \textbf{\bibinfo{volume}{3}}, \bibinfo{pages}{52--68} (\bibinfo{year}{2018}).

\bibitem{pal2022josephson}
\bibinfo{author}{Pal, B.} \emph{et~al.}
\newblock \bibinfo{title}{Josephson diode effect from cooper pair momentum in a
  topological semimetal}.
\newblock \emph{\bibinfo{journal}{Nat. Phys.}} \textbf{\bibinfo{volume}{18}},
  \bibinfo{pages}{1228--1233} (\bibinfo{year}{2022}).

\bibitem{jeon2022zero}
\bibinfo{author}{Jeon, K.-R.} \emph{et~al.}
\newblock \bibinfo{title}{Zero-field polarity-reversible josephson supercurrent
  diodes enabled by a proximity-magnetized pt barrier}.
\newblock \emph{\bibinfo{journal}{Nat. Mat.}} \textbf{\bibinfo{volume}{21}},
  \bibinfo{pages}{1008--1013} (\bibinfo{year}{2022}).

\bibitem{nadeem2023superconducting}
\bibinfo{author}{Nadeem, M.}, \bibinfo{author}{Fuhrer, M.~S.} \&
  \bibinfo{author}{Wang, X.}
\newblock \bibinfo{title}{The superconducting diode effect}.
\newblock \emph{\bibinfo{journal}{Nature Reviews Physics}}
  \bibinfo{pages}{1--20} (\bibinfo{year}{2023}).

\bibitem{roditchev2015direct}
\bibinfo{author}{Roditchev, D.} \emph{et~al.}
\newblock \bibinfo{title}{Direct observation of josephson vortex cores}.
\newblock \emph{\bibinfo{journal}{Nat. Phys.}} \textbf{\bibinfo{volume}{11}},
  \bibinfo{pages}{332--337} (\bibinfo{year}{2015}).

\bibitem{tsuei1994pairing}
\bibinfo{author}{Tsuei, C.} \emph{et~al.}
\newblock \bibinfo{title}{Pairing symmetry and flux quantization in a
  tricrystal superconducting ring of
  $\mathrm{YBa}_2\mathrm{Cu}_3\mathrm{O}_{7-\delta}$}.
\newblock \emph{\bibinfo{journal}{Phys. Rev. Lett.}}
  \textbf{\bibinfo{volume}{73}}, \bibinfo{pages}{593} (\bibinfo{year}{1994}).

\bibitem{hilgenkamp2003ordering}
\bibinfo{author}{Hilgenkamp, H.} \emph{et~al.}
\newblock \bibinfo{title}{Ordering and manipulation of the magnetic moments in
  large-scale superconducting $\pi$-loop arrays}.
\newblock \emph{\bibinfo{journal}{Nature}} \textbf{\bibinfo{volume}{422}},
  \bibinfo{pages}{50--53} (\bibinfo{year}{2003}).

\bibitem{gutfreund2023direct}
\bibinfo{author}{Gutfreund, A.} \emph{et~al.}
\newblock \bibinfo{title}{Direct observation of a superconducting vortex
  diode}.
\newblock \emph{\bibinfo{journal}{Nat. Comm.}} \textbf{\bibinfo{volume}{14}},
  \bibinfo{pages}{1630} (\bibinfo{year}{2023}).

\bibitem{tinkham1996introduction}
\bibinfo{author}{Tinkham, M.}
\newblock \emph{\bibinfo{title}{Introduction to superconductivity}}
  (\bibinfo{publisher}{McGraw-Hill}, \bibinfo{year}{1996}).

\bibitem{gross1994low}
\bibinfo{author}{Gross, R.} \& \bibinfo{author}{Koelle, D.}
\newblock \bibinfo{title}{Low temperature scanning electron microscopy of
  superconducting thin films and josephson junctions}.
\newblock \emph{\bibinfo{journal}{Rep. on Prog. in Phys.}}
  \textbf{\bibinfo{volume}{57}}, \bibinfo{pages}{651} (\bibinfo{year}{1994}).

\bibitem{dremov2019local}
\bibinfo{author}{Dremov, V.~V.} \emph{et~al.}
\newblock \bibinfo{title}{Local josephson vortex generation and manipulation
  with a magnetic force microscope}.
\newblock \emph{\bibinfo{journal}{Nat. Comm.}} \textbf{\bibinfo{volume}{10}},
  \bibinfo{pages}{4009} (\bibinfo{year}{2019}).

\bibitem{hovhannisyan2021lateral}
\bibinfo{author}{Hovhannisyan, R.~A.}, \bibinfo{author}{Grebenchuk, S.~Y.},
  \bibinfo{author}{Baranov, D.~S.}, \bibinfo{author}{Roditchev, D.} \&
  \bibinfo{author}{Stolyarov, V.~S.}
\newblock \bibinfo{title}{Lateral josephson junctions as sensors for magnetic
  microscopy at nanoscale}.
\newblock \emph{\bibinfo{journal}{Journal of Phys. Chem, Lett.}}
  \textbf{\bibinfo{volume}{12}}, \bibinfo{pages}{12196--12201}
  (\bibinfo{year}{2021}).

\bibitem{grebenchuk2020observation}
\bibinfo{author}{Grebenchuk, S.~Y.} \emph{et~al.}
\newblock \bibinfo{title}{Observation of interacting josephson vortex chains by
  magnetic force microscopy}.
\newblock \emph{\bibinfo{journal}{Phys. Rev. Res.}}
  \textbf{\bibinfo{volume}{2}}, \bibinfo{pages}{023105} (\bibinfo{year}{2020}).

\bibitem{stolyarov2022revealing}
\bibinfo{author}{Stolyarov, V.~S.} \emph{et~al.}
\newblock \bibinfo{title}{Revealing josephson vortex dynamics in proximity
  junctions below critical current}.
\newblock \emph{\bibinfo{journal}{Nano Lett.}} \textbf{\bibinfo{volume}{22}},
  \bibinfo{pages}{5715--5722} (\bibinfo{year}{2022}).

\bibitem{vool2021imaging}
\bibinfo{author}{Vool, U.} \emph{et~al.}
\newblock \bibinfo{title}{Imaging phonon-mediated hydrodynamic flow in
  $\mathrm{WTe}_2$}.
\newblock \emph{\bibinfo{journal}{Nat. Phys.}} \textbf{\bibinfo{volume}{17}},
  \bibinfo{pages}{1216--1220} (\bibinfo{year}{2021}).

\bibitem{tolpygo1996critical}
\bibinfo{author}{Tolpygo, S.~K.} \& \bibinfo{author}{Gurvitch, M.}
\newblock \bibinfo{title}{Critical currents and josephson penetration depth in
  planar thin-film high-$t_c$ josephson junctions}.
\newblock \emph{\bibinfo{journal}{Appl. Phys. Lett.}}
  \textbf{\bibinfo{volume}{69}}, \bibinfo{pages}{3914--3916}
  (\bibinfo{year}{1996}).

\bibitem{rowell1963magnetic}
\bibinfo{author}{Rowell, J.}
\newblock \bibinfo{title}{Magnetic field dependence of the josephson tunnel
  current}.
\newblock \emph{\bibinfo{journal}{Phys. Rev. Lett.}}
  \textbf{\bibinfo{volume}{11}}, \bibinfo{pages}{200} (\bibinfo{year}{1963}).

\bibitem{clem2010josephson}
\bibinfo{author}{Clem, J.~R.}
\newblock \bibinfo{title}{Josephson junctions in thin and narrow rectangular
  superconducting strips}.
\newblock \emph{\bibinfo{journal}{Phys. Rev. B}} \textbf{\bibinfo{volume}{81}},
  \bibinfo{pages}{144515} (\bibinfo{year}{2010}).

\bibitem{dynes1971supercurrent}
\bibinfo{author}{Dynes, R.} \& \bibinfo{author}{Fulton, T.}
\newblock \bibinfo{title}{Supercurrent density distribution in josephson
  junctions}.
\newblock \emph{\bibinfo{journal}{Phys. Rev. B}} \textbf{\bibinfo{volume}{3}},
  \bibinfo{pages}{3015} (\bibinfo{year}{1971}).

\bibitem{mayer1993magnetic}
\bibinfo{author}{Mayer, B.}, \bibinfo{author}{Schuster, S.},
  \bibinfo{author}{Beck, A.}, \bibinfo{author}{Alff, L.} \&
  \bibinfo{author}{Gross, R.}
\newblock \bibinfo{title}{Magnetic field dependence of the critical current in
  $\mathrm{YBa_2Cu_3O_{7-\delta}}$ bicrystal grain boundary junctions}.
\newblock \emph{\bibinfo{journal}{Appl. Phys. Lett.}}
  \textbf{\bibinfo{volume}{62}}, \bibinfo{pages}{783--785}
  (\bibinfo{year}{1993}).

\bibitem{holm1995novel}
\bibinfo{author}{Holm, J.} \& \bibinfo{author}{Mygind, J.}
\newblock \bibinfo{title}{A novel cryogenic scanning laser microscope tested on
  josephson tunnel junctions}.
\newblock \emph{\bibinfo{journal}{Rev. of Sci. Inst.}}
  \textbf{\bibinfo{volume}{66}}, \bibinfo{pages}{4547--4551}
  (\bibinfo{year}{1995}).

\bibitem{kuzmin2023tuning}
\bibinfo{author}{Kuzmin, R.}, \bibinfo{author}{Mehta, N.},
  \bibinfo{author}{Grabon, N.} \& \bibinfo{author}{Manucharyan, V.~E.}
\newblock \bibinfo{title}{Tuning the inductance of josephson junction arrays
  without squids}.
\newblock \emph{\bibinfo{journal}{Appl. Phys. Lett.}}
  \textbf{\bibinfo{volume}{123}}, \bibinfo{pages}{182602}
  (\bibinfo{year}{2023}).

\bibitem{hilgenkamp2002grain}
\bibinfo{author}{Hilgenkamp, H.} \& \bibinfo{author}{Mannhart, J.}
\newblock \bibinfo{title}{Grain boundaries in high-t c superconductors}.
\newblock \emph{\bibinfo{journal}{Rev. of Mod. Phys.}}
  \textbf{\bibinfo{volume}{74}}, \bibinfo{pages}{485} (\bibinfo{year}{2002}).

\bibitem{boris2013evidence}
\bibinfo{author}{Boris, A.~A.} \emph{et~al.}
\newblock \bibinfo{title}{Evidence for nonlocal electrodynamics in planar
  josephson junctions}.
\newblock \emph{\bibinfo{journal}{Phys. Rev. Lett.}}
  \textbf{\bibinfo{volume}{111}}, \bibinfo{pages}{117002}
  (\bibinfo{year}{2013}).

\bibitem{owen1967vortex}
\bibinfo{author}{Owen, C.} \& \bibinfo{author}{Scalapino, D.}
\newblock \bibinfo{title}{Vortex structure and critical currents in josephson
  junctions}.
\newblock \emph{\bibinfo{journal}{Phys. Rev.}} \textbf{\bibinfo{volume}{164}},
  \bibinfo{pages}{538} (\bibinfo{year}{1967}).

\bibitem{pagano1991magnetic}
\bibinfo{author}{Pagano, S.}, \bibinfo{author}{Ruggiero, B.} \&
  \bibinfo{author}{Sarnelli, E.}
\newblock \bibinfo{title}{Magnetic-field dependence of the critical current in
  long josephson junctions}.
\newblock \emph{\bibinfo{journal}{Phys. Rev. B}} \textbf{\bibinfo{volume}{43}},
  \bibinfo{pages}{5364} (\bibinfo{year}{1991}).

\bibitem{kuplevakhsky2006static}
\bibinfo{author}{Kuplevakhsky, S.} \& \bibinfo{author}{Glukhov, A.}
\newblock \bibinfo{title}{Static solitons of the sine-gordon equation and
  equilibrium vortex structure in josephson junctions}.
\newblock \emph{\bibinfo{journal}{Phys. Rev. B}} \textbf{\bibinfo{volume}{73}},
  \bibinfo{pages}{024513} (\bibinfo{year}{2006}).

\bibitem{kuplevakhsky2007exact}
\bibinfo{author}{Kuplevakhsky, S.} \& \bibinfo{author}{Glukhov, A.}
\newblock \bibinfo{title}{Exact analytical solution of the problem of
  current-carrying states of the josephson junction in external magnetic
  fields}.
\newblock \emph{\bibinfo{journal}{Phys. Rev. B}} \textbf{\bibinfo{volume}{76}},
  \bibinfo{pages}{174515} (\bibinfo{year}{2007}).

\bibitem{kuplevakhsky2010exact}
\bibinfo{author}{Kuplevakhsky, S.} \& \bibinfo{author}{Glukhov, A.}
\newblock \bibinfo{title}{Exact analytical solution of a classical josephson
  tunnel junction problem}.
\newblock \emph{\bibinfo{journal}{Low Temp. Phys.}}
  \textbf{\bibinfo{volume}{36}}, \bibinfo{pages}{1012--1021}
  (\bibinfo{year}{2010}).

\bibitem{moll2023evolution}
\bibinfo{author}{Moll, P.~J.} \& \bibinfo{author}{Geshkenbein, V.~B.}
\newblock \bibinfo{title}{Evolution of superconducting diodes}.
\newblock \emph{\bibinfo{journal}{Nat. Phys.}} \textbf{\bibinfo{volume}{19}},
  \bibinfo{pages}{1379–1380} (\bibinfo{year}{2023}).

\bibitem{zhang2022general}
\bibinfo{author}{Zhang, Y.}, \bibinfo{author}{Gu, Y.}, \bibinfo{author}{Li,
  P.}, \bibinfo{author}{Hu, J.} \& \bibinfo{author}{Jiang, K.}
\newblock \bibinfo{title}{General theory of josephson diodes}.
\newblock \emph{\bibinfo{journal}{Phys. Rev. X}} \textbf{\bibinfo{volume}{12}},
  \bibinfo{pages}{041013} (\bibinfo{year}{2022}).

\bibitem{Abrikosov1957Magnetic}
\bibinfo{author}{Abrikosov, A.~A.}
\newblock \bibinfo{title}{Magnetic properties of superconductors of the second
  group}.
\newblock \emph{\bibinfo{journal}{Sov. Phys. - JETP}}
  \textbf{\bibinfo{volume}{5}}, \bibinfo{pages}{1174--1182}
  (\bibinfo{year}{1957}).

\bibitem{hess1989scanning}
\bibinfo{author}{Hess, H.}, \bibinfo{author}{Robinson, R.},
  \bibinfo{author}{Dynes, R.}, \bibinfo{author}{Valles~Jr, J.} \&
  \bibinfo{author}{Waszczak, J.}
\newblock \bibinfo{title}{Scanning-tunneling-microscope observation of the
  abrikosov flux lattice and the density of states near and inside a fluxoid}.
\newblock \emph{\bibinfo{journal}{Phys. Rev. Lett.}}
  \textbf{\bibinfo{volume}{62}}, \bibinfo{pages}{214} (\bibinfo{year}{1989}).

\bibitem{wallraff2003quantum}
\bibinfo{author}{Wallraff, A.} \emph{et~al.}
\newblock \bibinfo{title}{Quantum dynamics of a single vortex}.
\newblock \emph{\bibinfo{journal}{Nature}} \textbf{\bibinfo{volume}{425}},
  \bibinfo{pages}{155--158} (\bibinfo{year}{2003}).

\bibitem{goldman1967meissner}
\bibinfo{author}{Goldman, A.} \& \bibinfo{author}{Kreisman, P.}
\newblock \bibinfo{title}{Meissner effect and vortex penetration in josephson
  junctions}.
\newblock \emph{\bibinfo{journal}{Phys. Rev.}} \textbf{\bibinfo{volume}{164}},
  \bibinfo{pages}{544} (\bibinfo{year}{1967}).

\bibitem{golod2022demonstration}
\bibinfo{author}{Golod, T.} \& \bibinfo{author}{Krasnov, V.~M.}
\newblock \bibinfo{title}{Demonstration of a superconducting diode-with-memory,
  operational at zero magnetic field with switchable nonreciprocity}.
\newblock \emph{\bibinfo{journal}{Nat. Comm.}} \textbf{\bibinfo{volume}{13}},
  \bibinfo{pages}{3658} (\bibinfo{year}{2022}).

\bibitem{zhou2017scanning}
\bibinfo{author}{Zhou, T.~X.}, \bibinfo{author}{St{\"o}hr, R.~J.} \&
  \bibinfo{author}{Yacoby, A.}
\newblock \bibinfo{title}{Scanning diamond nv center probes compatible with
  conventional afm technology}.
\newblock \emph{\bibinfo{journal}{App. Phys. Lett.}}
  \textbf{\bibinfo{volume}{111}}, \bibinfo{pages}{163106}
  (\bibinfo{year}{2017}).

\bibitem{biercuk2011dynamical}
\bibinfo{author}{Biercuk, M.}, \bibinfo{author}{Doherty, A.} \&
  \bibinfo{author}{Uys, H.}
\newblock \bibinfo{title}{Dynamical decoupling sequence construction as a
  filter-design problem}.
\newblock \emph{\bibinfo{journal}{Journal of Phy. B: Atomic, Mole. and Opt.
  Phys.}} \textbf{\bibinfo{volume}{44}}, \bibinfo{pages}{154002}
  (\bibinfo{year}{2011}).

\bibitem{pham2012enhanced}
\bibinfo{author}{Pham, L.~M.} \emph{et~al.}
\newblock \bibinfo{title}{Enhanced solid-state multispin metrology using
  dynamical decoupling}.
\newblock \emph{\bibinfo{journal}{Phys. Rev. B}} \textbf{\bibinfo{volume}{86}},
  \bibinfo{pages}{045214} (\bibinfo{year}{2012}).

\bibitem{blakely1996potential}
\bibinfo{author}{Blakely, R.~J.}
\newblock \emph{\bibinfo{title}{Potential theory in gravity and magnetic
  applications}} (\bibinfo{publisher}{Cambridge university press},
  \bibinfo{year}{1996}).

\bibitem{lima2009obtaining}
\bibinfo{author}{Lima, E.~A.} \& \bibinfo{author}{Weiss, B.~P.}
\newblock \bibinfo{title}{Obtaining vector magnetic field maps from
  single-component measurements of geological samples}.
\newblock \emph{\bibinfo{journal}{Journal of Geo. Res.: Solid Earth}}
  \textbf{\bibinfo{volume}{114}} (\bibinfo{year}{2009}).

\bibitem{casola2018probing}
\bibinfo{author}{Casola, F.}, \bibinfo{author}{Van Der~Sar, T.} \&
  \bibinfo{author}{Yacoby, A.}
\newblock \bibinfo{title}{Probing condensed matter physics with magnetometry
  based on nitrogen-vacancy centres in diamond}.
\newblock \emph{\bibinfo{journal}{Nat. Rev. Mat.}}
  \textbf{\bibinfo{volume}{3}}, \bibinfo{pages}{1--13} (\bibinfo{year}{2018}).

\bibitem{ZenodoLink}
\bibinfo{author}{Chen, S.} \& \bibinfo{author}{Park, S.}
\newblock \bibinfo{title}{Current induced hidden states in josephson
  junctions}.
\newblock \bibinfo{note}{Zenodo.
  \url{https://doi.org/10.5281/zenodo.13256436}}.

\bibitem{rohner2018real}
\bibinfo{author}{Rohner, D.} \emph{et~al.}
\newblock \bibinfo{title}{Real-space probing of the local magnetic response of
  thin-film superconductors using single spin magnetometry}.
\newblock \emph{\bibinfo{journal}{Sensors}} \textbf{\bibinfo{volume}{18}},
  \bibinfo{pages}{3790} (\bibinfo{year}{2018}).

\bibitem{ilin2014critical}
\bibinfo{author}{Ilin, K.} \emph{et~al.}
\newblock \bibinfo{title}{Critical current of $\mathrm{Nb}$, $\mathrm{NbN}$,
  and $\mathrm{TaN}$ thin-film bridges with and without geometrical
  nonuniformities in a magnetic field}.
\newblock \emph{\bibinfo{journal}{Phy. Rev. B}} \textbf{\bibinfo{volume}{89}},
  \bibinfo{pages}{184511} (\bibinfo{year}{2014}).

\bibitem{charaev2017proximity}
\bibinfo{author}{Charaev, I.} \emph{et~al.}
\newblock \bibinfo{title}{Proximity effect model of ultranarrow $\mathrm{NbN}$
  strips}.
\newblock \emph{\bibinfo{journal}{Phys. Rev. B}} \textbf{\bibinfo{volume}{96}},
  \bibinfo{pages}{184517} (\bibinfo{year}{2017}).

\bibitem{hu2003sputter}
\bibinfo{author}{Hu, R.}, \bibinfo{author}{Kerber, G.~L.},
  \bibinfo{author}{Luine, J.}, \bibinfo{author}{Ladizinsky, E.} \&
  \bibinfo{author}{Bulman, J.}
\newblock \bibinfo{title}{Sputter deposition conditions and penetration depth
  in nbn thin films}.
\newblock \emph{\bibinfo{journal}{IEEE Tran. on Appl. Superconductivity}}
  \textbf{\bibinfo{volume}{13}}, \bibinfo{pages}{3288--3291}
  (\bibinfo{year}{2003}).

\bibitem{kamlapure2010measurement}
\bibinfo{author}{Kamlapure, A.} \emph{et~al.}
\newblock \bibinfo{title}{Measurement of magnetic penetration depth and
  superconducting energy gap in very thin epitaxial nbn films}.
\newblock \emph{\bibinfo{journal}{Appl. Phys. Lett.}}
  \textbf{\bibinfo{volume}{96}}, \bibinfo{pages}{072509}
  (\bibinfo{year}{2010}).

\bibitem{mints2002nonlocal}
\bibinfo{author}{Mints, R.} \& \bibinfo{author}{Papiashvili, I.}
\newblock \bibinfo{title}{Nonlocal electrodynamics of josephson junctions in
  thin films and fractional vortices}.
\newblock \emph{\bibinfo{journal}{Superconductor Sci. and Tech.}}
  \textbf{\bibinfo{volume}{15}}, \bibinfo{pages}{307} (\bibinfo{year}{2002}).

\bibitem{dubos2001josephson}
\bibinfo{author}{Dubos, P.} \emph{et~al.}
\newblock \bibinfo{title}{Josephson critical current in a long mesoscopic
  $\mathrm{SNS}$ junction}.
\newblock \emph{\bibinfo{journal}{Phys. Rev. B}} \textbf{\bibinfo{volume}{63}},
  \bibinfo{pages}{064502} (\bibinfo{year}{2001}).

\bibitem{roth1989using}
\bibinfo{author}{Roth, B.~J.}, \bibinfo{author}{Sepulveda, N.~G.} \&
  \bibinfo{author}{Wikswo~Jr, J.~P.}
\newblock \bibinfo{title}{Using a magnetometer to image a two-dimensional
  current distribution}.
\newblock \emph{\bibinfo{journal}{Journal of Appl. Phys.}}
  \textbf{\bibinfo{volume}{65}}, \bibinfo{pages}{361--372}
  (\bibinfo{year}{1989}).

\bibitem{meltzer2017direct}
\bibinfo{author}{Meltzer, A.~Y.}, \bibinfo{author}{Levin, E.} \&
  \bibinfo{author}{Zeldov, E.}
\newblock \bibinfo{title}{Direct reconstruction of two-dimensional currents in
  thin films from magnetic-field measurements}.
\newblock \emph{\bibinfo{journal}{Phys. Rev. App.}}
  \textbf{\bibinfo{volume}{8}}, \bibinfo{pages}{064030} (\bibinfo{year}{2017}).

\bibitem{dubois2022untrained}
\bibinfo{author}{Dubois, A.} \emph{et~al.}
\newblock \bibinfo{title}{Untrained physically informed neural network for
  image reconstruction of magnetic field sources}.
\newblock \emph{\bibinfo{journal}{Phys. Rev. App.}}
  \textbf{\bibinfo{volume}{18}}, \bibinfo{pages}{064076}
  (\bibinfo{year}{2022}).

\bibitem{fulton1972quantum}
\bibinfo{author}{Fulton, T.}, \bibinfo{author}{Dunkleberger, L.} \&
  \bibinfo{author}{Dynes, R.}
\newblock \bibinfo{title}{Quantum interference properties of double josephson
  junctions}.
\newblock \emph{\bibinfo{journal}{Phys. Rev. B}} \textbf{\bibinfo{volume}{6}},
  \bibinfo{pages}{855} (\bibinfo{year}{1972}).

\bibitem{Reinhardt2024NatCommun}
\bibinfo{author}{Reinhardt, S.} \emph{et~al.}
\newblock \bibinfo{title}{Link between supercurrent diode and anomalous
  {{Josephson}} effect revealed by gate-controlled interferometry}.
\newblock \emph{\bibinfo{journal}{Nat. Comm.}} \textbf{\bibinfo{volume}{15}},
  \bibinfo{pages}{4413} (\bibinfo{year}{2024}).

\bibitem{ando2020observation}
\bibinfo{author}{Ando, F.} \emph{et~al.}
\newblock \bibinfo{title}{Observation of superconducting diode effect}.
\newblock \emph{\bibinfo{journal}{Nature}} \textbf{\bibinfo{volume}{584}},
  \bibinfo{pages}{373--376} (\bibinfo{year}{2020}).

\bibitem{baumgartner2022supercurrent}
\bibinfo{author}{Baumgartner, C.} \emph{et~al.}
\newblock \bibinfo{title}{Supercurrent rectification and magnetochiral effects
  in symmetric josephson junctions}.
\newblock \emph{\bibinfo{journal}{Nat. Nano.}} \textbf{\bibinfo{volume}{17}},
  \bibinfo{pages}{39--44} (\bibinfo{year}{2022}).

\bibitem{wu2022field}
\bibinfo{author}{Wu, H.} \emph{et~al.}
\newblock \bibinfo{title}{The field-free josephson diode in a van der waals
  heterostructure}.
\newblock \emph{\bibinfo{journal}{Nature}} \textbf{\bibinfo{volume}{604}},
  \bibinfo{pages}{653--656} (\bibinfo{year}{2022}).

\bibitem{lin2022zero}
\bibinfo{author}{Lin, J.-X.} \emph{et~al.}
\newblock \bibinfo{title}{Zero-field superconducting diode effect in
  small-twist-angle trilayer graphene}.
\newblock \emph{\bibinfo{journal}{Nat. Phys.}} \textbf{\bibinfo{volume}{18}},
  \bibinfo{pages}{1221--1227} (\bibinfo{year}{2022}).

\bibitem{hou2023ubiquitous}
\bibinfo{author}{Hou, Y.} \emph{et~al.}
\newblock \bibinfo{title}{Ubiquitous superconducting diode effect in
  superconductor thin films}.
\newblock \emph{\bibinfo{journal}{Phys. Rev. Lett.}}
  \textbf{\bibinfo{volume}{131}}, \bibinfo{pages}{027001}
  (\bibinfo{year}{2023}).

\bibitem{moler1998images}
\bibinfo{author}{Moler, K.~A.}, \bibinfo{author}{Kirtley, J.~R.},
  \bibinfo{author}{Hinks, D.}, \bibinfo{author}{Li, T.} \& \bibinfo{author}{Xu,
  M.}
\newblock \bibinfo{title}{Images of interlayer $\mathrm{J}$osephson vortices in
  $ \mathrm{Tl_2Ba_2CuO_{6+ \delta}}$}.
\newblock \emph{\bibinfo{journal}{Science}} \textbf{\bibinfo{volume}{279}},
  \bibinfo{pages}{1193--1196} (\bibinfo{year}{1998}).

\bibitem{krasnov1997fluxon}
\bibinfo{author}{Krasnov, V.}, \bibinfo{author}{Oboznov, V.} \&
  \bibinfo{author}{Pedersen, N.~F.}
\newblock \bibinfo{title}{Fluxon dynamics in long josephson junctions in the
  presence of a temperature gradient or spatial nonuniformity}.
\newblock \emph{\bibinfo{journal}{Physical Review B}}
  \textbf{\bibinfo{volume}{55}}, \bibinfo{pages}{14486} (\bibinfo{year}{1997}).

\bibitem{sundaresh2023diamagnetic}
\bibinfo{author}{Sundaresh, A.}, \bibinfo{author}{V{\"a}yrynen, J.~I.},
  \bibinfo{author}{Lyanda-Geller, Y.} \& \bibinfo{author}{Rokhinson, L.~P.}
\newblock \bibinfo{title}{Diamagnetic mechanism of critical current
  non-reciprocity in multilayered superconductors}.
\newblock \emph{\bibinfo{journal}{Nat. Comm.}} \textbf{\bibinfo{volume}{14}},
  \bibinfo{pages}{1628} (\bibinfo{year}{2023}).

\bibitem{bishop2023pytdgl}
\bibinfo{author}{Bishop-Van~Horn, L.}
\newblock \bibinfo{title}{pytdgl: Time-dependent ginzburg-landau in python}.
\newblock \emph{\bibinfo{journal}{Comp. Phys. Comm.}} \bibinfo{pages}{108799}
  (\bibinfo{year}{2023}).

\bibitem{watts1981nonequilibrium}
\bibinfo{author}{Watts-Tobin, R.~J.}, \bibinfo{author}{Kr{\"a}henb{\"u}hl, Y.}
  \& \bibinfo{author}{Kramer, L.}
\newblock \bibinfo{title}{Nonequilibrium theory of dirty, current-carrying
  superconductors: Phase-slip oscillators in narrow filaments near
  $\mathrm{T_c}$}.
\newblock \emph{\bibinfo{journal}{Journal of Low Temp. Phys.}}
  \textbf{\bibinfo{volume}{42}}, \bibinfo{pages}{459--501}
  (\bibinfo{year}{1981}).

\end{thebibliography}

\section*{Acknowledgments}
We thank Y. Xie, J. Cremer, A. Hamo, T. Werkmeister for inspiring discussions. A.Y. acknowledges support from the Army Research Office under Grant numbers: W911NF-22-1-0248 and W911NF-21-2-0147, the Gordon and Betty Moore Foundation through Grant GBMF 9468, and by the Quantum Science Center (QSC), a National Quantum Information Science Research Center of the U.S. Department of Energy (DOE). S.C. and S.P. acknowledges partial support from the Harvard Quantum Initiative in Science and Engineering. A.S. Acknowledges support by the European Union’s Horizon 2020 research and innovation program (Grant Agreement LEGOTOP No. 788715), the German Research Foundation DFG (CRC/Transregio 183, EI 519/7-1), and the Israel Science Foundation Quantum Science and Technology (2074/19). P.M. acknowledges financial support through SNSF project No. 188521. B. I. H. acknowledges support from NSF grant DMR 1231319.

\section*{Author Contributions Statement}
S.C., S.P., U.V. and A.Y. conceived and designed the experiments; S.C. and S.P. prepared the devices, performed the electrical transport and magnetometry measurements, analyzed and visualized the data with input from U.V. and A.Y.; S.C. and S.P. carried out the simulation and analysis with input from A.S., B.I.H. and A.Y.; D.A.B., M.F. and P.M. carried out the current reconstruction using machine learning method; T.Z. fabricated the diamond probes; S.C. and S.P. wrote the manuscript with input from U.V., A.S., B.I.H., A.Y., and contributions from all co-authors.

\section*{Competing Interests Statement}
A.Y., S.C., E.P., U.V., N.M., A.S., and B.I.H. have applied for a patent partially based on this work. The other authors declare no competing interests.




\clearpage

\onecolumngrid

\section*{Main Text Figures}
\begin{figure*}[ht]
\includegraphics[width=6.5in]{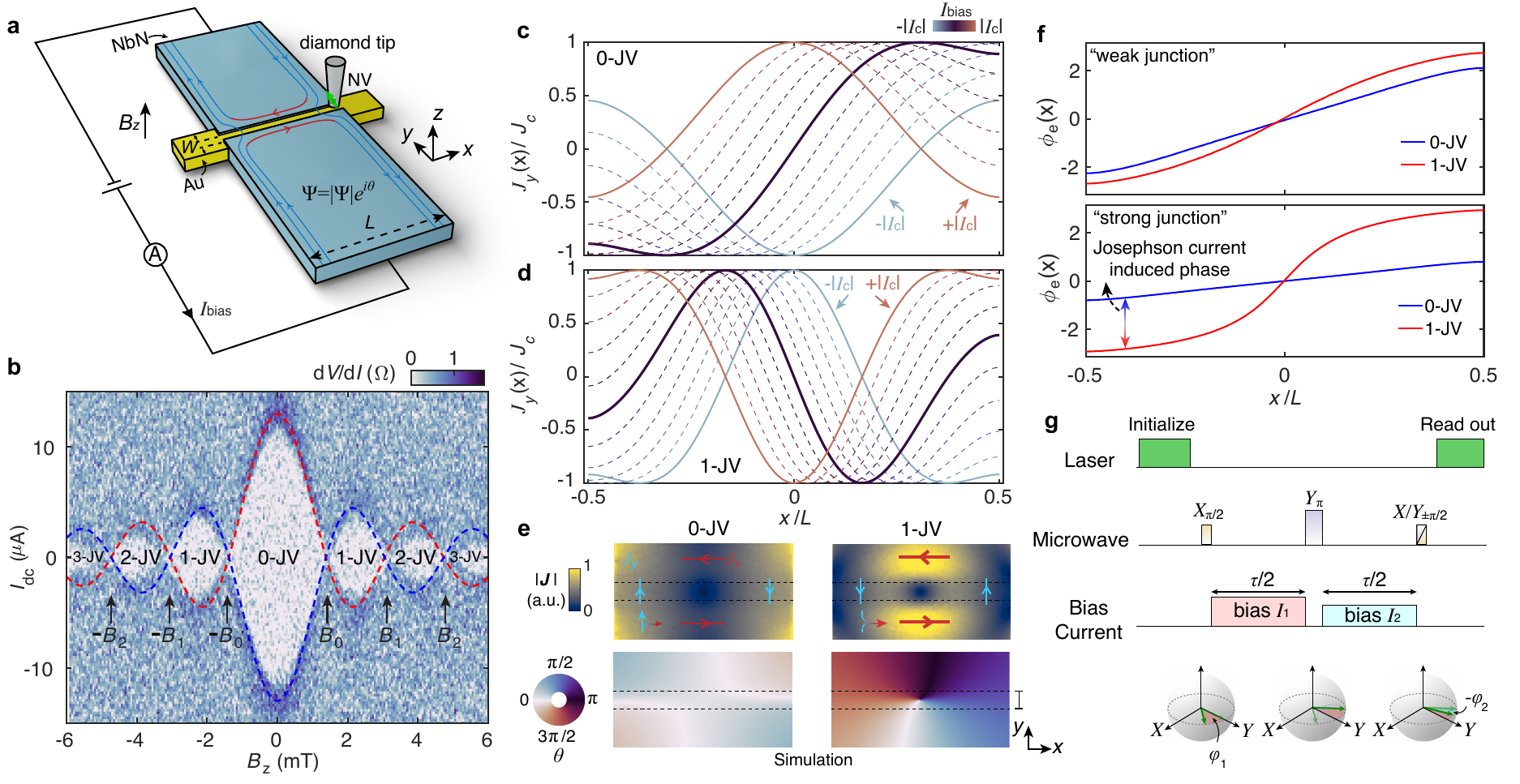} 
\centering
\caption{\textbf{Measurement setup and expected Josephson current flow.}
\textbf{a,} Schematics showing SC-normal-SC junction measured by scanning NV center embedded in a diamond tip. The SC wave function can be described by an amplitude and phase $\Psi=|\Psi|e^{i\theta}$. Under external magnetic field $B_z$, the screening current near the JJ (red lines) induces a phase difference $\phi_e(x)$. The bias current causes phase difference between the SC electrodes $\phi_\mathrm{bias}$.
\textbf{b,} Measured differential resistance $dV/dI$ versus perpendicular magnetic field $B_z$ and bias current $I_{\mathrm{dc}}$, at $T=7$ K. Dashed lines are the expected critical current (see \ref{SI:JJ_externalfield}), where red is $\phi_\mathrm{bias}= \pi/2$, blue is $\phi_\mathrm{bias}=-\pi/2$. The $I_c$ nodes are denoted as $\pm B_n$.
\textbf{c-d,} Calculated Josephson current normalized by critical current density for 0- and 1-JV states, at external $B_z = 1.10$ mT in \textbf{(c)}, and $B_z=1.91$ mT in \textbf{(d)}. The current flow is sine-like at zero bias current (bold lines), shifts along $x$ direction at finite bias current and becomes cosine-like at the critical current. 
\textbf{e,} Simulations showing the Josephson current flow (top) and local SC phase (bottom) of the 0- and 1-JV states. The screening current near the junction $J_x$ (red arrows) is reduced by the Josephson current $J_y$ (cyan arrows) in the 0-JV state, and enhanced in the 1-JV state. This causes the Josephson current induced phase. $I_\mathrm{bias}=0$, $B_z \approx1.2$ mT in this simulation.
\textbf{f,} Simulated $\phi_e(x)$ for 0- and 1-JV states, at the same $B_z$ as \textbf{(e)}. 
$\phi_e(x)$ is the difference of $\theta$ taken along the two dashed lines in each sub-panel of \textbf{(e)}.
\textbf{g,} NV control and current bias sequence based on the ac magnetometry protocol. The X(Y) microwave (MW) pulses rotate the qubit around the X(Y) axis by $\frac{\pi}{2}$ or $\pi$. NV qubit is put on the equator of the Bloch sphere and rotated by the magnetic field generated by current flow. Pulses of different bias current are synced with the MW pulses such that the final signal is the difference between the two current flow patterns.
}
\label{fig:1}
\end{figure*}

\begin{figure*}[ht]
\includegraphics[width=7in]{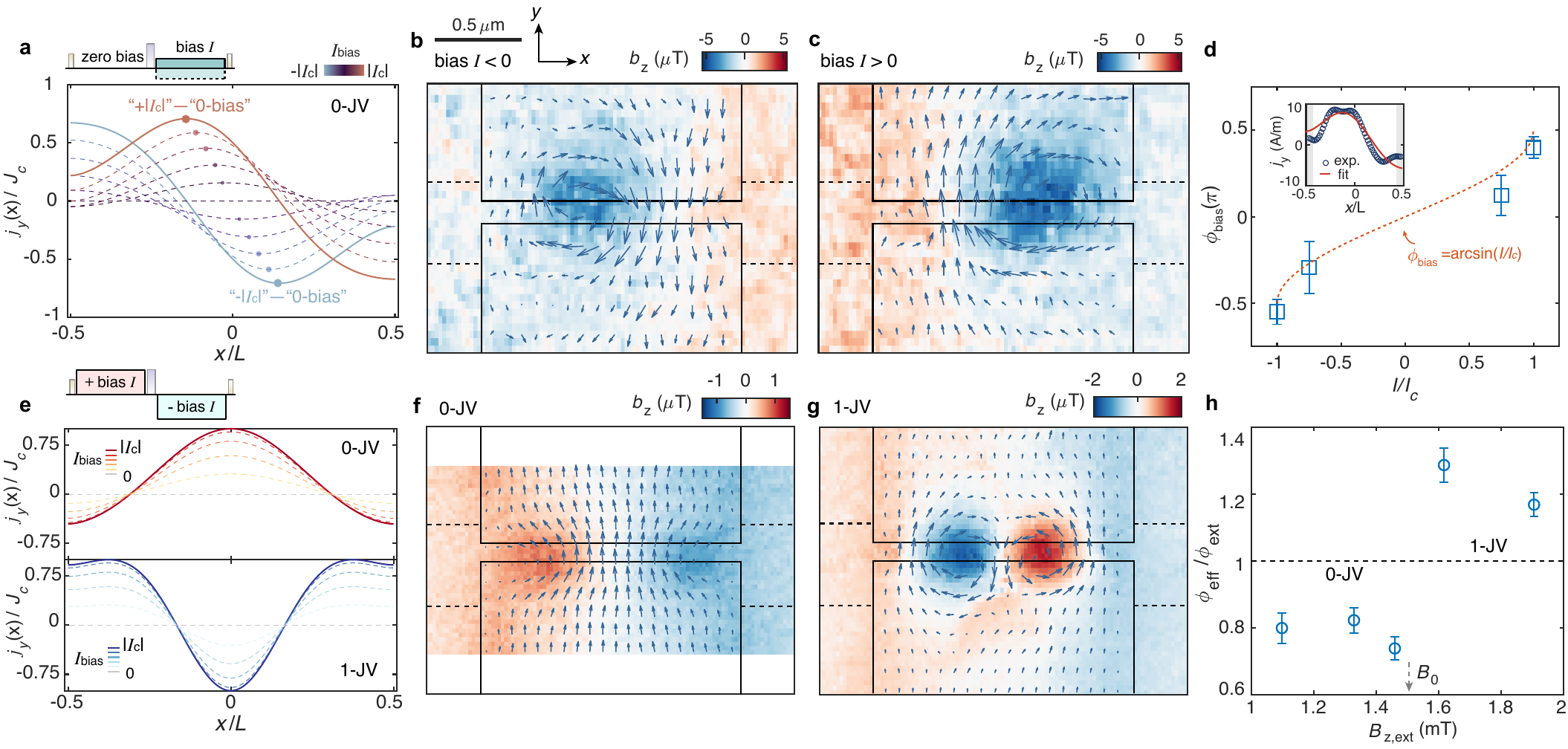} 
\centering
\caption{\textbf{Visualizing Josephson current response to bias current and magnetic field.}
\textbf{a,} Top sketch shows the measurement sequence that takes the difference between zero and finite current bias. Expected current flow show changing line shapes at various bias current $I_{\mathrm{bias}}$, with the opposite bias current leading to inverted current flow pattern around the center of the junction. 
\textbf{b-c,} Colour map shows $z$-component of the current-generated magnetic field $b_z$. The arrows show the reconstructed current flow vector. Results are measured using the sequence shown in \textbf{(a)}. External magnetic field is $B_z = 0.95$ mT. The SC electrodes are marked by solid lines and normal metal part is marked by dashed lines. 
\textbf{d,} Phase difference caused by bias current $\phi_{\mathrm{bias}}$ extracted from the current flow profile at the junction. The result agrees with the sinusoidal current phase relation. Inset shows the measured $j_y(x)$ at $I_\mathrm{bias}=I_c$, and the fitting results using Supplementary Eqn. \ref{eqn:deltaphifit}. The data in the gray area is excluded in the fitting.
\textbf{e,} Top sketch shows the measurement sequence that takes the difference between symmetric positive and negative $I_\mathrm{bias}$. The expected signals change sign when measuring the 0- and 1-JV states at their respective $|I_c|$. All signals show cosine-like shape, with amplitude growing with $I_{\mathrm{bias}}$.
\textbf{f-g,} $b_z$ and current flow vector maps measured using the sequence shown in \textbf{e)}. The Josephson current changes sign between 0- and 1-JV states. External magnetic field is $B_z=1.10$ mT in \textbf{(f)}, and $B_z= 1.91$ mT in \textbf{(g)}.
\textbf{h,} Effective phase difference across the junction $\phi_{\eff}=\phi_e|_{x=W/2}$ deviates from the external field contribution $\phi_\mathrm{ext}$ (indicated by the shaded area), as a result of induced phase from the Josephson current. 
Scale bar is 0.5 $\mu$m for panels \textbf{(b),(c),(f),(g)}. The measurements are taken at $T=7$ K. Error bars in \textbf{(d),(h)} represent standard deviation from fitting.
}
\label{fig:2}
\end{figure*}

\begin{figure*}[ht]
\includegraphics[width=6.5in]{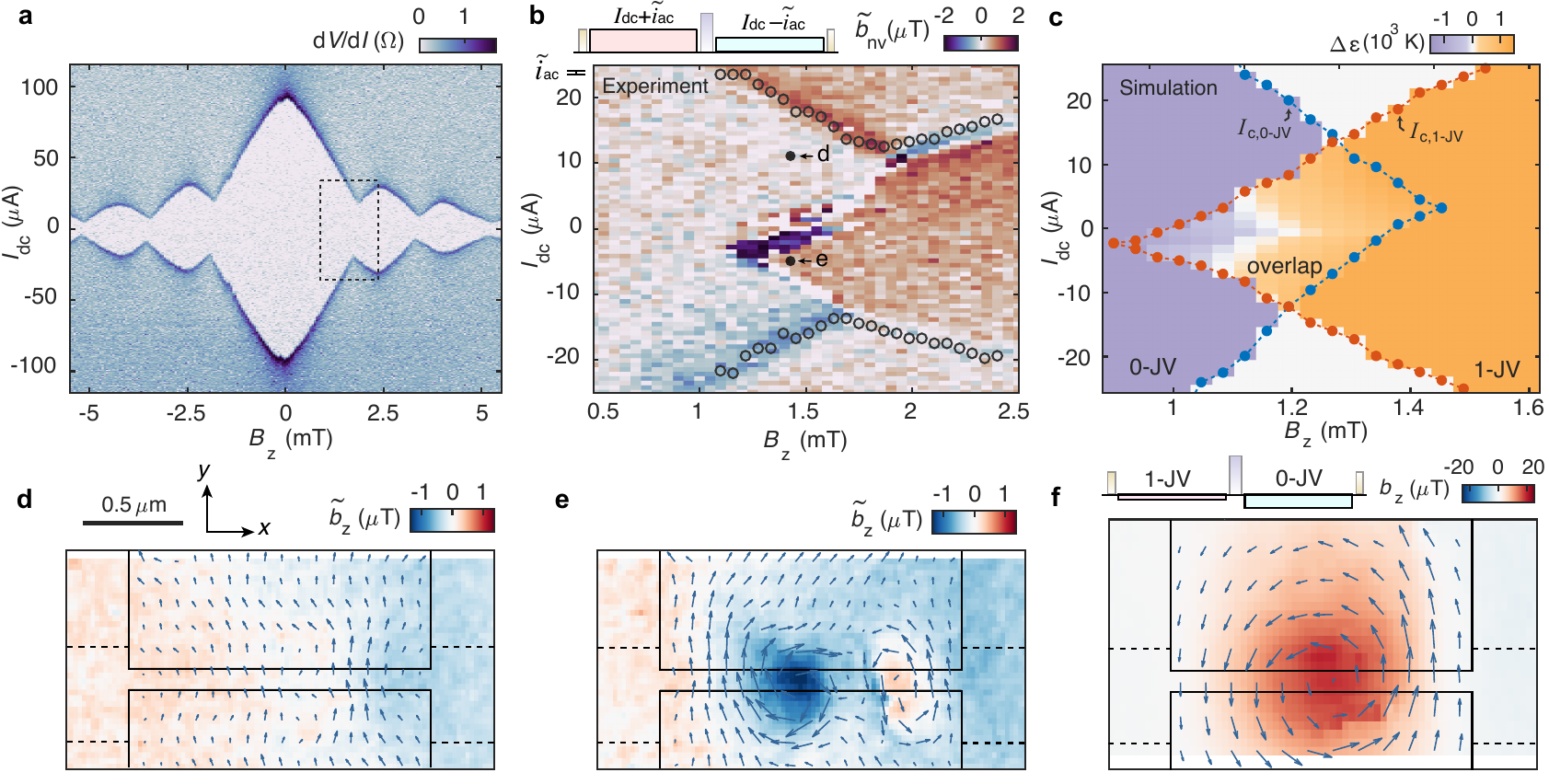} 
\centering
\caption{\textbf{Electric control of JV ground state below critical current.}
\textbf{a,} Differential resistance $dV/dI$ measured in the same device as Fig. \ref{fig:1}c but at $T=4$ K, showing order-of-magnitude increase and oscillation of $I_c$ that does not reach zero. The dashed box shows the measurement range in \textbf{(b)}.
\textbf{b,} Top sketch shows measurement sequence. Main panel shows the differential magnetic field projected along NV axis $\Tilde{b}_{\mathrm{nv}}$, generated by the current flow response to the small ac current $\Tilde{i}_{\mathrm{ac}}\approx0.8 \mu$A. $\Tilde{b}_{\mathrm{nv}}$ is shown versus dc bias current $I_{\mathrm{dc}}$ and perpendicular magnetic field $B_z$. The NV is fixed at $(x,y)\approx(-500,0)$ nm, where $x=y=0$ is the center of JJ. Circles are critical current extracted from transport result in \textbf{(a)}. 
\textbf{c,} Difference of Gibbs free energy between the 0- and 1-JV states below $I_c$, $\Delta \varepsilon = \varepsilon_\mathrm{0V}-\varepsilon_\mathrm{1V}$, from TDGL simulations.  Blue (red) circles show simulated critical current of state with 0- (1-) JV. The colour in the non-overlap region is saturated to indicate only 0-and 1-JV state is present.
\textbf{d-e,} Spatial maps of differential $\Tilde{b}_z$ and current flow vector measured with sequence shown in \textbf{(b)}, showing both \textbf{(d)} 0-JV and \textbf{(e)} 1-JV states at the same external field $B_z=1.4$ mT. The features in \textbf{(e)} are positioned asymmetrically because the JV is at right-of-center of the junction due to finite $I_{\mathrm{dc}}$.
\textbf{f,} Full vortex profile is observed when using sequence that takes the difference between 0- and 1-JV states.
}
\label{fig:3}
\end{figure*}

\begin{figure*}[ht]
\includegraphics[width=3.95in]{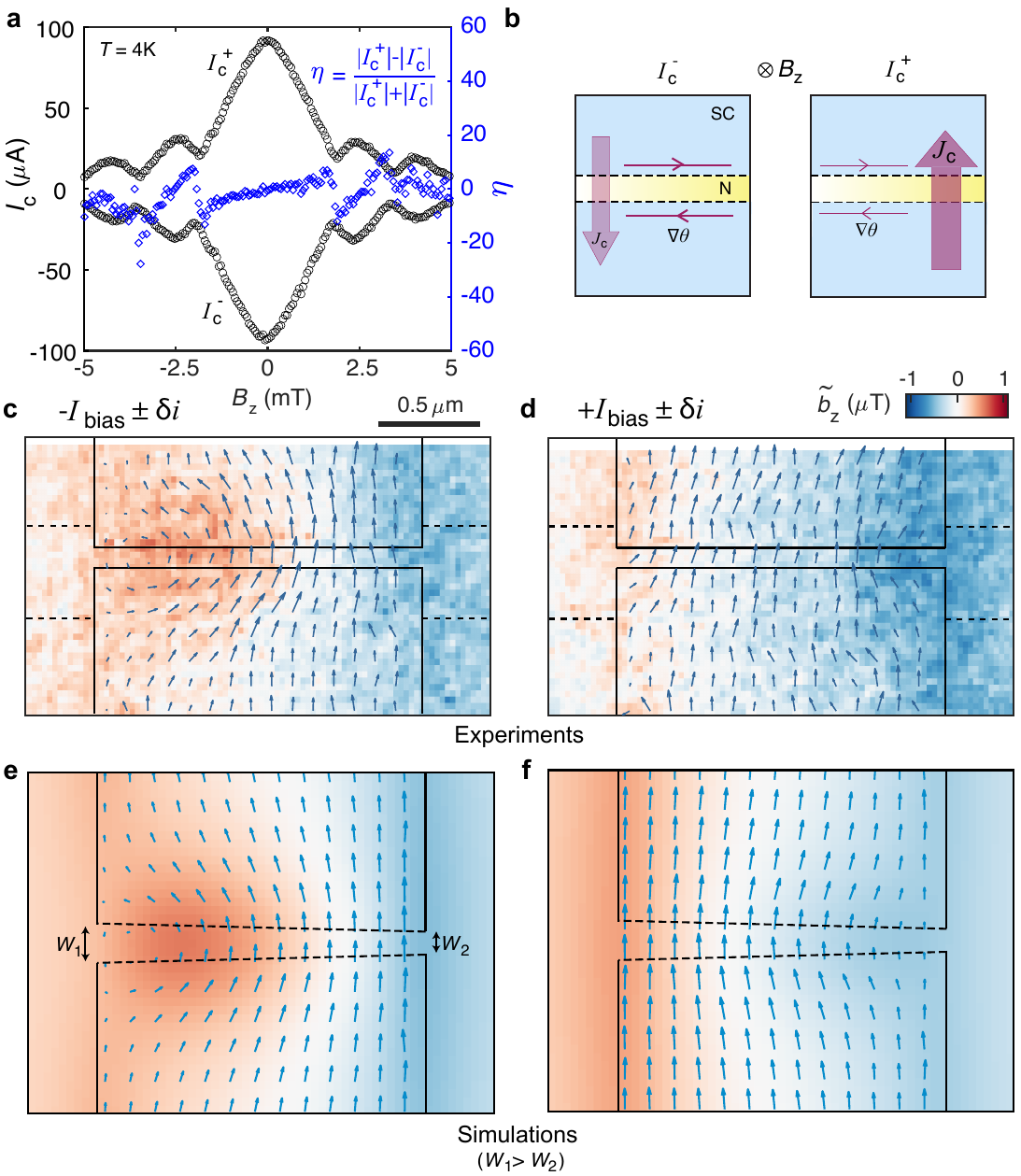} 
\centering
\caption{\textbf{A new mechanism for the Josephson diode effect.}
\textbf{a,} Left axis, forward and backward critical current extracted from results measured in Fig. \ref{fig:3}a. Right axis, asymmetry parameter $\eta$ showing JDE when time-reversal symmetry is broken by $B_z$.
\textbf{b,} Schematics showing inversion symmetry breaking (non-uniform $J_c$) at the junction can lead to different forward and backward critical current.
\textbf{c-d,} Differential magnetic field $\Tilde{b}_z$ and current flow vector measured at symmetric $\pm I_{\mathrm{bias}}$ show inversion asymmetric patterns. Measurements are taken at $B_z=0.5$ mT, using the same sequence as described in Fig. \ref{fig:3}b.
\textbf{e-f,} Simulated results corresponding to \textbf{(c)} and \textbf{(e)}, when inversion symmetry at the junction is broken, modeled as non-uniform junction width $W_1>W_2$. 
The measurements in \textbf{(a)}, \textbf{(c)} and \textbf{(d)} are taken at $T= 4$ K.
}
\label{fig:4}
\end{figure*}
\clearpage

\renewcommand{\thefigure}{\arabic{figure}}
\renewcommand{\figurename}{Supplementary Figure}
\renewcommand{\thesection}{Supplementary Note \arabic{section}}
\renewcommand{\thesubsection}{\arabic{subsection}}
\renewcommand{\theequation}{\arabic{equation}}
\renewcommand{\tablename}{Supplementary Table }
\renewcommand{\thetable}{\arabic{table}}
\setcounter{figure}{0} 
\setcounter{equation}{0} 
\setcounter{page}{1}
\onecolumngrid
\setcounter{secnumdepth}{1}
\section*{Supplementary Information}

\section{Penetration length measured by NV magnetometry}
\label{SI:PearlLength}

\begin{figure*}[ht]
\centering
\includegraphics[width=4.75in]{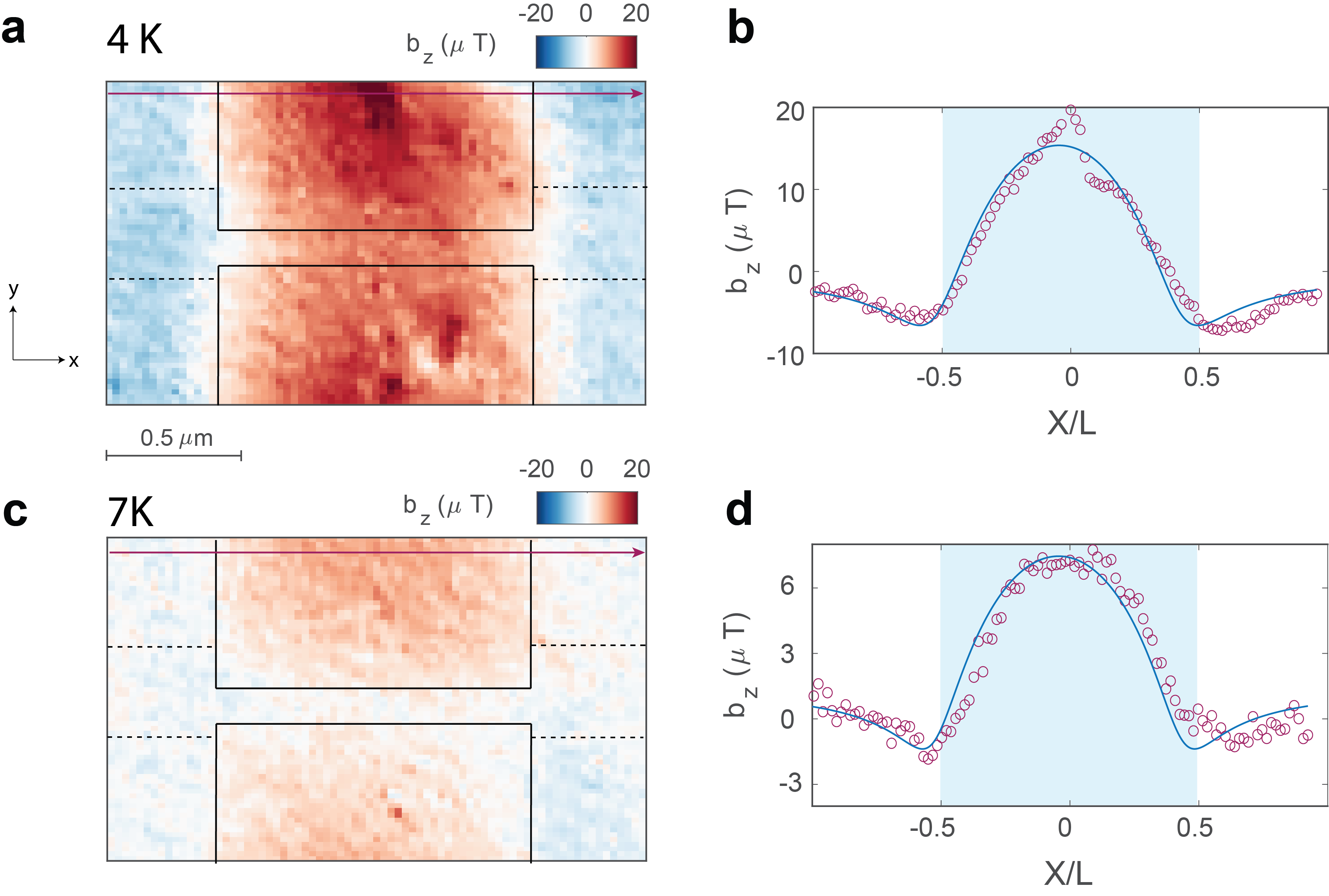} 
\caption{\textbf{Penetration length measured with NV magnetometry.} 
\textbf{a,} $z$-component of the Meissner screening field $b_z$ over the JJ and the SC electrodes. The measurement is done at external field $B_z=0.5$ mT, and $T=4$ K. The solid lines indicate SC electrodes and the dashed line is the normal metal. 
\textbf{b,} Circles show the measured Meissner field $b_z$ along the red line in (A), which is far away from the junction region. 
The line is the calculated magnetic field $b_z$ at the NV, generated by the Meissner current using Biot-Savart law. The Meissner current is obtained using the fitting result of $\lambda_p$ and Supplementary Eqn. \ref{eq:pearlfit}. Blue shaded area indicate the extent of the SC electrode. The kink in $b_z$ near $x=0$ is a measurement artifact.
\textbf{c-d,} are similar results as \textbf{a-b}, but measured at $T=7$ K.
}
\label{fig:S_Pearl}
\end{figure*}

In this section, we measure the Meissner screening of NbN electrodes with NV magnetometry and extract the penetration length. We utilize a NV control sequence that only has the $\frac{\pi}{2}$-pulses at the start and the end (Ramsey sequence), which directly measures the Meissner field generated by the SC, projected along the NV axis. The measurements are done at external magnetic field $B_z = 0.5$ mT where no vortex is present in either the JJ or the SC. The bias current is zero. We show the $z$-component of the Meissner field around the junction at 4K (Supplementary Fig. \ref{fig:S_Pearl}a), and 7K (Supplementary Fig. \ref{fig:S_Pearl}c). Stronger screening can be seen at 4K compared with 7K, indicating a shorter penetration length. 

To extract the absolute value of the Pearl length $\lambda_p$, the Meissner field $b_z$ is fit by numerically solving the 1D London equation, which applies a SC strip that is narrow in the $x$-direction and long in the $y$-direction \cite{rohner2018real}. The total magnetic field consists of the external field $B_{z,ext}$, and the Meissner screening field. Using the second London equation,

\begin{equation}\label{eq:pearlfit}
  B_{z,ext} + \frac{\mu_0}{2\pi} \int_{-\frac{L}{2}}^{\frac{L}{2}} \frac{J_y (x')}{x'-x} \, dx' = - \mu_0 \lambda_p \frac{\partial J_y(x)}{ \partial x}
\end{equation}
where $\lambda_p = \lambda_L^2/t$ is the pearl length, $\lambda_L$ is the London penetration length, $t$ is thickness of NbN, $J_y(x)$ is the Meissner sheet current of the NbN film, $\mu_0$ is the vacuum permeability. The integral equation is solved by discretizing the variables, and the Meissner field is compared with the measurement. We take a line cut of $b_z$ at about 500 nm away from normal metal area, to avoid influence from the Josephson current. The $b_z$ is fit with Supplementary Supplementary Eqn. \ref{eq:pearlfit} to extract $\lambda_p$, using $B_{z,ext}=0.5 $ mT, $L=1.5~\mu$m. 
We find $\lambda_p = 4.7 \pm 0.4\mu$m at 4 K (Supplementary Fig. \ref{fig:S_Pearl}b), and 
13.7 $\pm$ 1.4 $\mu$m at 7 K (Supplementary Fig. \ref{fig:S_Pearl}d). The corresponding London penetration length is $\lambda_L = 410 \pm$ 20 nm at 4 K and 690 $\pm$ 35 nm at 7 K. In comparison, previous indirect measurements of $\lambda_p$ range from 1 to 5 $\mu$m at 4 K \cite{ilin2014critical,charaev2017proximity,hu2003sputter,kamlapure2010measurement}.

With these results we can calculate the Josephson penetration length, $\lambda_J = \sqrt{\frac{\Phi_0 Lt}{4\pi\mu_0J_c \lambda_L^2}}$. In our device, $L=1.5~\mu$m, thickness $t=35$ nm, assuming uniform $J_c = I_c/L$ at zero magnetic field, we find $\lambda_J  = 780 \pm 40$ nm at $T = 4$ K, and $\lambda_J  = 1470 \pm 75$ nm at $T = 7$ K.

\clearpage

\section{Evolution of Josephson current flow with external magnetic field in weak junctions} \label{SI:JJ_externalfield}

In this section, we evaluate how external magnetic field $B_z$ contributes to the evolution of Josephson current flow and JVs. As discussed in the main text, $B_z$ affects $\phi_e(x)$, the profile of the super current, and modify the number of vortices trapped inside the junction. In the thin film, 1-D line junction ($W \ll L$), and weak junction limit, $\phi_e(x) \propto B_z \cdot \sigma(x)$, where $\sigma(x)$ is a non-linear function shown in Supplementary Eqn. \ref{eqn:nonlocalphie} \cite{clem2010josephson}. 

\begin{figure*}[ht]
\centering
\includegraphics[width=5.25in]{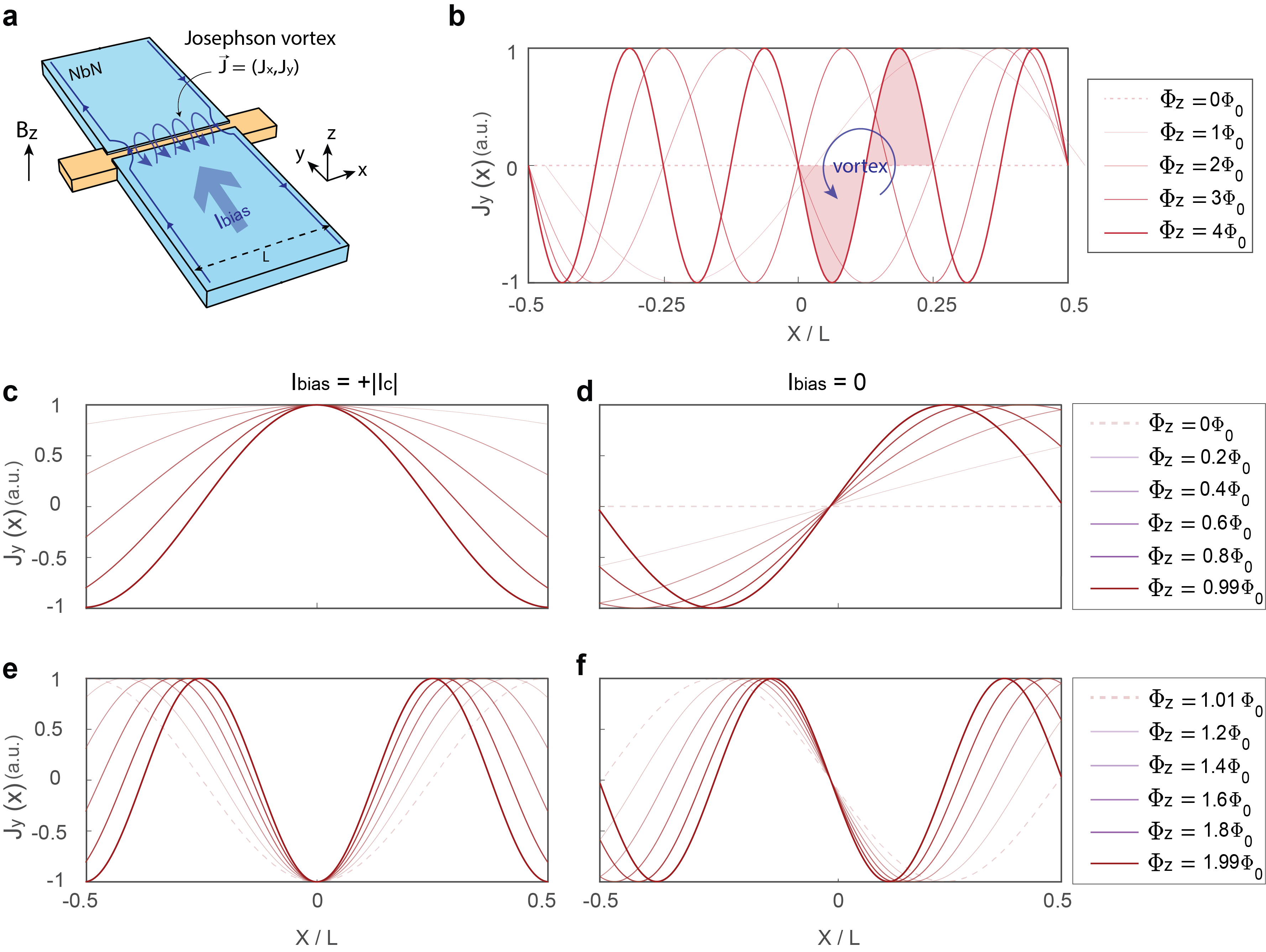} 
\caption{\textbf{Evolution of Josephson current profile with external magnetic flux.}
\textbf{a,} Schematics of Josephson current flow for $\Phi_z = 4 \Phi_0$. Blue arrows on the device is how the current flows for $I_\mathrm{bias}=0$. 
\textbf{b,} Line cut of $J_y$ at the junction with various fields from zero-flux to $\Phi_z = 4\Phi_0$. Shaded area indicates a JV.
\textbf{c,e} $I_\mathrm{bias} = +|I_c|$ and \textbf{d,f} $I_\mathrm{bias} = 0$ current flow in different magnetic flux. \textbf{c-d} is flux between 0 to 1$\Phi_0$ and \textbf{e-f} is flux between 1 to 2$\Phi_0$.
}
\label{fig:S_extfluxprofile}
\end{figure*}

Here, we simplify $\sigma(x)$ to a linear function to more intuitively show the magnetic field effect, $\phi_e(x) = 2\pi\frac{\Phi_z}{\Phi_0} \frac{x}{L}$. $\Phi_z = B_z\cdot A$ is the magnetic flux through the junction, $A$ is the junction area. This applies to JJs made with bulk superconductors. Nevertheless it still captures the changes of the Josephson current flow with $B_z$. In this model, when $B_z$ reaches the critical current nodes $\Phi_z = n\Phi_0$, the $n$-th JV enters the JJ (Supplementary Fig. \ref{fig:S_extfluxprofile}b). 

Consider the Gibbs free energy of the junction without external bias current,

\begin{equation}
    G (\phi_\mathrm{bias}) = \frac{\Phi_0}{2\pi}\cdot  [ I_c(\Phi_z) (1-\cos{\phi_\mathrm{bias}}) ]
    \label{Gibbs1}
\end{equation}

$I_c(\Phi_z)$ is the critical current when the external magnetic flux is $\Phi_z$, which changes sign at $\Phi_z=n\Phi_0$ (see Supplementary Eqn. \ref{eqn:S_cpr_totalI} and accompanying text). As a result, the $\phi_\mathrm{bias}$ which corresponds to the free energy minimum shifts by $\pi$, and the local current density, $\propto \sin{[\phi_e(x)+\phi_\mathrm{bias}]}$ changes sign when a Josephson vortex enters/exits the junction.

The periodicity of the oscillating Josephson current $J_y(x)$ shrinks with increasing $\Phi_z$, as seen by the current profile at the critical current (Supplementary Fig. \ref{fig:S_extfluxprofile}c, e), and at zero bias current (Supplementary Fig. \ref{fig:S_extfluxprofile}d, f). The $J_y(x)$ profile changes sign as $\Phi_z$ crosses the node from $0.99 \Phi_0$ to $1.01 \Phi_0$, and around every $n\Phi_0$ thereafter (Supplementary Fig. \ref{fig:S_extfluxprofile}d, f). 

We note that in the thin film limit and weak junctions, the JVs enter the junction at critical current nodes $B_n$ as mentioned in the main text. But at the nodes, the magnetic flux through the effective area $A_{\eff}=L^2/1.842$, $B_n\cdot A_{\eff}$ is not exactly $n\Phi_0$ \cite{clem2010josephson}. For example, at $B_0$ the magnetic flux through the effective area is 0.8173 $\Phi_0$, as shown in Supplementary Table \ref{table:Bn}. Nevertheless, the $J_y(x)$ periodicity and sign changes with $B_z$ still apply.

In summary, external magnetic flux manipulates Josephson current flow by changing the current profile and the number of JV, making it an important control knob in engineering SC devices.

\clearpage

\section{Josephson current flow and SC phase difference $\phi_{\mathrm{bias}}$ at finite magnetic flux} \label{SI:currenteffectinphase}

In this section, we derive how the bias current $I_\mathrm{bias}$ controls the phase difference between SC electrodes $\phi_\mathrm{bias}$, at a finite external magnetic field $B_z$. We show (i) the bias current - $\phi_\mathrm{bias}$ relation at finite $B_z$ and (ii) the $\pi$ phase shift associated with each JV in the junction.

We consider the junction spans between $x \in [-\frac{L}{2}, \frac{L}{2}]$, so $\phi_\mathrm{bias}$ is the phase difference between SC electrodes at $x=0$. We assume the critical current density is constant $J_c$, and start with the simplified case, $\phi_e(x) = 2\pi\frac{\Phi_z}{\Phi_0} \frac{x}{L}$. $\Phi_z = B_z\cdot A$ is the magnetic flux through the junction, $A$ is the junction area. This applies to JJs made with bulk superconductors. The external bias current is 

\begin{equation} 
I_\mathrm{bias}=\int_{-L/2}^{L/2} J(x) \,dx = \int_{-L/2}^{L/2}  J_c \sin{\left(2\pi\frac{\Phi_z}{\Phi_0} \frac{x}{L}+\phi_\mathrm{bias}\right)} \,dx = J_cL \mathrm{sinc}\left( \pi\frac{\Phi_z}{\Phi_0} \right) \cdot \sin{\phi_\mathrm{bias}}
\label{eqn:S_cpr_totalI}
\end{equation}

This shows the sinusoidal current-phase relation still applies at finite field. Here $I_c(\Phi_z) = J_cL \cdot \mathrm{sinc}\left( \pi\frac{\Phi_z}{\Phi_0} \right)$ is the critical current at finite flux $\Phi_z$. As a result, $\phi_\mathrm{bias}$ can be controlled by $I_\mathrm{bias}$ via
\begin{equation} 
    \phi_\mathrm{bias} = \arcsin\left(\frac{I_\mathrm{bias}}{I_c(\Phi_z)}\right) + n\pi
\label{eqn:S_cpr_dphi_I}
\end{equation}
where $n$ is the number of JV, which shifts the phase difference by $n\pi$. Intuitively, each JV has $2\pi$ phase winding around itself and this leads to $\pi$ phase difference at the center of the junction $x=0$. 

The phase shift due to JV can also be understood from an effective Gibbs free energy of the junction. The bias current adds a term to Eq. (\ref{Gibbs1}), giving
\begin{equation}
    G = \frac{\Phi_z}{2\pi}\cdot  [ I_c(\Phi_z) (1-\cos{\phi_\mathrm{bias}}) - I_\mathrm{bias} \phi_\mathrm{bias}]
\end{equation}
This is the well known ``washboard'' potential for biased JJ, and for the over-damped junction, the equilibrium $\phi_\mathrm{bias}$ occurs at the local minima of the free energy ($\frac{\partial G}{\partial \phi_\mathrm{bias}}=0$ and $\frac{\partial^2 G}{\partial \phi_\mathrm{bias} ^2}>0$). For odd number of JV at the junction, $I_c(\Phi_z)<0$, which leads to the $\pi$ phase shift when JV enters or exits the junction.

In the thin film weak-junction limit, $\phi_e(x)$ is given by Supplementary Eqn. \ref{eqn:nonlocalphie}. The total current is then given by
\begin{equation} 
\begin{split}
I_\mathrm{bias} &= \mathrm{Im} \int_{-L/2}^{L/2} J_c (x) e^{i [\phi_{e0} \sigma (x)+\phi_\mathrm{bias}]} \,dx \\
&\equiv I_c (B_z) \,\sin [\phi_\mathrm{bias} + \phi_c]
\end{split}
\label{eqn:S_cpr_totalI_clem-BH}
\end{equation}
where
\begin{equation}
I_c(B_z) = \left| \int_{-L/2}^{L/2} J_c (x) e^{i \phi_{e0} \sigma (x)} \,dx \right| ,
\end{equation}
\begin{equation}
 \phi_c =  \arg  \left(\int_{-L/2}^{L/2} J_c (x) e^{i \phi_{e0} \sigma (x)} \,dx \right).
\end{equation}
Equation \ref{eqn:S_cpr_dphi_I} can still apply in the thin film limit. We see that in the weak-junction limit, the dependence of the current on phase remains sinusoidal even if $J_c$ depends on $x$. In particular, when $J_c$ is a constant, $\phi_c=0$, and the current phase relation in Supplementary Eqn. \ref{eqn:S_cpr_dphi_I} can apply.

\clearpage

\section{Thin film Josephson junction and extracting $\Delta \phi$, $\phi_\mathrm{eff}$}

In this section we show (i) transport evidence of the junction being in the thin film limit, and (ii) the fitting methods to extract $\Delta \phi$ and $\phi_\mathrm{eff}$ shown in Fig. \ref{fig:2}d, h in the main text. 

\textbf{i. Thin film SC}
In a junction with $W \ll L$, $\phi_e(x)$ can be derived from the $x$-direction screening currents in the thin film SC leads, treated as semi-infinite strips with the boundary condition of zero Josephson current, $J_y(x)=0$ at $y=0$ (center of the junction). We assume the external contact electrodes are located at positions $y = \pm H$, with $H \gg L$. 
To leading order all screening currents flow within the SC electrodes and hence the boundary condition. One then finds, following Ref. \cite{mints2002nonlocal,clem2010josephson} 

\begin{equation}
\begin{aligned}
\phi_e(x,B_z) &= \frac{16B_zL^2}{\pi^2\Phi_0}\cdot \sigma(\pi x/L) \equiv \phi_{e0} \cdot \sigma(\pi x/L),\\
\sigma(\zeta)&=\sum_{n=0}^{n=\infty}(-1)^n\frac{\sin{(2n+1)\zeta}}{(2n+1)^3}
\end{aligned}
\label{eqn:nonlocalphie}
\end{equation} 

where $\Phi_0$ is the flux quantum. $\sigma(\zeta)$ is an odd function of its argument and may be reasonably approximated by $\sigma(\zeta) \approx \sin \zeta $. The scale of $\phi_e$ is set by the quantity $\phi_{e0} \equiv \phi_e|_{x=L/2} \approx 
1.7 B_z L^2 / \Phi_0$. In this model, the $\phi_e(x)$ is induced by the screening current in the SC electrodes. Its shape is determined by the $\sigma(\zeta)$ function and its amplitude $\phi_{e0}$, is proportional to the magnetic field $B_z$. As mentioned in the main text, this model does not include the Josephson current induced phase in strong junctions. 

Ref. \cite{clem2010josephson} derived the critical current nodes $B_n$ in the thin film limit. The $B_z$ periodicity for the critical current oscillation, in the limit of large magnetic field, is $\Delta B_{\infty} = 1.842 \Phi_0/L^2$. 
Refs. \cite{clem2010josephson, boris2013evidence} showed that in the thin film junction, the nodes $B_n$ are not evenly spaced. In our device, the lithographically defined dimension is $L=1.5\mu$m, thus $\Delta B_{\infty}$ should equal 1.88 mT. The $\Delta B_n = B_{n+1}-B_n$ extracted from our measurement in Fig. \ref{fig:1}b is given in the table \ref{table:Bn}. The normalized $\Delta B_n/\Delta B_{\infty}$ values are close to the theoretical values in Ref. \cite{clem2010josephson}.

\begin{table}[h!]
\begin{tabular}{|c||c|c|c|} 
 \hline
 units & $\Delta B_0$ & $\Delta B_1$ & $\Delta B_2$\\
 \hline
 mT & 1.48 & 1.76 & 1.78\\
 \hline
 Normalized & 0.79 & 0.94 & 0.95\\
\hline
\hline
Theory \cite{clem2010josephson} & 0.8173 & 0.9866 & 0.9946 \\
\hline
\end{tabular}
\caption{Spacing between the $I_c$ nodes $\Delta B_n = B_{n+1}-B_n$. Upper table, first line is in units of mT, second line is normalized by $\Delta B_{\infty}$. Lower table shows theoretical values from Ref. \cite{clem2010josephson}.}
\label{table:Bn}
\end{table}

\begin{figure*}[ht]
\centering
\includegraphics[width=5in]{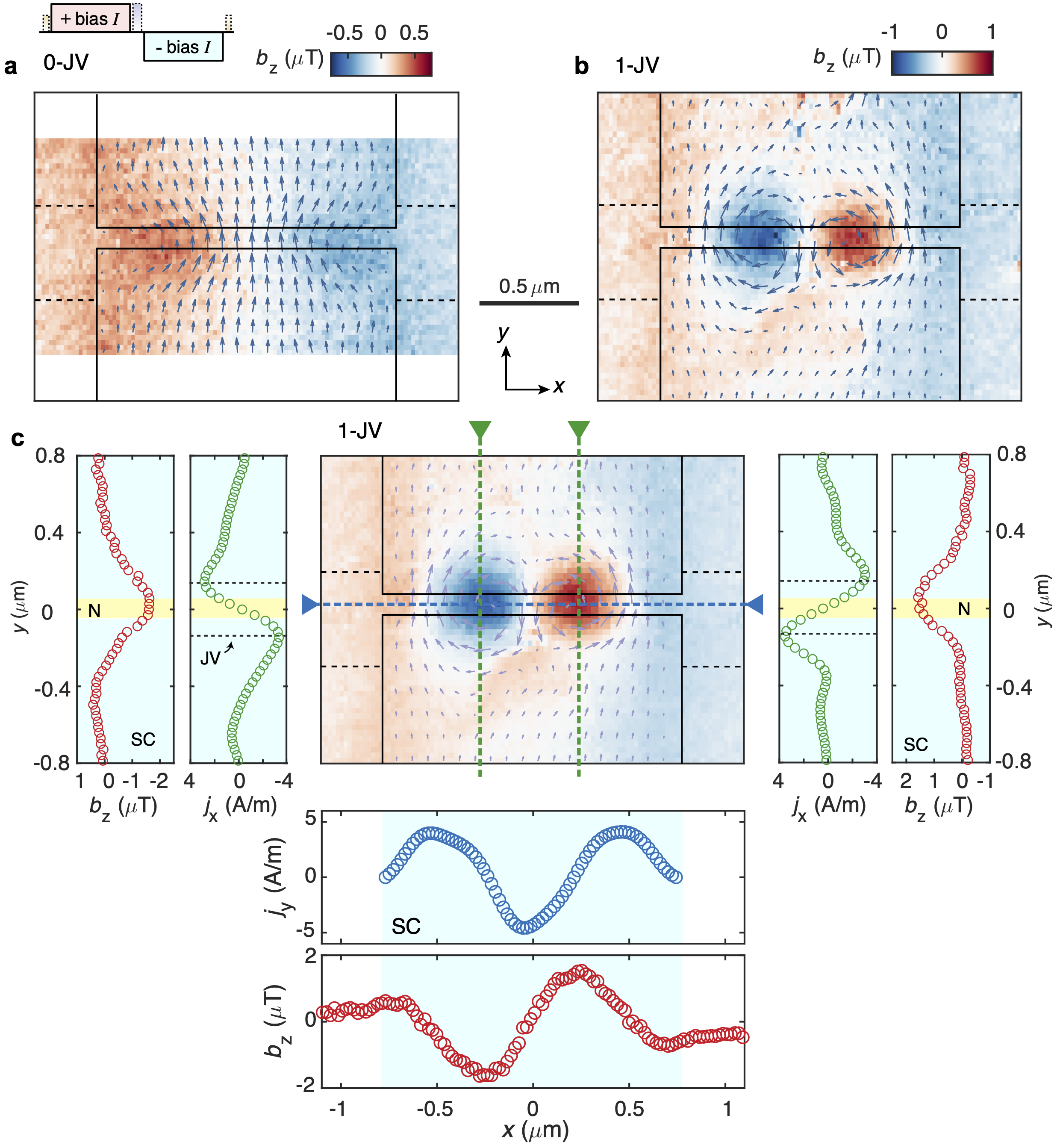} 
\caption{\textbf{Additional measurements using symmetric bias $\pm I_\mathrm{bias}$ sequence.}
\textbf{a-b,} show spatial maps of current flow and $z$-direction magnetic field measured at the same external $B_z$ as Fig. \ref{fig:2}f-g in the main text, but using $I_\mathrm{bias}=0.5I_c$ instead of $I_c$. Here we use colour scales with half the range, and the quiver with double the length per unit current density as in Fig. \ref{fig:2}f-g. The shape of the current flow is almost the same, while the amplitude is half of those in Fig. \ref{fig:2}f-g, as expected. The measurement is done at $T=7$ K.
\textbf{c,} shows the current flow and $b_z$ line cut from Fig. \ref{fig:2}g in the main text. The $j_x(y)$ line trace along the vertical direction shows the JV extends in to the SC electrodes by $\delta W \approx 350$ nm, making the effective area of the junction $A = LW'$, where $W' = W + 2\delta W \approx 850$ nm. This is consistent with the effective area $L^2/1.842$ as derived in Ref. \cite{clem2010josephson}.
}
\label{fig:E_0.5Ic}
\end{figure*}

We note that in most of the literature, a simplified model is used to estimate the periodicity of the Fraunhofer map. It assumes magnetic field penetration through an area $A = LW'=L(W+2\lambda_L)$, where $\lambda_L$ is the London penetration length. So the magnetic field periodicity is $\Delta B_\mathrm{sim} A = \Phi_0$. This applies to JJs made with bulk superconductors ($\lambda_L \gg L$). Translating this to the the thin film SC limit, we get an effective area $A_{\eff}=L^2/1.842$, and $W'_{\eff} = L/1.842$. Supplementary Fig. \ref{fig:E_0.5Ic} shows the size of the JV in the $y$-direction agrees with this $W'_{\eff}$.

\textbf{ii. Extracting the $\phi_\mathrm{bias}$ in Fig. \ref{fig:2}d}

In Fig. \ref{fig:2}b-c, the measurement is taken by subtracting the $I_\mathrm{bias}$ case by the zero bias case. To extract the effective phase difference between SC electrodes $\phi_\mathrm{bias}$, the experimental results are fit to the following equation,
\begin{equation}
j_y(x)=J_c \cdot \left[ \sin(\phi_e(x,B_z) + \phi_\mathrm{bias}) - \sin(\phi_e(x,B_z))\right]
\label{eqn:deltaphifit}
\end{equation}
Here $J_c$ (critical current density) and $\phi_\mathrm{bias}$ are the two fitting parameters, while $L=1.5~\mu$m, $B_z = B_\mathrm{z,ext} = 0.95$ mT are fixed parameters. The fitting results at each $I_\mathrm{bias}$ are shown in Supplementary Fig. \ref{fig:E_iphilines}.

\textbf{iii. Extracting the $\phi_{\eff}$ in Fig. \ref{fig:2}h}

In Fig. \ref{fig:2}f-g, the measurement is taken by subtracting the $I_\mathrm{bias}$ case by the $-I_\mathrm{bias}$ case. The $\phi_\mathrm{eff}$ shown in Fig. \ref{fig:2}h is $\phi_{\eff} = \frac{16B_{\eff}L^2}{\pi^2\Phi_0}$. The external magnetic field induced phase is $\phi_\mathrm{ext} = \frac{16B_z L^2}{\pi^2\Phi_0}$. The experimental results are fit to the following equation,
\begin{equation}
\begin{split}
j_y(x) &= J_c \cdot \left[ \sin(\phi_e(x,B_z) + \phi_\mathrm{bias}) - \sin(\phi_e(x,B_z)- \phi_\mathrm{bias})\right]\\
&= 2J_c \sin \phi_{\mathrm{bias}} \cdot \cos[\phi_e(x,B_\mathrm{eff})]\\
&\equiv J_0 \cdot \cos[\phi_e(x,B_\mathrm{eff})]
\end{split}
\label{eqn:phiefffit}
\end{equation}
Here $J_0$ and $B_\mathrm{eff}$ are the fitting parameters, $L=1.5~\mu$m is fixed. The fitting results are shown in Supplementary Fig. \ref{fig:E_7Kallmap}f-j.

In both cases of fitting $\phi_\mathrm{bias}$ and $\phi_{\eff}$, the portions of the reconstructed $j_y(x)$ with the distance to the edge of the SC smaller than the NV stand-off distance are excluded from the fitting process, to avoid the ringing and distortion effects of the reconstructed result near the edge.

\begin{figure*}[ht]
\centering
\includegraphics[width=6.3in]{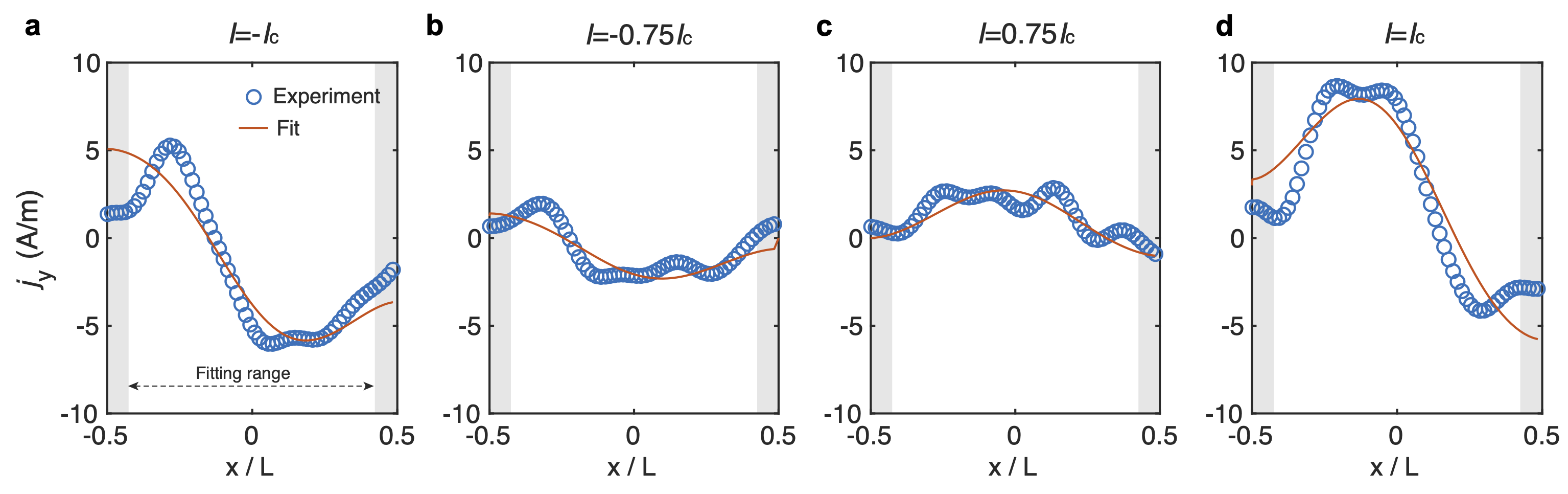} 
\caption{\textbf{Extracting SC phase difference $\phi_\mathrm{bias}$ from current profile measured at different $I_\mathrm{bias}$.} Current flow profile at the center of the JJ measured using the finite to zero bias current sequence, as described in Fig. \ref{fig:2}b-c in the main text. The grey areas which corresponds to regions closer to the JJ edge by the stand-off distance of the NV ($\approx$180 nm), are excluded from the fitting. The bias current in each panel is \textbf{(a)} $-I_c$, \textbf{(b)} $-0.75\cdot I_c$, \textbf{(c)} $0.75\cdot I_c$ and \textbf{(d)} $I_c$. The blue circles represent the reconstructed $j_y$ at the junction, and the red lines represent the fit using sinusoidal current-phase relation and $\phi_\mathrm{bias}$ as fitting parameter. The extracted $\phi_\mathrm{bias}$ is shown in Fig. \ref{fig:2}d in the main text.}
\label{fig:E_iphilines}
\end{figure*}

\begin{figure*}[ht]
\centering
\includegraphics[width=7in]{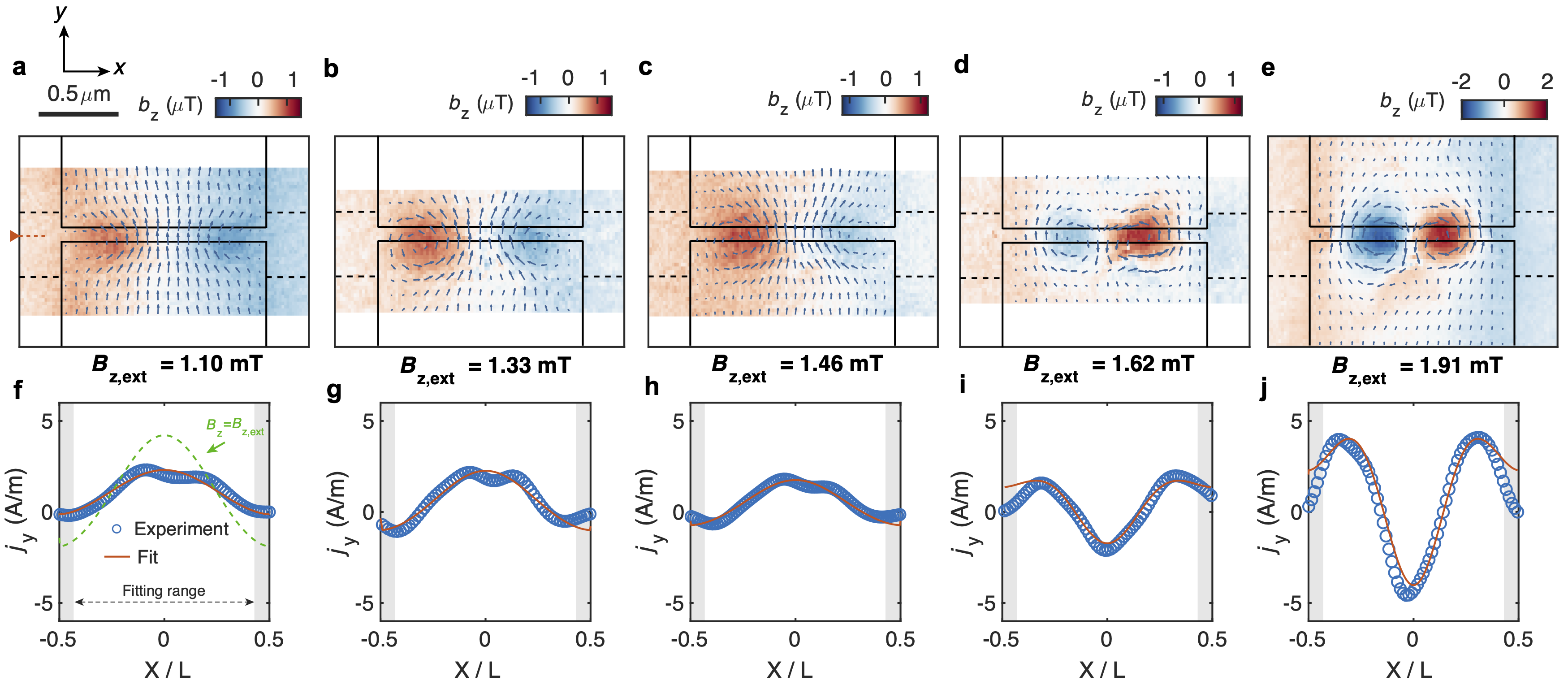} 
\caption{\textbf{Josephson current flow at various magnetic flux around $B_z=B_0$.}
\textbf{a-e,} show the spatial maps of current flow and $z$-direction magnetic field measured using symmetric $\pm I_\mathrm{bias}$ sequence as Fig. \ref{fig:2}f-g in the main text, at external field values as shown by the labels above. \textbf{(a)-(c)} are $B_z<B_0$ and 0-JV; \textbf{(d)-(e)} are $B_z>B_0$ and 1-JV. The Josephson current flow switches sign from 0- to 1-JV. \textbf{(a)} and \textbf{(e)} are the same as Fig. \ref{fig:2}f and g in the main text. The bias current used during the measurement in \textbf{(a)} $I_\mathrm{bias}/I_c \approx 0.7$; \textbf{(b)} $I_\mathrm{bias}/I_c \approx 0.7$; \textbf{(c)} $I_\mathrm{bias}/I_c \approx 0.9$; \textbf{(d)} $I_\mathrm{bias}/I_c \approx 0.6$; \textbf{(e)} $I_\mathrm{bias}/I_c \approx 0.8$. We emphasize again that the normalized shape of $j_y$ is not expected to depend on $I_\mathrm{bias}/I_c$, as shown in Supplementary Fig. \ref{fig:E_0.5Ic}. So the result presented in the main text is insensitive to the exact value of $I_\mathrm{bias}/I_c$.
\textbf{(f-j)} Circles show the reconstructed current flow at the center of JJ extracted from \textbf{(a)-(e)}, and the lines show the fitting to extract effective magnetic field $B_{z,\eff}$, as shown in Fig. \ref{fig:2}h in the main text. The dashed green line in \textbf{(f)} shows that the $j_y(x)$ profile expected from the $\phi_{\mathrm{ext}}$ induced by the external field, which does not match our measurement. The red lines show the fitting results to extract $\phi_{\mathrm{eff}}$. The grey areas which corresponds to regions closer to the JJ edge by the stand-off distance of the NV ($\approx150$ nm), are excluded from the fitting.
}
\label{fig:E_7Kallmap}
\end{figure*}
\clearpage


\begin{figure*}[ht]
\centering
\includegraphics[width=4.5in]{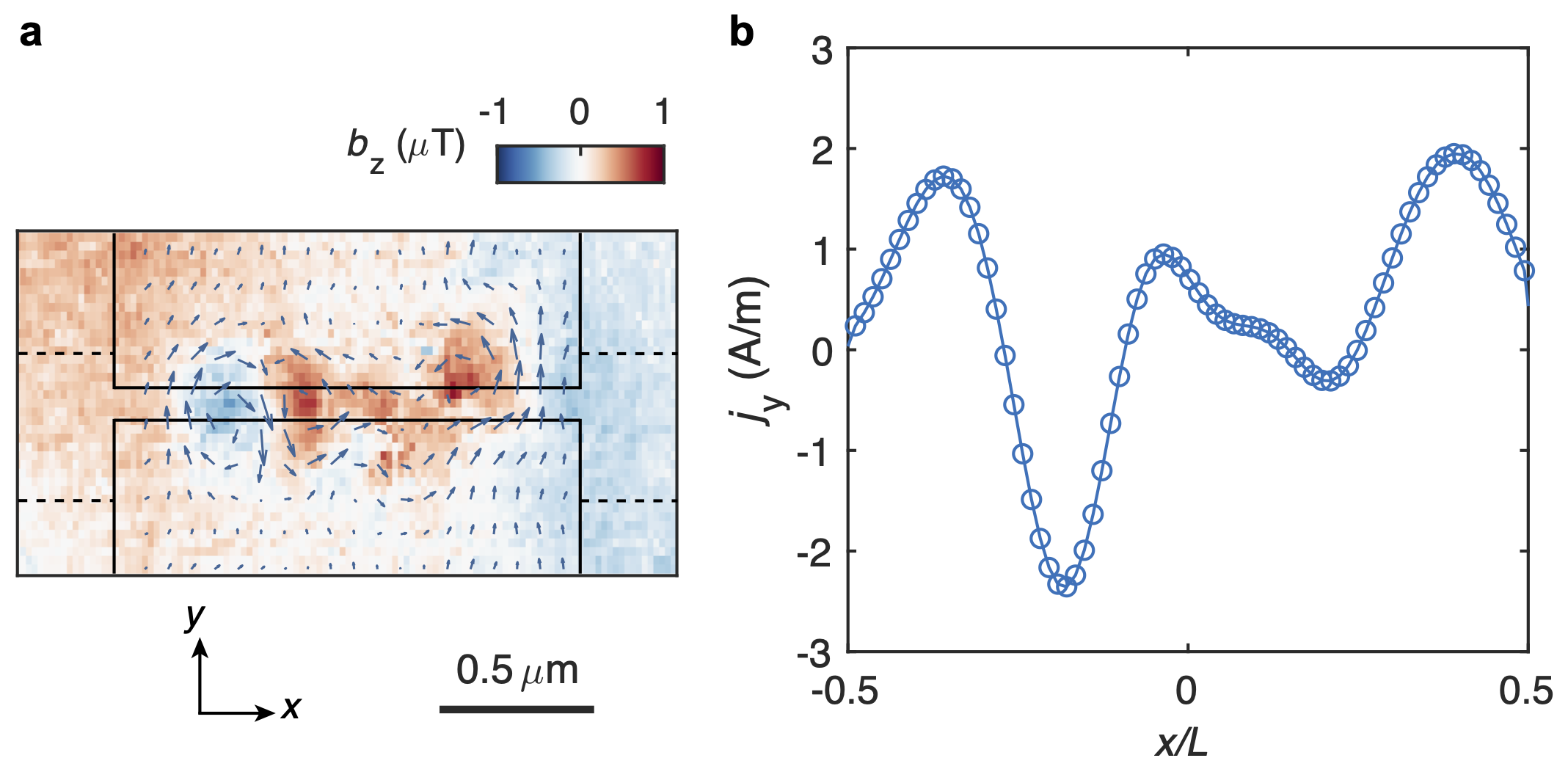} 
\caption{\textbf{Current flow for 2-JV state.}
\textbf{a,} Spatial maps of current flow and $z$-direction magnetic field measured using the symmetric $\pm I_\mathrm{bias}$ sequence as Fig. \ref{fig:2}f-g in the main text, measured at $B_z \approx 4$ mT and $T = 7$ K.
\textbf{b,} Line cut of current flow at the center of the JJ showing cosine-like current profile with twice the oscillations as in Fig. \ref{fig:2}g, indicating 2 JVs at the junction. The circles show the reconstructed current value, the line is a guide for the eye connecting the circles.
}
\label{fig:E_phi2}
\end{figure*}

\begin{figure*}[ht]
\centering
\includegraphics[width=6in]{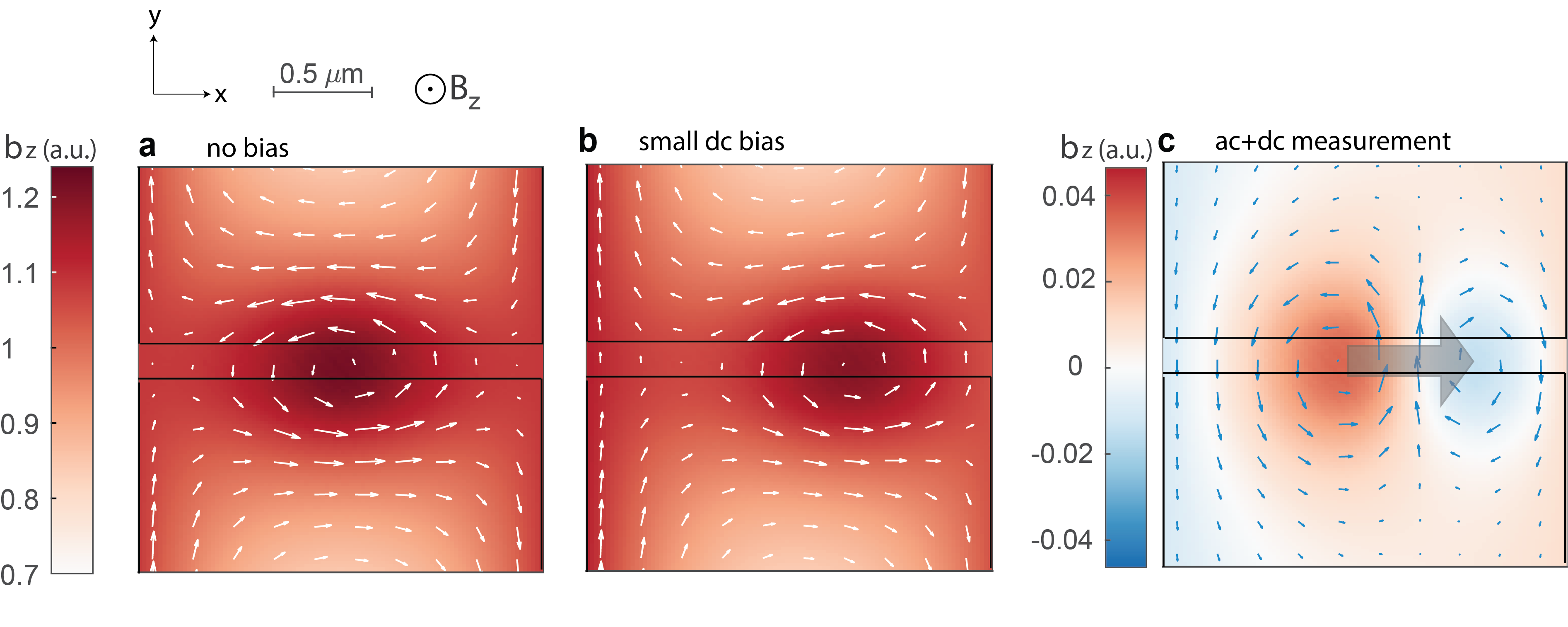} 
\caption{\textbf{JV response to small changes in bias current.}
\textbf{a-b,} Schematic drawings of current flow and and $z$-component of the magnetic field at two slightly different bias current, showing the position of JV controlled by the bias current. Colour scale is the same for both maps, in the unit of flux-quantum.
\textbf{c,} The difference between \textbf{(a)} and \textbf{(b)} shows feature similar to Fig. \ref{fig:3}d in the main text. As the bias current moves the JV along $x$-direction, the distance between the two current loops in \textbf{(c)} represents the size of the JV.
}
\label{fig:S_acvortexmoving}
\end{figure*}

\begin{figure*}[ht]
\centering
\includegraphics[width=3in]{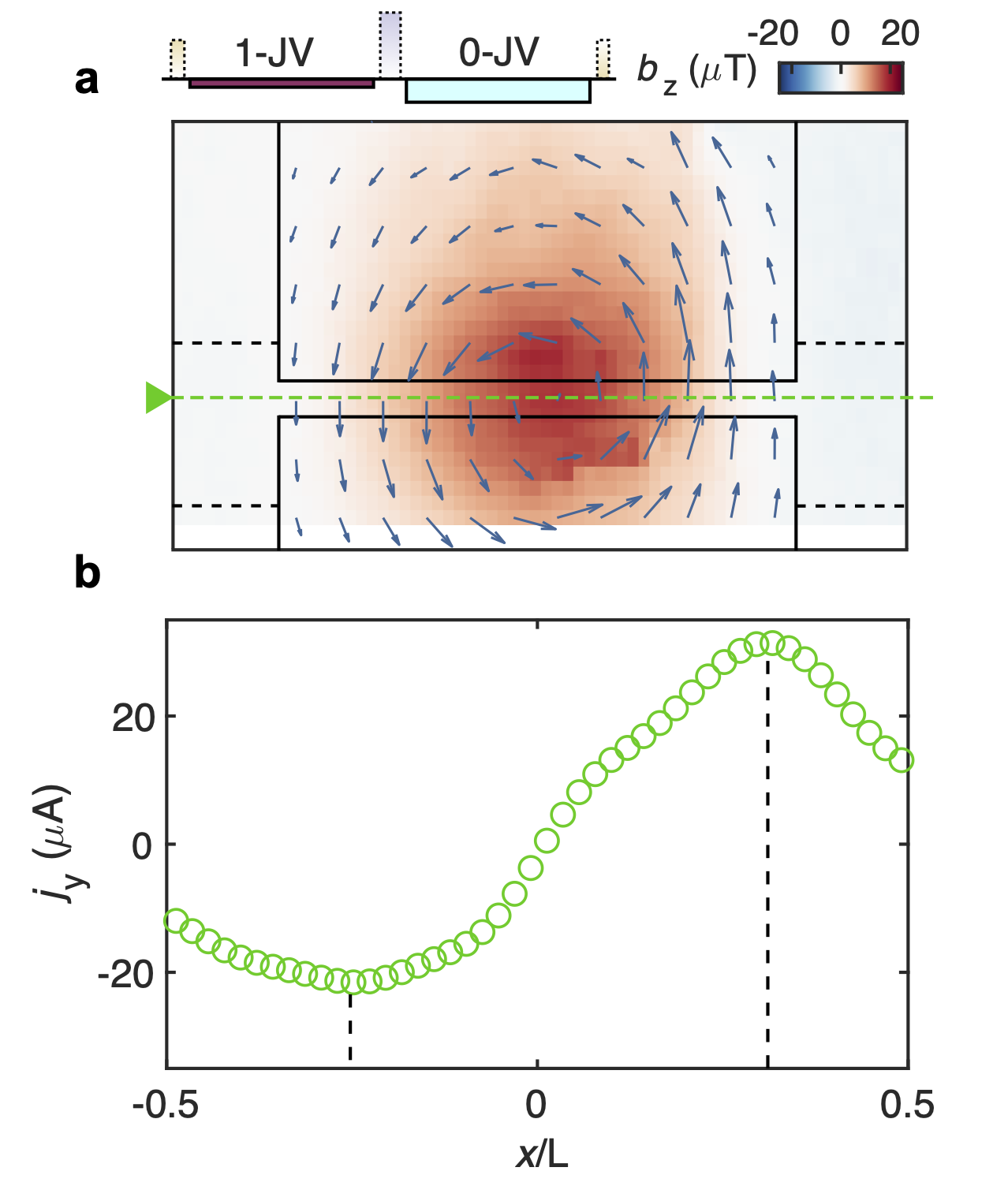} 
\caption{\textbf{Lateral size of the JV.}
\textbf{a,} Same map as Fig. \ref{fig:3}f as the main text. The green dashed line indicate the position of the line cut shown in \textbf{(b)}.
\textbf{b,} Line trace of $j_y$ taken at the center of the JJ. The distance between the peak and valley of $j_y$ indicates the lateral size of the JV. 
The size of the JV along $x$ direction is about 500 nm. It is consistent with the distance between the center of the loops in Fig. \ref{fig:3}e, which effectively measures $\partial b_z/\partial x$ of the JV (Supplementary Fig. \ref{fig:S_acvortexmoving}). The measured JV size is slightly smaller than the theoretical size $\lambda_J\approx780$ nm (see Methods), because the JV is constrained by $L$ which is comparable to $2\lambda_J$.}

\label{fig:E_JVsize}
\end{figure*}

\begin{figure*}[ht]
\centering
\includegraphics[width=5.35in]{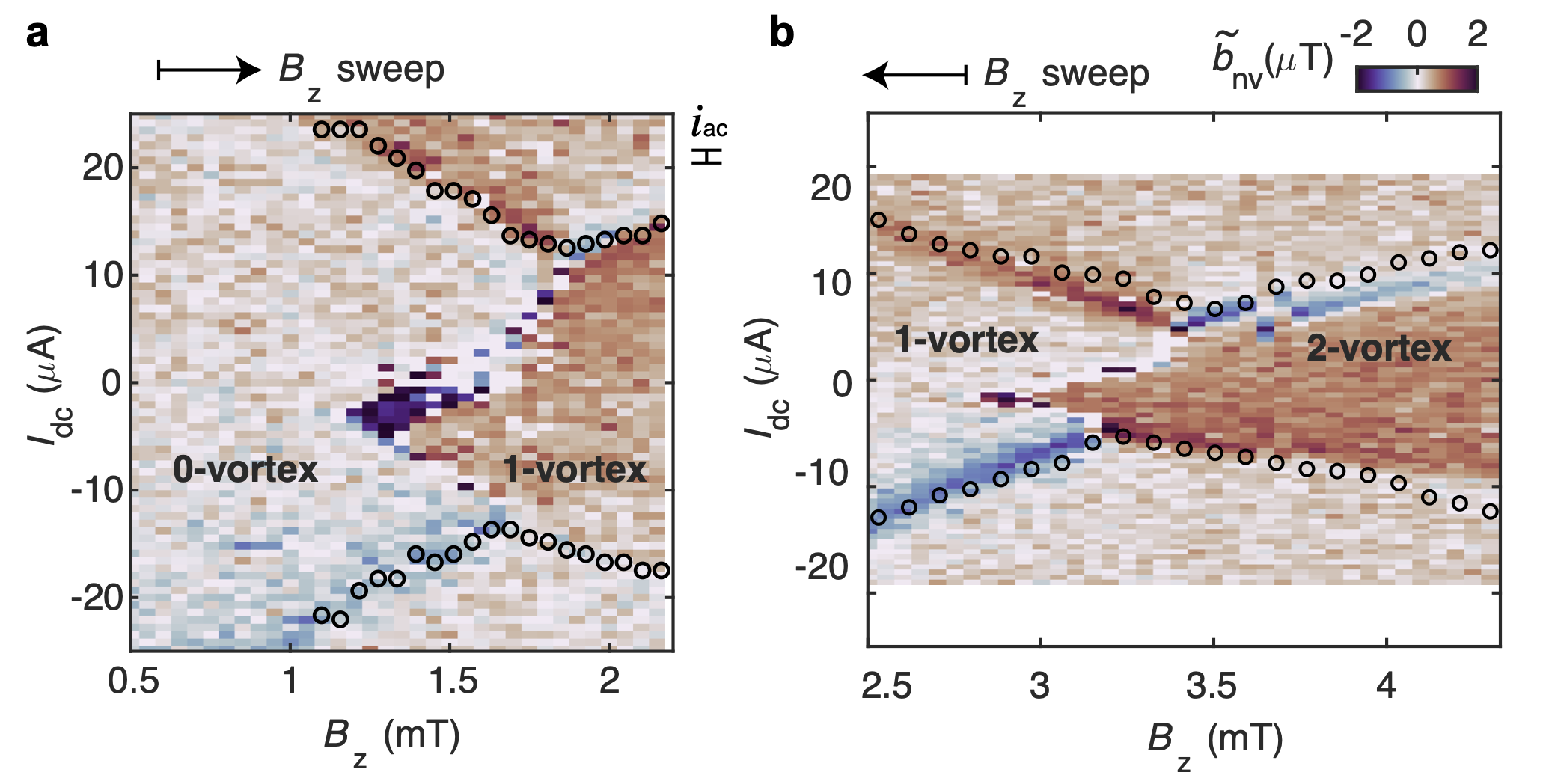} 
\caption{\textbf{Additional measurements showing the competition between ground state configurations.}
\textbf{a,} differential ac magnetic field measurement similar to Fig. \ref{fig:3}b in the main text, when external $B_z$ is swept from low to high field. The result shows the phase diagram is insensitive to $B_z$ sweeping direction. 
\textbf{b,} differential ac magnetic field measurement at the range when 1- and 2-JV states overlap, showing the phase boundary below $I_c$ extends only from the 2-JV state.
}
\label{fig:E_bzac_extra}
\end{figure*}

\begin{figure*}[ht]
\centering
\includegraphics[width=2.5in]{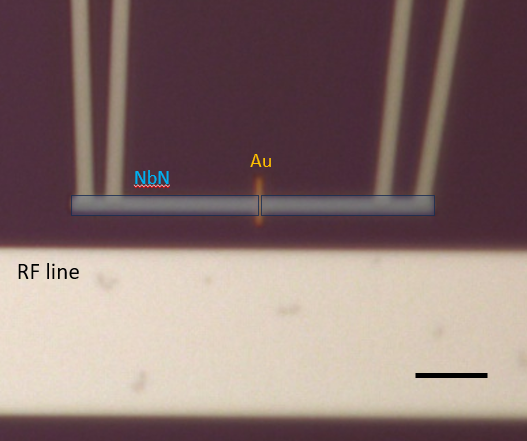} 
\caption{\textbf{Optical image of the JJ device.}
Optical micro-graph showing one of the JJ devices used in the paper. The SC electrodes made with NbN are false coloured. The RF line is used to deliver the microwave pulses to manipulate the NV. Scale bar is 5 $\mu$m. 
}
\label{fig:S_deviceimage}
\end{figure*}

\begin{figure*}[ht]
\centering
\includegraphics[width=3in]{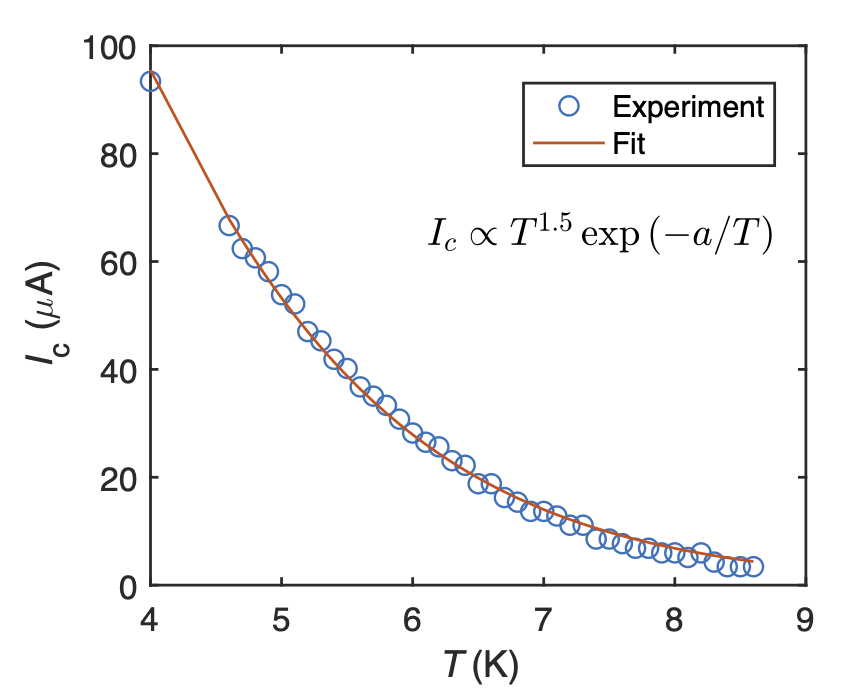} 
\caption{\textbf{Temperature dependence of the critical current.}
Temperature-dependent critical current at zero magnetic field versus fitted curve for a diffusive junction \cite{dubos2001josephson}. 
}
\label{fig:E_IcT}
\end{figure*}

\begin{figure*}[ht]
\centering
\includegraphics[width=3in]{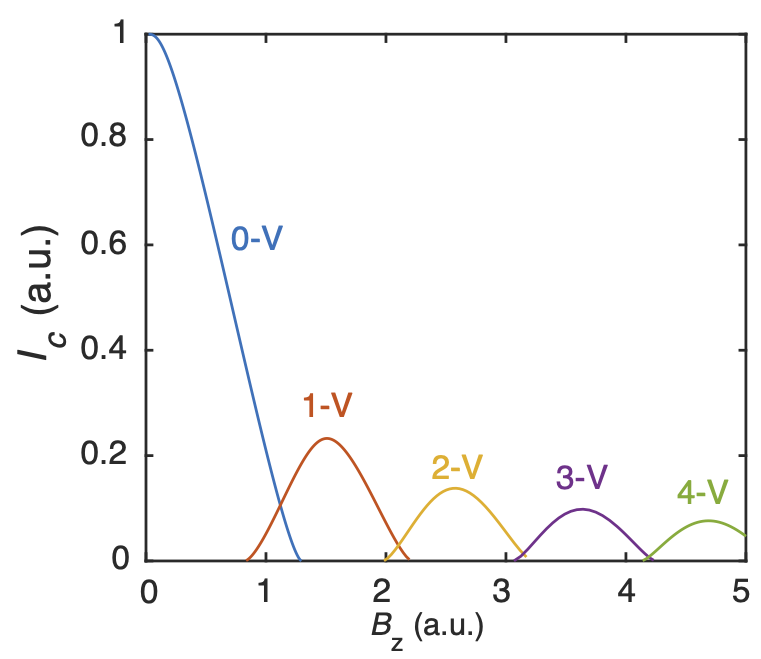} 
\caption{\textbf{Overlapping ground state solutions in JJs with strong self field effect.}
Analytical solutions of the non-linear sine-Gordon equation \cite{kuplevakhsky2007exact} showing critical current $I_c$ for ground state configurations with different number of JVs, as indicated by the line traces of different colours. The traces are generated with junction length $L = 3 \lambda_J$ to highlight the overlap regions.
}
\label{fig:E_TDGLcompare2D3D}
\end{figure*}
\clearpage

\section{Detail of Reconstructing Current Flow from Magnetic Field}
\label{SI:recondetail}

We show examples of this process with data taken for 0- and 1-JV states in Supplementary Fig. \ref{fig:E_fourierbxyz}. Two NV centers with different axis were used when taking these data-sets, as indicated by the arrows in Supplementary Fig. \ref{fig:E_fourierbxyz}a and e.

\begin{figure*}[ht]
\centering
\includegraphics[width=6.9in]{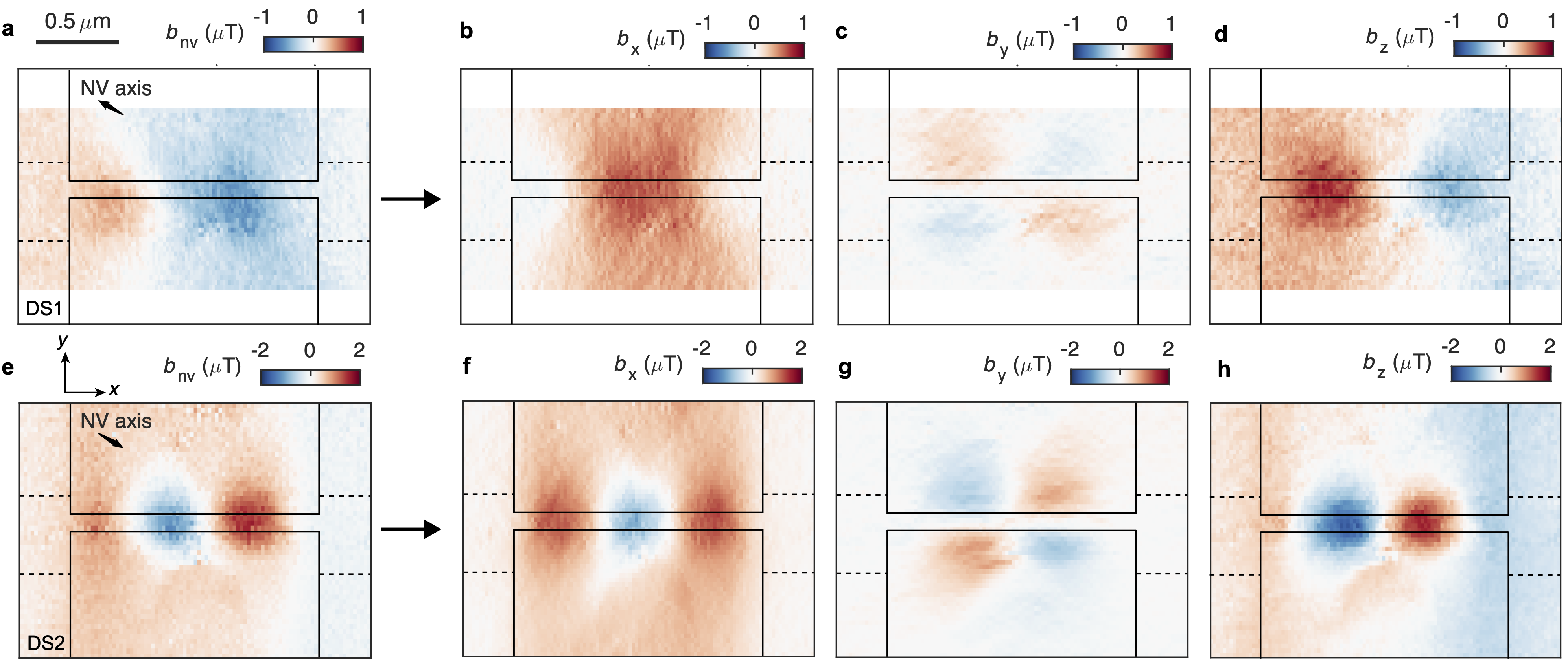} 
\caption{\textbf{Examples of converting $b_\mathrm{nv}$ to $b_{x,y,z}$.}
\textbf{a and e} show magnetic field projected along NV axis measured at two different external $b_z$, which we refer to as dataset1 (DS1) and dataset2 (DS2). 
\textbf{b-d,} show the vector magnetic field $b_{x,y,z}$ reconstructed from \textbf{(a)}. 
\textbf{f-h,} show the vector magnetic field $b_{x,y,z}$ reconstructed from \textbf{(e)}. 
DS1 is measured at external field $B_z = 1.46$ mT. DS2 is measured at external field $B_z = 1.91$ mT. The NV directions point partially out of plane, and their in-plane projections are shown in the insets of \textbf{(a)} and \textbf{(e)}. The scale bar is shared by all colour maps.
}
\label{fig:E_fourierbxyz}
\end{figure*}

We employ the Fourier \cite{roth1989using}, regularization \cite{meltzer2017direct} and machine learning \cite{dubois2022untrained} methods to reconstruct the current flow. For the Fourier method, the full padded data is used in the current reconstruction. For the regularization and machine learning methods, the padded $b_z$ is cut down to 2 times of the measurement window due to computational constraints. In the following, $h$ is the stand-off distance of the NV sensor from the sample plane, and we first describe the methods and then show the results.

\begin{enumerate}
    \item Fourier method \cite{roth1989using}. The current and the in-plane components of the magnetic field are related in the Fourier space via
    \begin{equation}
    \begin{aligned}
    j_y (\textbf{k}) = b_x (\textbf{k},h) \cdot \frac{2}{\mu_0} e^{h k} \\
    j_x (\textbf{k}) = -b_y(\textbf{k},h) \cdot \frac{2}{\mu_0} e^{h k}
    \end{aligned}
    \end{equation}
    here $\mu_0$ is the vacuum permeability. A low-pass Hanning filter $\mathcal{W}$ is applied to $j_{x,y} (\textbf{k})$ before Fourier transforming back to the real space $j_{x,y} (x,y)$,
    \begin{equation}
    \mathcal{W} =
     \begin{cases}
          0.5 \cdot [1+\cos{(k\alpha h/2)}], & \text{for $k<2\pi/\alpha h$}\\
          0, & \text{for $k<2\pi/\alpha h$}
     \end{cases}       
    \end{equation}
    here $\alpha$ sets the cut-off wavelength in the reconstruction. The reconstructed current outside the device area is small and set to zero. We compare the effect of $\alpha \in [1,2]$ in Supplementary Fig. \ref{fig:E_fourieralphacomp}. Although increasing $\alpha$ mitigates the ringing (spatial features oscillating faster than $h$), large $\alpha$ also smears out the result and reduces the amplitude of the reconstructed current. All of the results in the main text are reconstructed with $\alpha=1.5$ to balance between these effects, but the conclusion about Josephson current induced phase in Fig. \ref{fig:2}d and h from the main text does not rely on the choice of $\alpha$.
    
    \item Regularization method. For this method we follow Ref. \cite{meltzer2017direct} and the code there-in. Briefly speaking, it uses kernels $K_1,~K_2$ that takes into account the finite thickness of the SC film $d=35$ nm, and $b_z$ is related to $j_{x,y}$ via
        \begin{equation}
        b_z(x,y,h) = K_1(x,y,h,d)*j_x(x,y)+K_2(x,y,h,d)*j_y(x,y) + N(x,y)
        \end{equation}
    here * represents the convolution integral, $N(x,y)$ is noise. To reconstruct the current, the following regularization functional is minimized 
        \begin{equation}
       \mathrm{min}(||K_1*j_x+K_2*j_y-b_z||^2 + \lambda (||\mathcal{L}j_x||^2+||\mathcal{L}j_y||^2)
        \end{equation}
    here $\mathcal{L}$ is the Laplacian $\nabla^2$, $\lambda$ is the regularization parameter. Compared with the Fourier method, this functional penalizes fast oscillations in the reconstructed current flow. To account for the current flow outside the field-of-view, reflection rule at the boundaries is applied to the padded data. The results using regularization method is shown in Supplementary Fig. \ref{fig:E_reconallcomp}b, f.
    
    \item Machine learning method. The neural network based reconstruction follows a similar construction to that performed in Ref. \cite{dubois2022untrained} with some modifications for reconstruction of current density. The magnetic field is passed to a fully connected neural network which has an output image $g(x,y)$. This is a stream function whose derivatives define the current density, 
        \begin{equation}
        \nabla \times [g(x,y)\hat{z}] = \textbf{j}(x,y)
        \end{equation}
    which enforces the final current density to have zero divergence. 

    To encode the spatial resolution of the reconstructed current density, which is limited by the NV to sample standoff distance ($d_\mathrm{nv}=150$ nm for DS1, $d_\mathrm{nv}=130$ nm for DS2), we model each pixel as a Gaussian distribution with a width of $\sigma = d_\mathrm{nv}/2$. This acts to broaden the output stream function and remove fast oscillating terms before the derivatives are determined. 

    The calculated current densities are then transformed into a single magnetic field image that is compared with the original measured magnetic field, which forms the loss function, and the neural network weights are updated accordingly. 
    The results using machine learning method is shown in Supplementary Fig. \ref{fig:E_reconallcomp}c, g.

\end{enumerate}
\begin{figure*}[t]
\centering
\includegraphics[width=6.9in]{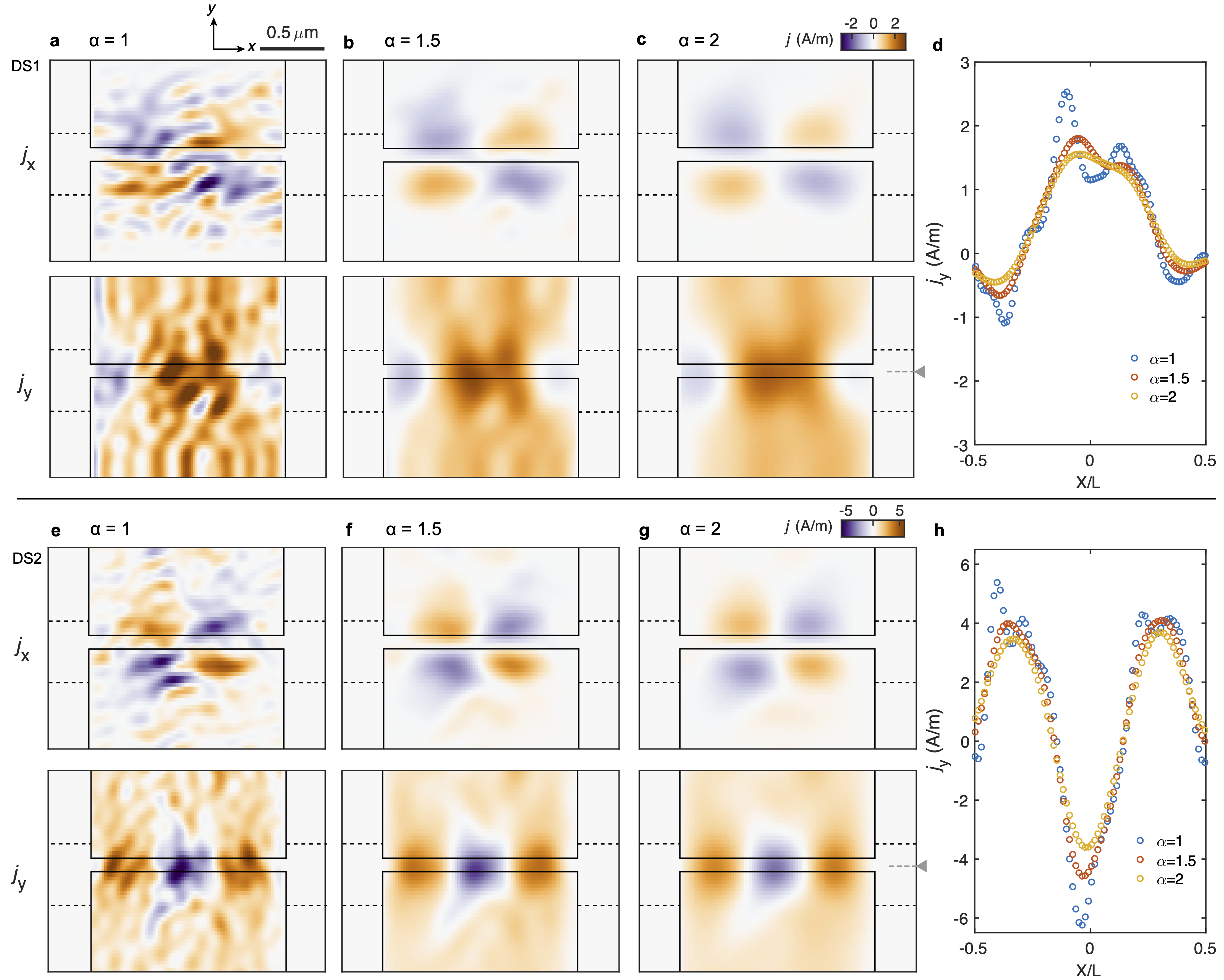} 
\caption{\textbf{Comparison of current flow reconstructed from Fourier methods with different filter functions.}
\textbf{a-c,} show the reconstructed current flow $j_{x,y}$ from DS1 using the Fourier method with cut-off parameter $\alpha = 1,~1.5,~2$.
\textbf{d,} show the line trace $j_y$ with different $\alpha$ at the center of JJ, as indicated by the arrow in \textbf{(c)}. 
\textbf{e-h,} show the corresponding results for DS2. In both cases, increasing $\alpha$ mitigates ringing in the reconstructed current, but also reduces the amplitude of the current. So the results in the main text are reconstructed with $\alpha=1.5$.
The scale bar is shared by all colour maps.
}
\label{fig:E_fourieralphacomp}
\end{figure*}

\begin{figure*}[ht]
\centering
\includegraphics[width=6.9in]{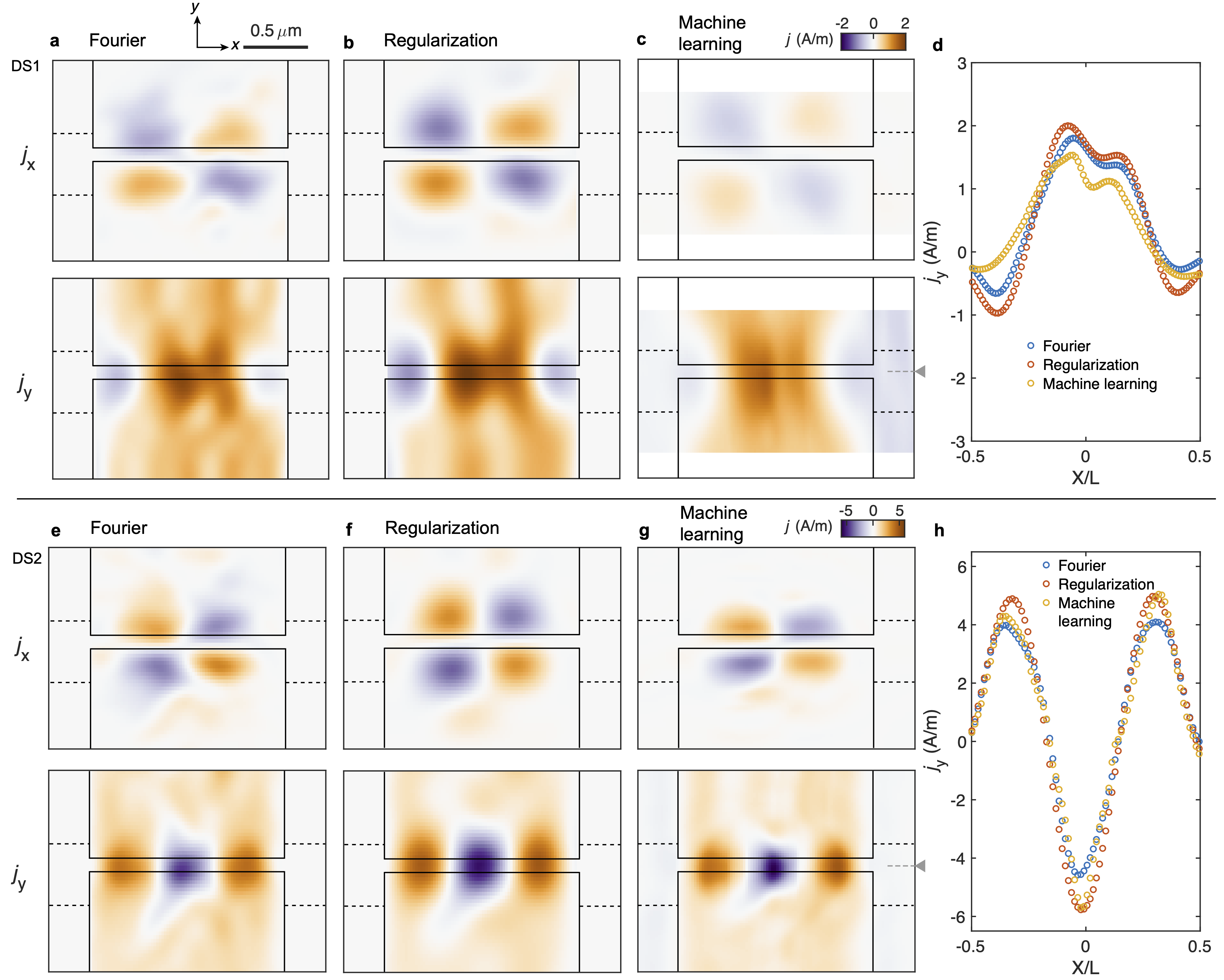} 
\caption{\textbf{Comparison of current flow reconstructed from different methods.}
\textbf{a-c,} show the reconstructed current flow $j_{x,y}$ from DS1 using the \textbf{(a)} Fourier ($\alpha=1.5$), \textbf{(b)} regularization, and \textbf{(c)} machine learning methods.
\textbf{d,} show the line trace $j_y$ with different methods at the center of JJ, as indicated by the arrow in \textbf{(c)}. 
\textbf{e-h,} show the corresponding results for DS2. The scale bar is shared by all colour maps.
}
\label{fig:E_reconallcomp}
\end{figure*}
\clearpage

\section{Lack of diode effect at 7 K}
\label{SI:nondiode7K}
\begin{figure*}[ht]
\centering
\includegraphics[width=5in]{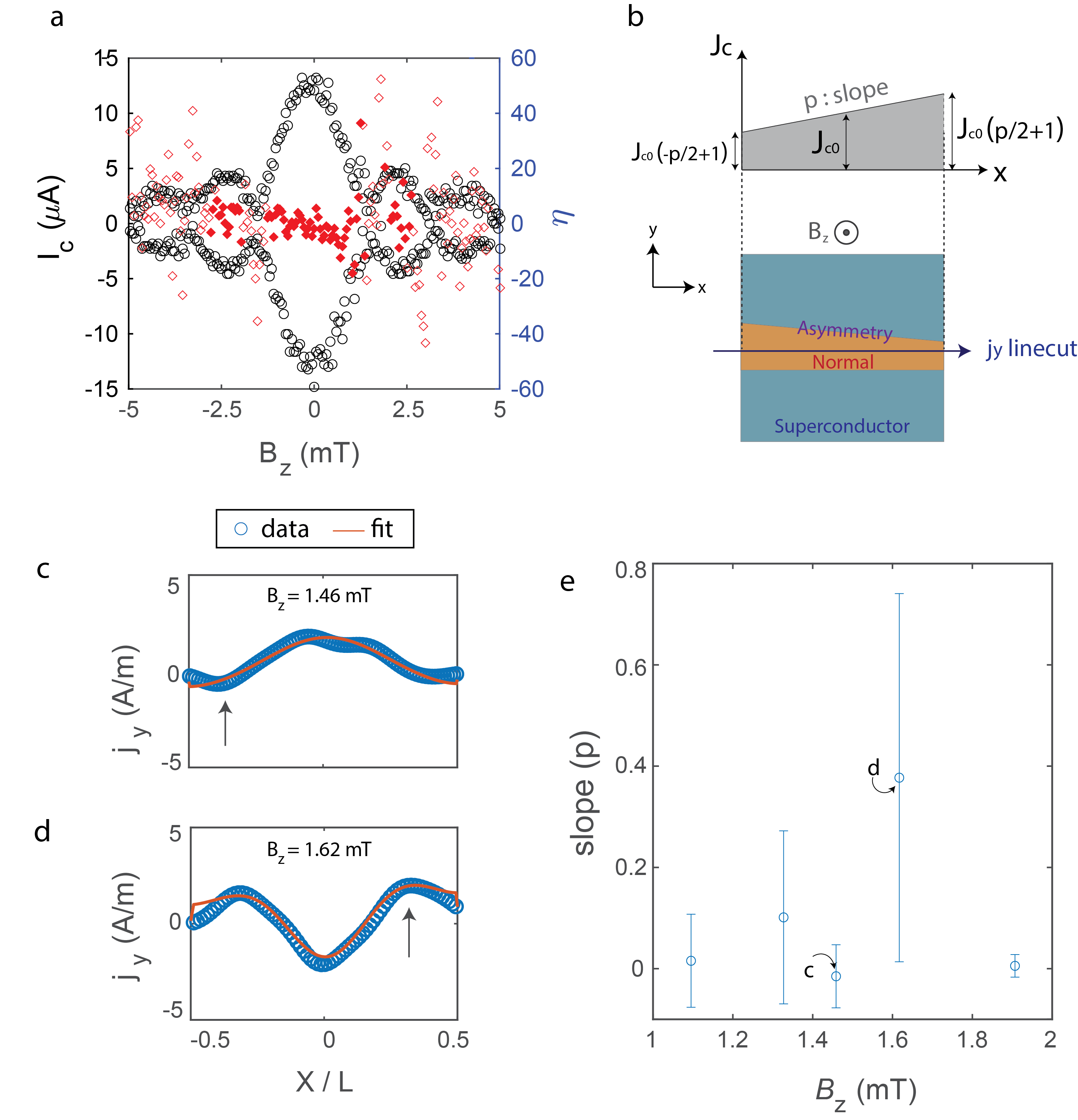} 
\caption{\textbf{Critical current and inversion symmetry breaking at $T=7$ K.}
\textbf{a,} Critical current $I_c$ and asymmetry factor $\eta=\frac{|I_c^+|-|I_c^-|}{|I_c^+|+|I_c^-|}$ versus perpendicular external magnetic field $B_z$. Compared with the $T=4$ K result in Fig. \ref{fig:4}a, $\eta$ here averages to zero and JDE is negligible at $T=$7 K. The black circles show the critical current, the red diamonds show $\eta$. The $\eta$ points where either of the $|I_c^{\pm}|< 3\mu$A is shown as unfilled diamonds, due to the large uncertainty arising from the small $I_c$.
\textbf{b,} Upper panel shows the non-uniform critical current $J_c(x) = J_{c0} (p\cdot \frac{x}{L}+1)$ used to fit the measured current profile in \textbf{(d)-(e)}. lower panel shows a schematic drawing showing of the inversion symmetry breaking at the junction, which could be caused by extrinsic factors such as non-uniform junction width.
\textbf{c-d,} Circles show the measured current profile at $y=0$ in the JJ measured with NV sequence as described in Fig. \ref{fig:2}f-g in the main text, at \textbf{(c)} $B_\mathrm{z,ext}=1.46$ mT (Supplementary Fig. \ref{fig:E_7Kallmap}h), \textbf{(d)} $B_\mathrm{z,ext}=1.62$ mT (Supplementary Fig. \ref{fig:E_7Kallmap}i). Lines are fitting results using the non-uniform $J_c(x)$ as shown in \textbf{(b)}. Arrows point at the extrema of current profile. The $j_y$ minima is smaller at $x<0$ for 0-JV, and the maxima is larger at $x>0$ for 1-JV. 
\textbf{e,} Fitting result of the slope $p$ at each $B_z$. All the results except $B_z=1.91$ mT show $p>0$, indicating larger $J_{c}$ for $x>0$. 
}
\label{fig:S_asymm7Kfit}
\end{figure*}

In our experiment, the Josephson diode effect (JDE) is not observed when the current induced phase is not strong enough. Supplementary Fig. \ref{fig:S_asymm7Kfit}a shows the critical current $I_c$ and related asymmetry parameter $\eta$ at $T=7$ K. Compared to the result in Fig. \ref{fig:4}a, $\eta$ vanishes to zero even for $B_z \neq 0$ (broken time-reversal symmetry). 

From the local measurement of the current flow, we find broken inversion symmetry at $T=7$ K. Using the sequence that measures the difference of symmetric $\pm I_\mathrm{bias}$ (Fig. \ref{fig:2}f-g in the main text), the current profile $j_y(x)$ at the center of JJ is not symmetric with $x=0$, suggesting the inversion symmetry breaking.

Supplementary Fig. \ref{fig:S_asymm7Kfit}c and d show two examples of $j_y(x)$. The measured $j_y(x)$ is fit with a non-uniform critical current profile. To the first order, the critical current density $J_c(x) = J_{c0} (p\cdot \frac{x}{L}+1)$ (Supplementary Fig. \ref{fig:S_asymm7Kfit}b). $p>0$ indicates larger critical current on the right side ($x>0$). This leads to lower $j_y$ minima for $x<0$ in the 0-JV case, and higher $j_y$ maxima for $x>0$ in the 1-JV case. These features are qualitatively observed in all five data sets measured (Supplementary Fig. \ref{fig:E_7Kallmap}f-j). 

Furthermore, we can quantitatively estimate the non-uniformity of the critical current. The measured current profile $j_y(x)$ is given by a modified version of Supplementary Eqn. \ref{eqn:phiefffit},

\begin{equation}
j_y(x) =  J_{c0} (p\cdot \frac{x}{L}+1) \cdot \left[ \sin(\phi_e(x,B_\eff) + \phi_{\mathrm{bias}1}) - \sin(\phi_e(x,B_\eff)+ \phi_{\mathrm{bias}2})\right]
\label{eqn:asymJcFit}
\end{equation}

During the fitting, $J_{c0}$, $p$ and $B_{\eff}$ are the fitting parameters. For given $p$ and $B_{\eff}$, we first find $\phi_{\mathrm{bias}1}$ ($\phi_{\mathrm{bias}2}$) that corresponds to $+I_c$ ($-I_c$), and then use Supplementary Eqn. \ref{eqn:asymJcFit} to obtain $j_y(x)$. We note that although not all the data sets shown in Supplementary Fig. \ref{fig:E_7Kallmap}f-j are measured at $|I_\mathrm{bias}|=|I_c|$, the difference should be small. The junction length $L=1.5 \mu$m is fixed during the process.

The fitting results of slope $p$ for all data sets are shown in Supplementary Fig. \ref{fig:S_asymm7Kfit}e. 
The linear $J_c(x)$ is just the first order correction, used as a toy model to highlight the non-uniform critical current density. In reality the critical current density could change along $x$ direction due to variations of the SC/N interface transparency, separation $W$ of the junction, etc. The fitting result in Supplementary Fig. \ref{fig:S_asymm7Kfit}e suggests $J_c(x>0) > J_c(x<0)$, consistent with the result from the current flow results measured at $T=4$ K (Fig. \ref{fig:4}). Overall, our local measurements show that inversion symmetry breaking at the JJ could be ubiquitous owing to extrinsic artefacts in the fabrication process, and may or may not manifest in the global measurement of asymmetric critical current.

\clearpage

\section{Josephson Diode Effect arising from symmetry breaking and Josephson current induced phase}
\label{SI:JDE_2JJ}

In this section we examine the roles of time-reversal and inversion symmetry, and current flow induced phase in realizing the JDE as explained in Fig. \ref{fig:4} of the main text, using a model with two lumped JJs in parallel. From a phenomenological perspective, the minimum requirement for JDE is that the current phase relation contains more than just the first harmonic term, plus a phase offset, $I(\phi) = a_1 \sin(\phi) + a_2 \sin (2\phi + \phi_0)$ \cite{pal2022josephson,jeon2022zero}. The second harmonic term could be due to the ballistic transport in the JJ or high transparency of the SC/N interface, which are difficult to verify experimentally. Using a two-junction model, we show that broken time reversal and inversion symmetry, combined with the Josephson current induced phase can effectively cause such a second harmonic term in current phase relation, even when starting with only the first harmonic term for the diffusive and low-transparency JJ. A similar model was proposed in Refs. \cite{fulton1972quantum,Reinhardt2024NatCommun}, here we present more analysis in the context of the Josephson diode effect.

\begin{figure*}[ht]
\centering
\includegraphics[width=7in]{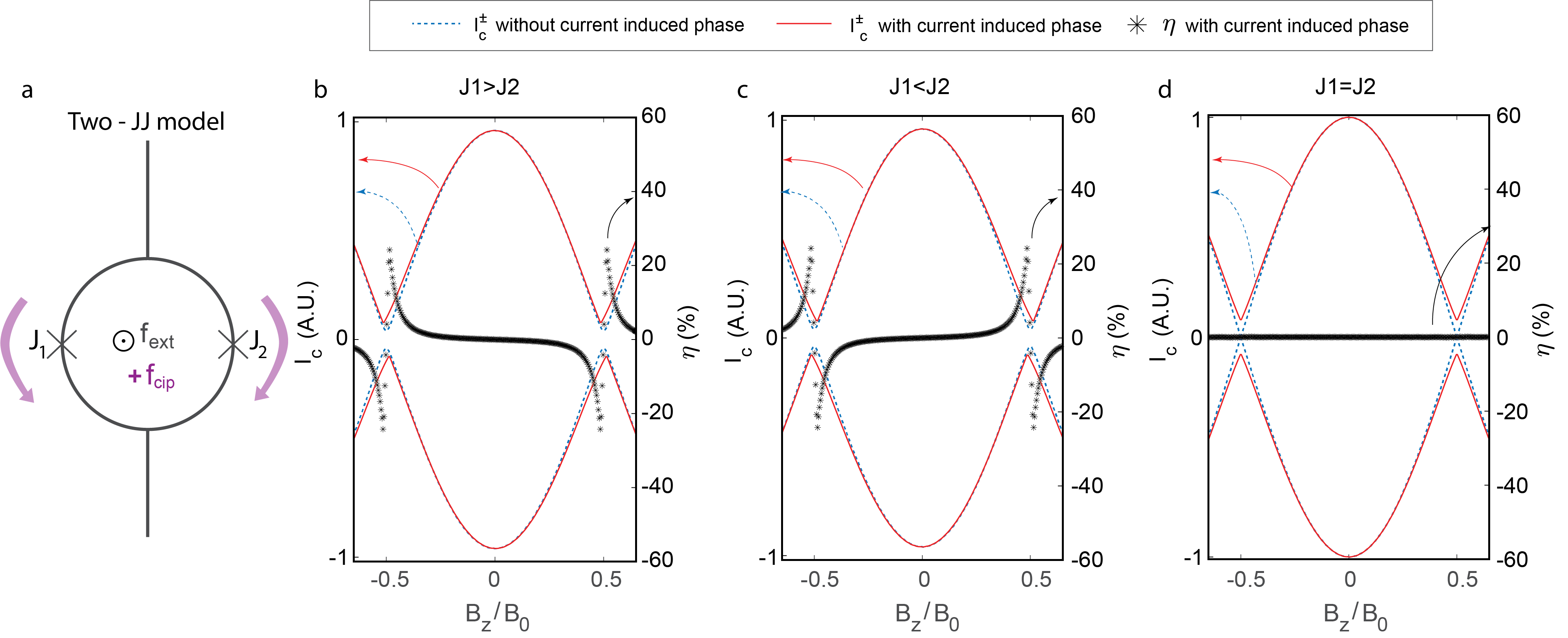} 
\caption{\textbf{The two-junction model.}
\textbf{a,} Schematics drawing of the model, showing left(right) JJs with critical current of $J_{1(2)}$ forming a loop. The total phase difference between the JJs comes from the external magnetic field $f_\mathrm{ext}$, and the Josephson current induced phase $f_\mathrm{cip}$.
\textbf{b-d,} Numerical simulation of forward/backward critical current $I_c^{\pm}$, and asymmetric parameter $\eta = \frac{|I_c^+|-|I_c^-|}{|I_c^+|+|I_c^-|}$ as a function of external flux when \textbf{(b)} $J_1 > J_2$, \textbf{(c)} $J_1<J_2$ and \textbf{(d)} $J_1 = J_2$. In \textbf{(b)} and \textbf{(c)}, the difference between $J_{1,2}$ is 10 \%. Red (Blue) lines show the critical current $I_c$ calculated with (without) the current induced phase. Black stars show asymmetric factor $\eta$ when the current flow induced phase is included. In the inversion symmetric case \textbf{(d)}, there is no diode effect. In the inversion symmetry broken cases $J_1\neq J_2$ in \textbf{(b)} and \textbf{(c)}, the diode effect is present when the current induced phase is included. $\eta$ changes sign for $J_1$ larger or smaller than $J_2$.
}
\label{fig:S_TwoJJmodel}
\end{figure*}

Supplementary Fig. \ref{fig:S_TwoJJmodel}a shows a schematic drawing of the two-junction model. The JJ studied in our paper could be regarded as a set of lumped JJs in parallel, so we consider the simplest case of two lumped JJs with critical current $J_{1,2}$. Inversion symmetry breaking is indicated by $J_1 \neq J_2$. The phase difference across the junction consists of the external magnetic field contribution $f_\mathrm{ext}$, and the Josephson current induced phase $f_\mathrm{cip}$. The total bias current across the junction is 
\begin{equation}
    I_\mathrm{bias} = J_1 \sin{(\Delta\phi - \frac{f_\mathrm{ext}+f_\mathrm{cip}}{2}}) + J_2\sin{(\Delta\phi+ \frac{f_\mathrm{ext}+f_\mathrm{cip}}{2})} ,
\label{eq:cip_Itot}
\end{equation}
where $\Delta \phi \in [-\pi,\pi]$ is the phase difference between the SC electrodes. The Josephson current induced phase of the left and right junctions is
\begin{equation}\label{eq:cip_find}
    f_\mathrm{cip} = \mathscr{L}_k\left[J_1\sin{(\Delta\phi-\frac{f_\mathrm{ext}+f_\mathrm{cip}}{2} )}-J_2\sin{(\Delta\phi+ \frac{f_\mathrm{ext}+f_\mathrm{cip}}{2}})\right],
\end{equation}
where $\mathscr{L}_k$ is proportional to the kinetic inductance. 

We discuss three representative scenarios.
\begin{enumerate}
    \item Neglecting Josephson current induced phase. If $f_\mathrm{cip} = 0$ in Supplementary Eqn. \ref{eq:cip_Itot}, the $I_\mathrm{bias}$ only contains the first harmonic term of $\Delta \phi$ with a phase offset, and JDE does not exist.
    \item $J_1 = J_2$. Supplementary Eqn. \ref{eq:cip_find} is reduced to $f_\mathrm{cip} =-2 \mathscr{L}_k \cos(\Delta \phi) \sin (f_\mathrm{ext}+f_\mathrm{cip})$, which yields the same solutions of $f_\mathrm{cip}$ when $\Delta \phi \leftrightarrow -\Delta \phi$. This means $I_\mathrm{bias} (\Delta \phi) = -I_\mathrm{bias} (-\Delta \phi)$, and JDE does not exist.
    \item $J_1 \neq J_2$. In this case, $f_\mathrm{cip}$ needs to be solved numerically. Taking the limit of $f_\mathrm{cip} \ll f_\mathrm{ext}, \Delta \phi$, i.e., $\mathscr{L}_k J_i \ll 1$, we expand terms in Supplementary Eqn. \ref{eq:cip_find} to first order of $f_\mathrm{cip}$, and get
    \begin{equation}
        f_\mathrm{cip} \simeq \mathscr{L}_k
        \left[ J_1\sin{(\Delta\phi-\frac{f_\mathrm{ext}}{2})}-J_2\sin{(\Delta\phi+\frac{f_\mathrm{ext}}{2})}-
        \frac{f_\mathrm{cip}}{2}\left[J_1\cos{(\Delta\phi-\frac{f_\mathrm{ext}}{2})}+J_2\cos{(\Delta\phi+\frac{f_\mathrm{ext}}{2})}\right] \right] .
    \label{eq:cip_approxfind}
    \end{equation}
    Combined with Supplementary Eqn. \ref{eq:cip_Itot}, we find
    \begin{equation}
    \begin{split}
    I_\mathrm{bias} \simeq& J_1\sin{(\Delta\phi-\frac{f_\mathrm{ext}}{2})} +J_2\sin{(\Delta\phi+\frac{f_\mathrm{ext}}{2}))} -
        \frac{f_\mathrm{cip}}{2}\left[ J_1\cos{(\Delta\phi-\frac{f_\mathrm{ext}}{2})}-J_2\cos{(\Delta\phi+\frac{f_\mathrm{ext}}{2})}\right]\\
        = &J_1\sin{(\Delta\phi-\frac{f_\mathrm{ext}}{2})} +J_2\sin{(\Delta\phi+\frac{f_\mathrm{ext}}{2}))} + \\
       & \frac{\mathscr{L}_k}{2} 
       \frac{J_1^2\sin(2\Delta\phi-f_\mathrm{ext})+J_2^2\sin(2\Delta\phi+f_\mathrm{ext})-2J_1J_2\sin(2\Delta\phi)}
       {2+\mathscr{L}_k \left[ J_1\cos(\Delta\phi-\frac{f_\mathrm{ext}}{2}) +J_2\cos(\Delta\phi+\frac{f_\mathrm{ext}}{2})\right]} .
    \label{eq:cip_approxfinal}
    \end{split}
    \end{equation}
Here the second harmonic term of $\Delta \phi$ with a phase shift is present, and JDE can be observed.
\end{enumerate}

We also numerically solve for the critical current in the two-JJ model when the current flow induced phase is included. At each external magnetic flux $f_\mathrm{ext}$, we first solve for $f_\mathrm{cip}$ at individual $\Delta \phi \in [-\pi, \pi]$ in Supplementary Eqn. \ref{eq:cip_find}. The ($f_\mathrm{cip}$, $\Delta \phi$) is then plugged into Supplementary Eqn. \ref{eq:cip_Itot} to find the maximum (minimum) $I_\mathrm{bias}$ as $I_c^+$ ($I_c^-$). Three cases of $J_{1(2)}$ are considered. When the system is inversion symmetric, i.e., $J_1=J_2$, the current flow induced phase only lifts the node and does not manifest JDE (Supplementary Fig. \ref{fig:S_TwoJJmodel}d). When $J_1\neq J_2$, $I_c^\pm$ becomes asymmetric when $f_\mathrm{ext} \neq 0$. In particular, the JDE changes polarity when exchanging $J_1$ and $J_2$, or changing the sign of external field (Supplementary Fig. \ref{fig:S_TwoJJmodel}b and c). The critical current nodes are near half-integer of $\Phi_0$ because of the two lumped JJs in the model (effectively a SQUID), but it does not affect the interpretation. 

We note that the above result also applies to the case with strong self field effect, by replacing the kinetic inductance with the geometric inductance of the junction, and replacing the $f_{\mathrm{cip}}$ with the phase induced by the current-generated magnetic field, as pointed out in Refs. \cite{goldman1967meissner,fulton1972quantum}.

The JDE has garnered much attention due to its application in low dissipation electronics \cite{ando2020observation,baumgartner2022supercurrent,wu2022field,pal2022josephson,lin2022zero,jeon2022zero,hou2023ubiquitous,nadeem2023superconducting}, and some of the more recent interest has focused on the connection between JDE and finite momentum pairing of the Cooper pairs in the JJ \cite{pal2022josephson,jeon2022zero,lin2022zero}. Here in our work we are able to pinpoint the origin of the observed JDE with a combination of measurements of electrical transport, and visualization of the current flow. The inversion symmetry breaking in our device likely arises from the non-uniform junction width or transparency ubiquitous in the nano-fabrication process. The Josephson current induced phase is revealed thanks to the local current flow mapping, because the JJ is not deep in the so-called ``strong-junction" regime by the conventional metric. In our device $L \approx 2\lambda_J$ even at $T=4$ K, and the calculation of $\lambda_J$ depends on an estimate of $\lambda_L$ which could vary from sample to sample. In this spirit, we summarize some additional ways to realize the JDE experimentally from the literature. 

\begin{enumerate}
    \item Trapped vortices in superconductors. The JDE requires breaking time reversal symmetry. This could come from the Abrikosov vortices (AV) trapped in thin film superconductors even after external magnetic field is retracted. When an AV is near the JJ, it causes a phase gradient along the transverse direction that mimics the effect of magnetic flux induced by external field \cite{golod2022demonstration}. In the case of layered SC, JVs could be trapped between layers due to history of an in-plane magnetic field \cite{moler1998images}. Additionally, the trapped vortices could be caused by magnetic materials at the JJ or nearby \cite{gutfreund2023direct}. 
    \item Asymmetric injection of bias current. The inversion symmetry of the JJ could be broken by non-uniform critical current density. This could be due to local defects as mentioned above, or local temperature gradient \cite{krasnov1997fluxon}. The effect could be further enhanced by engineering electrodes to intentionally inject the current asymmetrically to the JJ \cite{golod2022demonstration}.
    \item Multi-layer SC. When multiple kinds of SC with different critical current is used, or in the case of heterogeneous film quality along the normal direction, JDE could develop when an in-plane external field perpendicular to the junction is applied \cite{sundaresh2023diamagnetic}. The mechanism is similar to the one described in our main text, for a JJ that exists in the $yz$ plane and the external field is $B_x$.
\end{enumerate}
 
\section{Time-dependent Ginzburg Landau simulation}
\label{SI:TDGL}

The total Ginzburg Landau (GL) free energy for a thin film S structure with thickness $t_\mathrm{SC}$, under external magnetic field $B_\mathrm{z,ext}$, in SI unit, is,
\begin{equation}
F_{\mathrm{GL}} = t_{SC}\cdot \int d^2\textbf{r} \left[-\alpha \eta(\textbf{r})|\Psi|^{2} +  \frac{\beta}{2} |\Psi|^{4} + \frac{1}{4m_e}|(-i \hbar \nabla - 2e\textbf{A})\Psi|^{2} \right] + \frac{1}{2\mu_0} \int d^3\textbf{r} |\nabla \times \textbf{A} - B_\mathrm{z,ext}|^{2} ,
\label{eqn:GLenergy_SI}
\end{equation}
where $m_e$ is the electron mass, $e$ is the electron charge, $\mu_0$ is the vacuum permeability, $\eta({\bm{r}})$ is the inhomogeneity factor; $\eta=1$ for the SC electrodes, and $\eta<0$ for the normal area of the junction. For the strong junction simulations shown in Figs. \ref{fig:3} and \ref{fig:4}, we use $\eta=-1$ for the normal region. The characteristic lengths here are the GL penetration length $\lambda= \sqrt{\frac{m_e \beta}{2\mu_0 e^2 |\alpha|}}$ and the GL coherence length $\xi =\sqrt{\frac{\hbar^2}{4m_e |\alpha|}}$. For $\Psi = |\Psi| e^{i\theta}$, the super current density is $\textbf{J} = \frac{e}{m_e} (\hbar \nabla \theta - 2e\textbf{A}) |\Psi|^2$, and the sheet current density is $\textbf{K} = t_{SC}\textbf{J}$. 

In thermal equilibrium, with no bias current, the functions 
$\textbf{A}(\textbf{r})$ and $\Psi (\textbf{r})$ should be chosen to minimize $F_{\mathrm{GL}}$, subject to suitable boundary conditions. Minimizing with respect to $\Psi$ leads to the Ginzburg-Landau differential equation for $\Psi$ in the vector potential $\textbf{A}$, and minimizing with respect to $\textbf{A}$ produces a vector potential resulting from the applied magnetic field and from the supercurrent associated with the wave function $\Psi$.  

If the vector potential $\textbf{A}(\textbf{r})$ is specified, the wave function $\Psi$ can be obtained using a two-dimensional time-dependent Ginzburg-Landau equation, which will cause $\Psi$ to relax at long times to at least a local minimum of $F_{\mathrm{GL}}$ in the given vector potential. 
We use the package in Ref. \cite{bishop2023pytdgl} to carry out the TDGL simulations for the junction, based on equations derived for dirty superconductors in Ref. \cite{watts1981nonequilibrium}. Briefly speaking, the package solves the following dimensionless TDGL equation,
\begin{equation}
\frac{u}{\sqrt{1+\gamma^2 |\Psi|^2}} \left( \frac{\partial}{\partial t} + i \mu + \frac{\gamma^2}{2} \frac{\partial |\Psi|^2}{\partial t}\right) \Psi = (\eta-|\Psi|^2)\Psi + (\nabla-i  \textbf{A})^2 \Psi .
\label{eqn:ptdgl_main}
\end{equation}
Here $t$ is time; $u\approx 5.79$, $\gamma=10$ are constants; $\mu (\textbf{r},t)$ is the electric potential. The variables $\Psi$, $\textbf{A}$, $\mu$ and $t$ in Supplementary Eqn. \ref{eqn:ptdgl_main} are in dimensionless units given in Ref. \cite{bishop2023pytdgl}. The electric potential evolution results from the current continuity equation, where the total current $\textbf{J}$ comprises of the super current and normal current,
\begin{equation}
\nabla \cdot \textbf{J} = \nabla \cdot \mathrm{Im}[\Psi^* (\nabla - i\textbf{A})\Psi] - \nabla^2 \mu = 0 .
\label{eqn:ptdgl_mu}
\end{equation}

On the SC/vacuum interface, the Neumann boundary conditions are used: 
\begin{equation}
    \begin{aligned}
        \hat{\bm{n}} \cdot (\nabla - i\textbf{A})\Psi&=0 \\
        \hat{\bm{n}}\cdot \nabla\mu &= 0 ,
    \end{aligned}
\end{equation}
where $\hat{\bm{n}}$ is the unit vector normal to the interface.
On the interfaces between SC and current terminals (which is used to apply the bias current), Dirichlet boundary conditions on $\Psi$ and Neumann boundary condition on $\mu$ are used,
\begin{equation}
    \begin{aligned}
        \Psi&=0 \\
        \hat{\bm{n}}\cdot \nabla\mu &= |{\bm{K}}_{\mathrm{bias}}| ,
    \end{aligned}
\end{equation}
where $|{\bm{K}}_{\mathrm{bias}}| = I_\mathrm{bias}/L$. In the case where $I_\mathrm{bias}=0$, the solution gives a chemical potential $\mu$ that is independent of position, leading to an equilibrium solution, where the supercurrent is divergence-free. When $I_\mathrm{bias} \neq 0$, provided that the total current is less than the critical current in the specified magnetic field, there will be a solution where the normal current decays rapidly near the normal contact, while $\mu$ is essentially a constant and the Ginzburg-Landau equation applies away from the contacts. 

Ideally, the vector potential should be determined self-consistently with the computed wave function $\Psi$. This can be done using an iterative procedure, which will be described below.  However, the iteration is computationally expensive, and the correction due to the self field is small in the thin-film limit. Consequently, the self field has been ignored in most of our calculations, and the vector potential was set by  $B_\mathrm{z,ext}$ via $\nabla \times \textbf{A} =B_\mathrm{z,ext}$.

\clearpage

\textbf{-Simulation parameters}

The results in Fig. \ref{fig:3}c and Fig. \ref{fig:4}e, f are simulated without the self field. A schematic drawing of the simulated device is shown in Supplementary Fig. \ref{fig:S_TDGL_initial}a and the parameters used are 

\begin{table}[h!]
\begin{tabular}{||c|c||} 
 \hline
 $\lambda$ & 400 nm\\
 \hline
 $\xi$ & 100 nm\\
 \hline
 $L$  & 1.5 $\mu$m\\
 \hline
 $W_1$ & 160 nm \\
 \hline
 $W_2$ & 140 nm\\
 \hline
 $t_{SC}$ & 35 nm\\
 \hline
\end{tabular}
\caption{Parameters used in the TDGL simulation.}
\label{table:TDGLpar}
\end{table}

Here we choose $\lambda = \lambda_L$, the London penetration length measured by the experiment (Supplementary Fig. \ref{fig:S_Pearl}), $\xi$ is chosen to realize similar critical current at zero magnetic field as the experiment at $T=4$ K,when $\eta= -1$ in the normal region, and $W_{1,2}$ are chosen to reproduce the asymmetric features in the experiment. The total length in $y$ direction of the simulated device is $Y=7.5~\mu$m, with the current terminals attached along the edges of $y=\pm Y/2$. The maximum grid edge size is 40 nm. Below the critical current, a steady state solution can be found such that $\frac{\partial \Psi}{\partial t}=0$. Above the critical current, a steady state cannot be found due to JV motion (ac Josephson effect), and is beyond the scope of this work.

\vspace{1cm}
\textbf{-Simulation results}

Below we describe how the results in the main text are obtained.

\begin{enumerate}
    \item \textbf{Overlapping 0- and 1-JV states.} The overlapping JV states are generated using different initial conditions; the initial condition for the 0-JV states is a uniform and real $\Psi$ (Supplementary Fig. \ref{fig:S_TDGL_initial}b); the initial condition for the 1-JV state is a seed solution from the non-overlap part of the 1-JV state (Supplementary Fig. \ref{fig:S_TDGL_initial}c). In the overlap area, the results is either 0- or 1-JV states depending on the initial condition; outside the overlap area, the results is independent of the initial condition. In other words, there are two local energy minima with respect to the spatial configuration of $\Psi$ in the overlap region; and only one local minimum outside the overlap region. Under such a scenario the expectation from thermodynamics is hysteresis in the overlap region, with respect to the direction of the $B_z$ or $I_\mathrm{bias}$ sweep. However, as we pointed out in the main text, this was not observed in the experiment and an open question for future work.
    
    The total Gibbs free energy shown in Fig. \ref{fig:3}c is $\varepsilon = F_{GL} - I_{\mathrm{bias}} \Phi_0 \frac{\phi}{2\pi}$ to account for the bias current. Here $\phi$ is the mean value of $\theta(x)|_{y=y_0} - \theta(x)|_{y=-y_0}$ averaged over $x$. The result shown in Supplementary Fig. \ref{fig:S_TDGL_initial}b-c is taken at $y_0 = 0.5 \mu$m and insensitive to $y_0 \gg W_{1,2}$.

    \begin{figure*}[ht]
    \centering
    \includegraphics[width=5.8in]{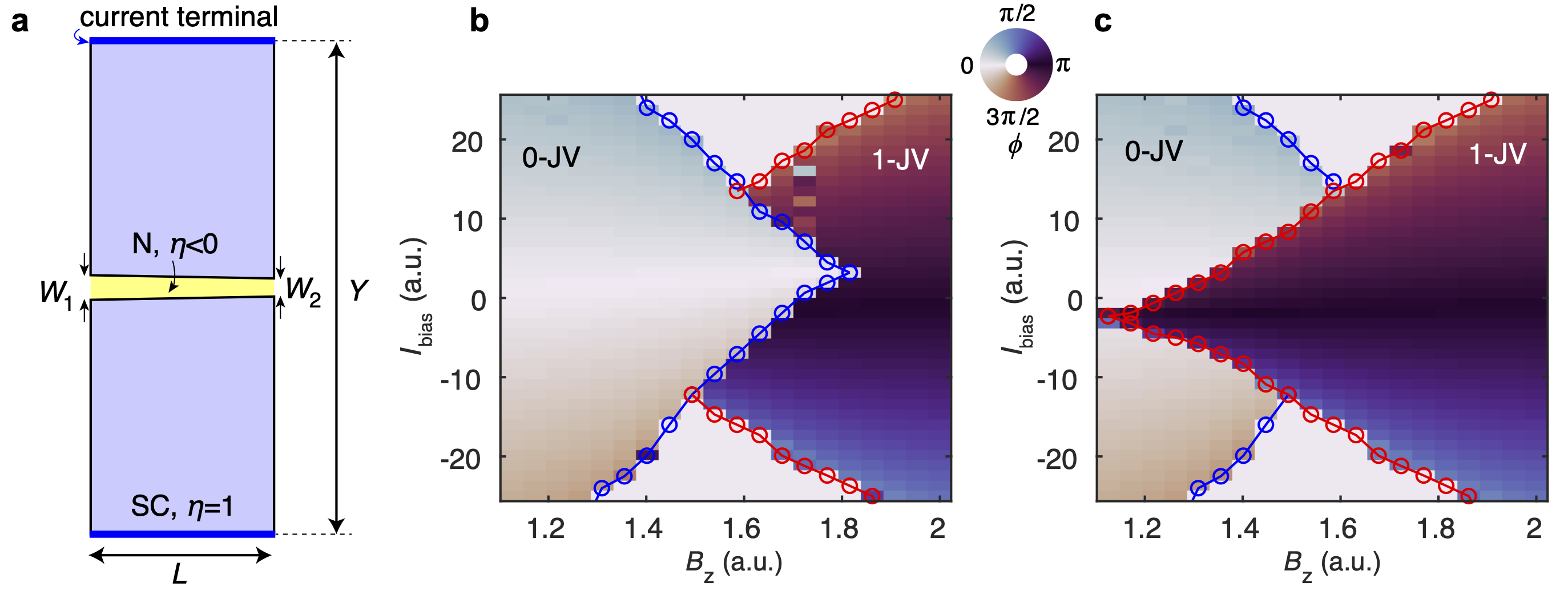} 
    \caption{\textbf{TDGL Simulation schematic and initial conditions}.
    \textbf{a,} Schematic drawing of the simulated device. The drawing is not to scale so as to highlight the junction area in the middle of the device.
    \textbf{b-c,} Phase difference across the junction the \textbf{(b)}0-JV and \textbf{(c)}1-JV state. The circles mark the critical current for the two solutions. The result beyond the critical current is blanked for clarity.
    }
    \label{fig:S_TDGL_initial}
    \end{figure*}
    
    \item \textbf{Varying the critical current.} The critical current of the JJ is tuned by $\eta$ of the normal region. For smaller critical current, such as the case at $T=7$ K, $\eta=-5$. The current induced phase is enhanced when critical current is large ($\eta=-1$). Specifically, in the 0-JV state the transverse current near the JJ, $J_x$ is reduced by the amount of the current flowing across the JJ; in the 1-JV state, however, $J_x$ is enhanced by the vortical current (Supplementary Fig. \ref{fig:S_TDGL_eta}). The phase difference $\phi(x)$ along the JJ in ref. \cite{clem2010josephson} was derived when neglecting the Josephson current across the junction. The change of $J_x$ when including the effect of the Josephson current leads to the current flow induced phase discussed in the main text.

    \begin{figure*}[ht]
    \centering
    \includegraphics[width=5.8in]{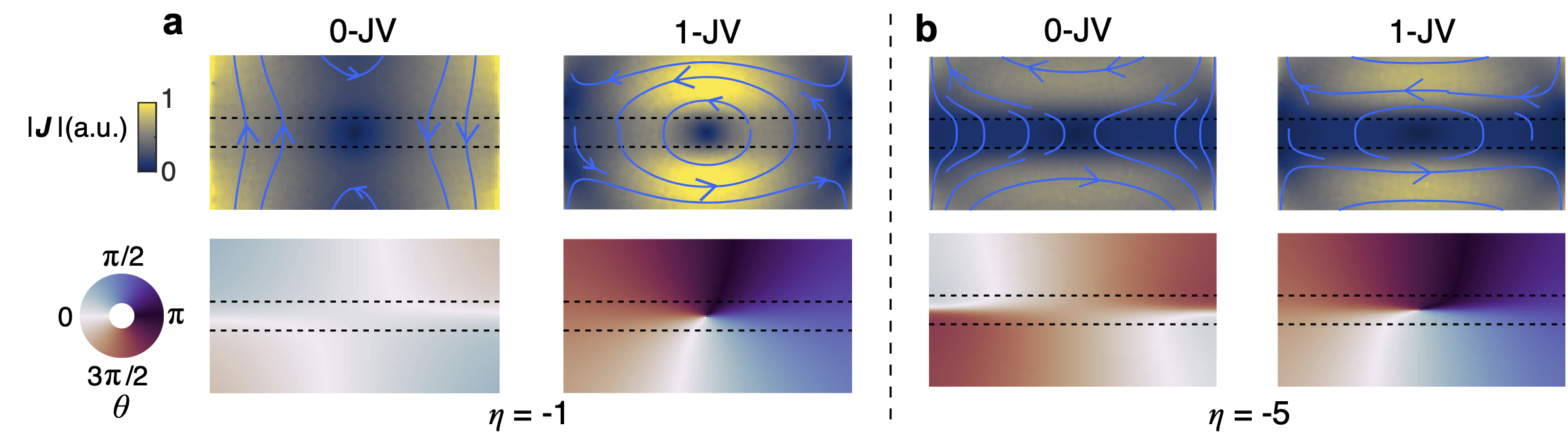} 
    \caption{\textbf{Varying critical current with $\eta$ of the normal region}.
    \textbf{a-b,} Simulated local current density and superconducting phase using \textbf{(a)} $\eta=-1$ and \textbf{(b)} $\eta=-5$ for the normal area of the junction. The simulations here are done in a symmetric junction for clarity, with $W_1=W_2=150$ nm. All cases are simulated at the same external magnetic field, at zero bias and neglecting self-field effect.
    }
    \label{fig:S_TDGL_eta}
    \end{figure*}

    \item \textbf{Effect of coupling to the self-field.} We have investigated the self-field effect in the case $I_{\mathrm{bias}}=0$, by correctly including the self-generated magnetic field in the vector potential felt by the superconductor. 
    The full vector potential can be written as
    $\textbf{A} = \textbf{A}_{\mathrm{ext}}  + \textbf{a}$, where
    \begin{equation}
    \label{afromK}
    \textbf{a} (\textbf{r}) = \frac{\mu_0}{4\pi} \int \frac{\textbf{K}}{ (\textbf{r}')}{|\textbf{r}-\textbf{r}'|} d^2\textbf{r}',
    \end{equation} 
    and $\textbf{k}$ is the sheet current obtained by solving the Ginzburg-Landau equation in the presence of the vector potential $\textbf{A}$.    As an initial step, we find a first order correction $\textbf{a}_0$ by substituting for $\textbf{k}$ in (\ref{afromK}) with the sheet current $\textbf{k}_0$ obtained for $\textbf{A} = \textbf{A}_{\mathrm{ext}} $.
    We then solve the Ginzburg-Landau equation with $\textbf{A} = \textbf{A}_{\mathrm{ext}}  + \textbf{a}_0$, calculate a new value of  $\textbf{k}$, and iterate until convergence is reached.

    We find that the 3D magnetic field energy (last term in Supplementary Eqn. \ref{eqn:GLenergy_SI}) is larger for the 0-JV than the 1-JV state inside the overlap region. However, the total energy difference between the 0- and 1-JV states $\Delta \varepsilon = \varepsilon_0 - \varepsilon_1$ decreases when the self field effect is included. The fact that  larger field energy results in lower total energy can be understood by the following argument. If one constrains $\Psi$ to have the form $\Psi_0 (\textbf{r})$ calculated with $\textbf{A}= \textbf{A}_{\mathrm{ext}}$ and substitutes it in Supplementary Eqn. \ref {eqn:GLenergy_SI}, one finds 
    \begin{equation}
    F_{\mathrm{GL}} = F_{A_{\mathrm{ext}}} - \int d^2 \textbf{r} \,\textbf{k}_0\cdot \textbf{a} + \frac{1}{2\mu_0} \int d^3\textbf{r} \, |\nabla \times \textbf{a}|^{2} + t_{SC}\frac{e^2}{m_e} \int d^2 \textbf{r} \, |\textbf{a}|^2 |\Psi|^2 .
    \label{eqn:GLenergy_sfe1}
    \end{equation}

    For a thin film SC, the second and third terms scale with $t_{SC}^2$, while the last term scales with $t_{SC}^3$ and can be neglected. If we then choose  $ \textbf{a}$  to minimize Supplementary Eqn. \ref{eqn:GLenergy_sfe1}, we find  
    $ \textbf{a}$=$ \textbf{a}_0$, and  the second term on the right-hand side of Supplementary Eqn. \ref{eqn:GLenergy_sfe1} is -2 times the third term. This gives us the approximation
    \begin{equation}
    F_{\mathrm{GL}} \approx F_{A_{\mathrm{ext}}} -  \frac{1}{2\mu_0} \int d^3\textbf{r} \,|\nabla \times \textbf{a}_0|^{2} .
    \label{eqn:GLenergy_1}
    \end{equation}
    This implies that coupling to the self field decreases the total energy by the amount of the 3D magnetic field energy, thereby favoring the 0-JV state over the 1-JV state. 
    
    A more complete calculation would allow for deviations of $\Psi$ from the starting form $\Psi_0$, which will further reduce the total free energy. We do not have a simple expression for this effect, but it seems reasonable to assume that the reduction will also be  greater when  the induced magnetic field is greater. Moreover, we expect that the change in $\Psi$ should be proportional to $\textbf{a}_0$, and the resulting change in the free energy will be of relative order $|\textbf{a}_0|^2$.  From  our numerical calculations, we find that the approximation (Supplementary Eqn. \ref{eqn:GLenergy_1}) gives about twice of the total reduction in energy arising from  the self-field coupling. The result is seen in Supplementary Fig.  \ref{fig:S_TDGL_sfe}a-b.  

    In order to further confirm this analysis, we have performed numerical simulations where we have artificially strengthened the coupling to the magnetic field by decreasing $\lambda$ while holding fixed the parameters $\xi,\, m_e$ and $ \alpha$. As the superfluid density is proportional to $\lambda^{-2}$, we expect the magnetic field energy to scale as $\lambda^{-4}$ while the energy without the self-field effect ($F_\mathrm{{A_{ext}}}$) scales as $\lambda^{-2}$. The results, plotted in Supplementary Fig. \ref{fig:S_TDGL_sfe}c, f, are in accord with this expectation. Finally, coupling to the self-field decreases the energy for the 0-JV state more than the 1-JV state (Supplementary Fig. \ref{fig:S_TDGL_sfe}d-f). This could be viewed as a result of the 0-JV state having larger magnetic field energy than the 1-JV state.

    \begin{figure*}[h!]
    \centering
    \includegraphics[width=6.5in]{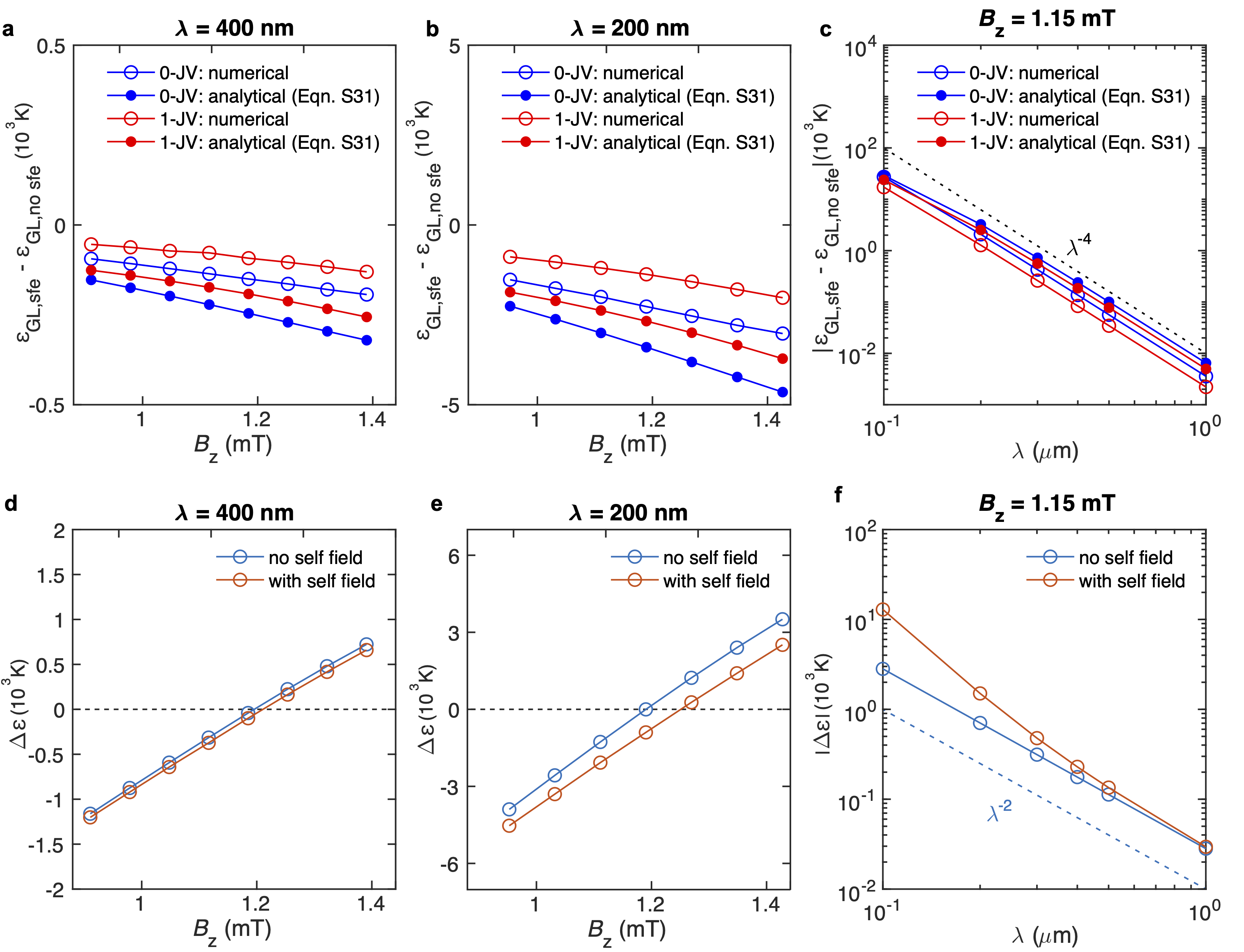} 
    \caption{\textbf{Energy difference including the self-field effect}.
    \textbf{a-b,} The energy difference of solutions with or without coupling to the self field, versus external magnetic field $B_z$. Results are taken at \textbf{(a)} $\lambda=400$ nm, and \textbf{(b)} $\lambda=200$ nm. The blue (red) colour points represent the 0- (1-) JV. The empty points are numerical results obtained by iterating both $\Psi$ and $\textbf{a}$. The filled points are results given by the approximate analytical formula Supplementary Eqn. \ref{eqn:GLenergy_1}, where the energy shift from coupling to the self field is the negative of the 3D magnetic field energy. 
    \textbf{c,} Absolute value of the energy difference of solutions with or without coupling to the self-field, plotted versus of $\lambda$. In particular, in the analytical approximation given by Supplementary Eqn. \ref{eqn:GLenergy_1}, the energy difference is the magnetic field energy which is expected to scale with $\lambda ^{-4}$ (shown by the dashed line as a guide for the eye).
    \textbf{d-e,} Energy difference of the 0 and 1-JV states $\Delta \varepsilon = \varepsilon_{0} - \varepsilon_{1}$ if the self field effect is included for \textbf{(d)} $\lambda=400$ nm, and \textbf{(e)} $\lambda=200$ nm. Smaller $\lambda$ represents larger self field effect.
    \textbf{f,} $|\Delta \varepsilon|$ as a function of $\lambda$ at fixed external magnetic field. The total energy difference between 0- and 1-JV states without coupling to the self-field is expected to scale with $\lambda^{-2}$ (shown by the dashed line as a guide for the eye).
    Panels C and F are simulated at $B_z=1.15$ mT.
    The simulations are done in a symmetric junction with $W_1 = W_2 = 150$ nm, and the external dc bias is zero.
    }
    \label{fig:S_TDGL_sfe}
    \end{figure*}

\end{enumerate}

\clearpage

\section{Simulations for an asymmetric geometry}

We present in Supplementary Fig. \ref{fig:S_asymmACDC} results of simulations comparing symmetric and asymmetric geometries.

\begin{figure*}[h]
\centering
\includegraphics[width=5.9in]{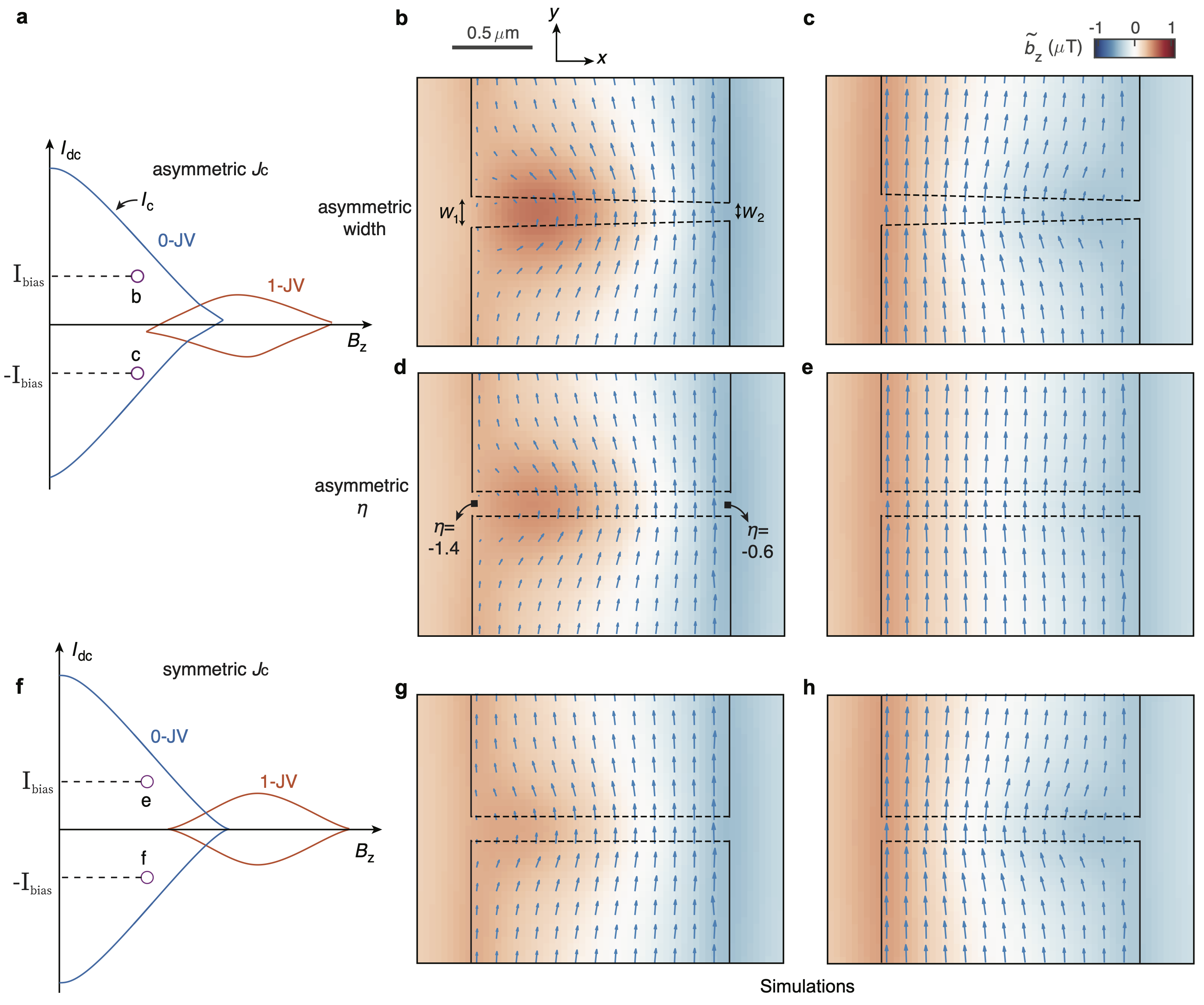} 
\caption{\textbf{TDGL simulations on JJs with symmetric or asymmetric geometry.}
\textbf{a, f,} Schematic drawing of critical current of 0- and 1-JV states versus external magnetic field $B_z$, for \textbf{(a)} inversion-asymmetric, and \textbf{(d)} inversion-symmetric junctions. 
\textbf{b-e,} Simulated current flow and $z$-direction of magnetic field from a differential measurement. The maps are taken in an asymmetric JJ with non-uniform critical current density along $x$ direction. \textbf{(b)-(c)} are simulated with tilted junction width $W_1>W_2$. \textbf{(d)-(e)} are simulated with non-uniform $\eta$ factor in the normal region ($\eta$ changes linearly with $x$ from -1.4 to -0.6). The dc bias current are symmetric with zero as shown in \textbf{(a)}. \textbf{(b)-(c)} are the same as Fig. \ref{fig:4}e-f in the main text.
\textbf{g-h,} Simulated current flow and z-direction of magnetic field from a symmetric JJ with uniform critical current density along $x$ direction. The dc bias current are shown in \textbf{(f)}. Current flow is inversion symmetric for $\pm I_\mathrm{bias}$ when inversion symmetry of the JJ is preserved.
}
\label{fig:S_asymmACDC}
\end{figure*} 
\clearpage

\end{document}